\documentclass[11pt]{article}
\usepackage{epsfig}
\usepackage{amstex}
\numberwithin{equation}{section}

\title{The geometry of spinors and the multicomponent BKP and DKP hierarchies}
\author{Victor Kac\footnote{VK is partially supported by NSF grant
DMS-9622870.}\\
\ \\
Department of Mathematics,\\
Massachusetts Institute of Technology\\
Cambridge, MA 02139\\
USA \and
Johan van de Leur\footnote{JvdL is financially supported by the
Netherlands Organization for Scientific Research (NWO).}\\
\ \\
Faculty of Applied Mathematics,\\
University of Twente\\
P.O. Box 217,
7500 AE Enschede\\
The Netherlands}

\date{}

\newtheorem{lemma}{Lemma}[section]
\newtheorem{proposition}{Proposition}[section]
\newtheorem{theorem}{Theorem}[section]

\newtheorem{remark}{Remark}[section]
\newtheorem{corollary}{Corollary}[section]

\begin{document}
\maketitle
\begin{abstract}
We develop a formalism of multicomponent BKP hierarchies using
elementary geometry of spinors. The multicomponent KP and the modified
KP hierarchy (hence all their reductions like KdV, NLS, AKNS or DS)
are reductions of the multicomponent BKP.
\end{abstract}

\section{Introduction}

The remarkable connection between the KP hierarchy and representation
theory was discovered in the early 1980's by Sato~\cite{Sa} and further
developed by Date, Jimbo, Kashiwara and Miwa~\cite{DJKM1}, ~\cite{DJKM2} using
the
spinor formalism and the classical boson-fermion correspondence. We
will now describe this in some detail.

Consider the vector space $V'=V^+\oplus V^-$, where
$V^\pm=\bigoplus_{i\in\mathbb{Z}+\frac{1}{2}}\mathbb{C}\psi_i^\pm $
with symmetric bilinear form
\begin{equation}
\label{0.-1}
(\psi^\pm_i,\psi_j^\pm)=0,\quad(\psi^\pm_i,\psi_j^\mp)=\delta_{i,-j}.
\end{equation}
Let $C\ell\; V'$ be the associated Clifford algebra with relations
($u,v\in V'$):
\begin{equation}
\label{0.0}
uv+vu=(u,v)1.
\end{equation}
Consider its spin representation in a vector space $F$ with vacuum
vector $|0\rangle$ such that
\begin{equation}
\label{0.-2}
\psi_j^\pm|0\rangle=0\quad\mbox{for }j>0.
\end{equation}
By introducing the notion of charge as follows:
\[
\mbox{charge}\;\;|0\rangle=0,\quad \mbox{charge}\;\;\psi_j^\pm=\pm 1,
\]
the space $F$ decomposes into charge sectors
\[
F=\bigoplus_{k\in\mathbb{Z}}F_k.
\]
The group $\overline{\mbox{Spin}\; V'}$ generated by all
elements of the form $1+a\psi_i^+\psi_j^-$ ($a\in\mathbb{C}$) acts on
$F$ and preserves the charge sectors.

The operators
\[
S^\pm=\sum_{j\in\mathbb{Z}+\frac{1}{2}}\psi_j^\pm \otimes
\psi_{-j}^\mp,
\]
commute with the diagonal action of $\overline{\mbox{Spin}\; V'}$
on $F\otimes F$. Since $S^\pm(|0\rangle\otimes|0\rangle)=0$, an element
$\tau$ of
$\overline{O_0}=\overline{\mbox{Spin}\; V'}|0\rangle$, the group
orbit of the vacuum vector, satisfies equations
\begin{equation}
  \label{0.1}
S^\pm(\tau\otimes\tau)=\sum_{j\in\mathbb{Z}+\frac{1}{2}}\psi_j^\pm\tau\otimes
\psi_{-j}^\mp\tau=0.\tag*{$(\ref{0.1})_\pm$}
\end{equation}
\addtocounter{equation}{1}

One can
show that equation $(\ref{0.1})_+$  and $(\ref{0.1})_-$ are equivalent
and that the converse also holds: if $\tau\ne 0$ satisfies
$(\ref{0.1})_+$, then $\tau\in \overline{O_0}$ (see~\cite{KacRaina} for a
proof). The same holds for $(\ref{0.1})_-$. Each of the equivalent
equations $(\ref{0.1})_+$  and $(\ref{0.1})_-$ is called the KP
hierarchy.

The orbit $\overline{O_0}$ can be identified with an algebraic
version of Sato's infinite Grassmannian. Let for $k\in\mathbb{Z}$,
$U^\pm_k=\bigoplus_{j>k}\mathbb{C}\psi^\pm_j$, the Grassmannian
$Gr^\pm$ consists of all subspaces $W^\pm \subset V^\pm$ such that for
all $k>>0$ $U^\pm_k\subset W^\pm$ and $\dim\;W^\pm/U^\pm_k=k$.
Let $\tau\in \overline{O_0}$, then one can identify the corresponding
$W^\pm_\tau\in Gr^\pm$ as follows;
\[
W^\pm_\tau=\{v\in V^\pm |v\tau=0\}.
\]

Recall the classical boson-fermion correspondence~\cite{KacRaina}.
Consider the following generating series,  called {\it charged fermionic
fields}:
\begin{equation}
  \label{2.1}
  \psi^{\pm}{(z)} = \sum_{i\in \frac{1}{2} + \mathbb{Z}} \psi^{\pm}_i
  z^{-i -\frac{1}{2}} .
\end{equation}
Then we have:
\begin{equation}
  \label{2.2}
  \psi^{\lambda} (y) \psi^{\mu} (z) + \psi^{\mu} (z) \psi^{\lambda}
  (y) = \delta_{\lambda,-\mu}\delta (y-z),\qquad\lambda,\mu=\pm,
\end{equation}
where
\[
\delta (y-z) = y^{-1} \sum_{n\in \mathbb{Z}} \left(\frac{y}{z} \right)^n.
\]

We split up the fields $\psi^{\pm}{(z)}$ in its positive and negative part:
\[
\begin{array}[h]{rcl}
\psi^{\pm}{(z)} &=& \psi^{\pm}(z)_+ + \psi^{\pm}(z)_-,\;
\;\;\mbox{where}\\[3mm]
\psi^{\pm}(z)_+ = \sum_{i\in \mathbb{Z}_+}
\psi^{\pm}_{-i-\frac{1}{2}} z^i&\mbox{and}&
\psi^{\pm}(z)_- = \psi^{\pm}(z) - \psi^{\pm}(z)_+ ,
\end{array}
\]
so that
\[
\psi^{\pm}(z) |0\rangle  = \psi^{\pm}(z)_+ |0\rangle .
\]
Define the bosonic fields
\begin{equation}
  \label{2.3}
  \alpha (z) = \sum_{k\in \mathbb{Z}} \alpha_k z^{-k-1} = : \psi^{+}
    (z) \psi^- (z):
\end{equation}
where the normally ordered product is defined, as usual, by
\[
: \psi^{\lambda} (y) \psi^{\mu} (z) : = \psi^{\lambda} (y)_+
\psi^{\mu} (z) - \psi^{\mu} (z) \psi^{\lambda} (y)_-.
\]
Then one has (using Wick's formula):
\begin{equation}
  \label{2.6}
  \begin{array}{lcl}
  [\psi^{\lambda} (y), \alpha (z)] &=&\displaystyle -\lambda
\delta(y-z),\qquad\lambda=\pm,
\\[3mm]
  [\alpha (y), \alpha(z)] &=&\delta^\prime_z (y-z).
  \end{array}
\end{equation}

Thus, the $\alpha_m$ form the oscillator algebra,
\[
[\alpha_m , \alpha_n ] = m \delta_{m, - n}, \;\;\mbox{with}\;\;
\alpha_k |0\rangle  = 0\;\;\mbox{for}\;\;k\geq 0 .
\]

The operator  $\alpha_0$ is diagonalizable in $F$, with eigenspaces the
{\it charge sectors} $F_k$, i.e. $\alpha_0 f_k = kf_k$ for $f_k \in
F_k$. For that reason we call $\alpha_0$ the {\it charge operator}.

In order to express the fermionic fields $\psi^{\lambda} (z)$ in terms
of the oscillator algebra, we need an additional operator $Q$ on $F$
defined by
\[
Q |0\rangle  = \psi^+_{-\frac{1}{2}} |0\rangle , \;\; Q\psi^+_k =
\psi^{\pm}_{k\mp 1} Q.
\]

\begin{theorem}
  \label{t2.1}(~\cite{DJKM1}, ~\cite{JM}, ~\cite{K2})
\[
\psi^{\pm} (z) = Q^{\pm 1} z^{\pm \alpha_0} \exp (\mp \sum_{k<0}
\frac{\alpha_k}{k} z^{-k}) \exp (\mp \sum_{k>0} \frac{\alpha_k}{k} z^{-k}).
\]
\end{theorem}

We now identify $F$ with the space $B=\mathbb{C} [q,q^{-1}, t_1, t_2 ,
\ldots ]$ via the vector space homorphism
\[
\sigma : F \tilde{\rightarrow} B
\]

given by

\[
\sigma (\alpha_{-m_1} \ldots \alpha_{-m_s} Q^k |0\rangle  ) = m_1 m_2 \ldots
m_s
t_{m_1} t_{m_2} \ldots t_{m_s} q^k .
\]

The transported charge is as follows
\[
\mbox{charge }(p(t)q^k)=k,
\]
and the transported charge decomposition is
\[
\displaystyle
B=\bigoplus_{m\in \mathbb{Z}} B_m, \;\;\mbox{where}\;\; B_m = \mathbb{C}
[t_1 , t_2 \ldots ]q^m .
\]

The transported action of the operators $\alpha_m$ and $Q$ on $B$ is as
follows:
\[
\begin{array}[h]{lcl}
\sigma \alpha_{-m}\sigma^{-1} (p(t) q^k) &=& mt_m p(t) q^k,\\[3mm]
\sigma \alpha_m \sigma^{-1} (p(t) q^k) &=& \frac{\partial p(t)}{\partial t_m}
q^k,\\[3mm]
\sigma \alpha_0 \sigma^{-1} (p(t) q^k) &=& q \frac{\partial}{\partial q} (p(t)
q^k) = k p(t) q^k,\\[3mm]
 \sigma Q \sigma^{-1} (p(t) q^k) &=& p(t) q^{k+1}.
\end{array}
\]

Hence
\[
 \sigma\psi^{\pm} (z) \sigma^{-1} = q^{\pm 1} z^{\pm q\frac{\partial}{\partial
    q}} e^{\pm \xi (t,z)} e^{\mp \eta(t,z)},
\]

where
\[
\xi (t,z) = \sum^{\infty}_{k=1} t_k z^k \quad \mbox{and}\quad \eta(t,z)
= \sum^{\infty}_{k=1} \frac{1}{k} \frac{\partial}{\partial t_k} z^{-k} .
\]

We now rewrite equations $(\ref{0.1})_\pm$ in the bosonic picture. First
notice that $(\ref{0.1})_\pm$ is equivalent to
\begin{equation}
\label{0.2}
\mbox{Res}_{z=0}dz\;\psi^\pm(z)
\tau\otimes\psi^\mp(z)\tau=0.\tag*{$(\ref{0.2})_\pm$}
\end{equation}
\addtocounter{equation}{1}

Here and further $\mbox{Res}_{z=0}dz\;\sum f_iz^i$, where the $f_i$
are independent of $z$, stands for $f_{-1}$. Now think of $B_0\otimes
B_0$ as the polynomial algebra
$\mathbb{C}[t'_1,t''_1,t'_2,t''_2,\ldots]$.
Then applying $\sigma$ to both sides of $(\ref{0.2})_\pm$, we obtain a
system of equations describing the orbit $\sigma(\overline{O_0})$:
\begin{equation}
\label{0.3}
\mbox{Res}_{z=0}dz\;e^{\mp\eta(t',z)}\tau(t')e^{\pm\xi(t',z)}
e^{\pm\eta(t'',z)}\tau(t'')e^{\mp\xi(t'',z)}=0.
\tag*{$(\ref{0.3})_\pm$}
\end{equation}
\addtocounter{equation}{1}
Divide $(\ref{0.3})_\pm$ by $\tau(t')\tau(t'')$, one obtains the bilinear
identity
\[
\mbox{Res}_{z=0}dz\;\Psi^\pm(t',z)\Psi^\mp(t'',z)=0,
\]
for the KP wave function $\Psi^+(t,z)$ and its adjoint
$\Psi^-(t,z)$ defined by
\[
\Psi^\pm(t,z)={{e^{\mp\eta(t,z)}\tau(t)}\over
{\tau(t)}}e^{\pm\xi(t,z)}.
\]

Instead of dividing by $\tau(t')\tau(t'')$, one can make a change of
variables
\begin{equation}
\label{0.3a}
t_k=\frac{1}{2}(t'_k+t''_k),\quad s_k=\frac{1}{2}(t'_k-t''_k).
\end{equation}
Then $(\ref{0.3})_\pm$ becomes
\begin{equation}
\label{0.4}
\mbox{Res}_{z=0}dz\;e^{\pm\xi(2s,z)}e^{\mp\eta(s,z)}\tau(s+t)\tau(t-s)=0.
\tag*{$(\ref{0.4})_\pm$}
\end{equation}
\addtocounter{equation}{1}

Introducing the elementary Schur functions
$P_k(t),\ k\in\mathbb{Z}$ by
\begin{equation}
\label{0.5}
\sum_{k\in\mathbb{z}}P_k(t)z^k=e^{\xi(t,z)},
\end{equation}
we rewrite $(\ref{0.4})_\pm$ as follows
\begin{equation}
\label{0.6}
\sum_{j=0}^\infty P_j(\pm 2s)P_{j+1}(\mp \tilde{\frac{\partial}{\partial s}})
\tau(s+t)\tau(t-s)=0,
\tag*{$(\ref{0.6})_\pm$}
\end{equation}
\addtocounter{equation}{1}
where
\[
\tilde{\frac{\partial}{\partial t}}=(\frac{\partial}{\partial t_1},
\frac{1}{2}\frac{\partial}{\partial
t_2},\frac{1}{3}\frac{\partial}{\partial t_3},
\ldots).
\]
Using Taylors formula we rewrite $(\ref{0.6})_\pm$ once more:
\begin{equation}
\label{0.7}
\sum_{j=0}^\infty P_j(\pm 2s)P_{j+1}(\mp \tilde{\frac{\partial}{\partial u}})
e^{\sum_{r=1}^\infty s_r\frac{\partial}{\partial u_r}}
\tau(t+u)\tau(t-u)|_{u=0}=0,
\tag*{$(\ref{0.7})_\pm$}
\end{equation}
\addtocounter{equation}{1}
which is a generating series of Hirota bilinear equations for the KP
hierarchy:
\begin{equation}
\label{0.8}
\sum_{j=0}^\infty P_j(\pm 2s)P_{j+1}(\mp\tilde{D})
e^{\sum_{r=1}^\infty s_rD_r}
\tau\cdot\tau=0.
\tag*{$(\ref{0.8})_\pm$}
\end{equation}
\addtocounter{equation}{1}

Recall that for a polynomial $p$, the corresponding Hirota bilinear
equation on $\sigma$ and $\rho$ is
\[
p(D)\sigma\cdot\rho=p(\frac{\partial}{\partial u})\sigma
(t+u)\rho(t-u)|_{u=0}=0.
\]
Calculating the coefficient of $s_3$ in $(\ref{0.8})_+$ one obtains
\begin{equation}
\label{0.9}
(D_1^4-4D_1D_3+3D_2^2)\tau\cdot\tau=0.
\end{equation}
Putting $x=t_1,\ y=t_2,\ t=t_3$ and
$u(x,y,t)=(2\log \tau(x,y,t,t_4,t_5,\ldots))_{xx}$, where
$t_4,t_5,\dots$ are viewed as parameters, then (\ref{0.9}) turns into
\[
\frac{3}{4}u_{yy}=(u_t-\frac{3}{2}uu_x-\frac{1}{4}u_{xxx})_x,
\]
which is the classical Kadomtsev-Petviashvili (KP)equation (hence the
name KP hierarchy).

We return now to $\tau\in\overline{O_0}$. Instead of considering only
$W^+_\tau\in Gr^+$
or
$W^-_\tau\in Gr^-$
one could consider $W_\tau=W^+_\tau\oplus W^-_\tau$. Notice that this
is a maximal isotropic subspace of $V'$. So instead of considering
$Gr^+$ one could consider another Grassmannian $Gr$, consisting of all
maximal isotropic subspaces $W\subset V'$ such that $U^\pm_k\subset W$
for $k>>0$. This Grassmannian should parametrize some other group
orbit. In order to describe this group, it will be convenient to add
one more basis element to $V'$. Let
$V=V'\oplus\mathbb{C}\psi_0$, with bilinear form given by (\ref{0.-1})
and
\[
(\psi_0,V')=0,\quad (\psi_0,\psi_0)=1.
\]
Let $C\ell\; V$ be the corresponding Clifford algebra, with relations
(\ref{0.0}),but now
for $u,v\in V$.
The  spin representation of $C\ell\; V$ in the vector space $F$ is given by
(\ref{0.-2}) and
\[
\psi_0|0\rangle=-\frac{1}{\sqrt 2}|0\rangle.
\]
Since $\psi_0^2=\frac{1}{ 2}$, the eigenvalues of $\psi_0$ on $F$
are $-\frac{1}{\sqrt 2}$ and $\frac{1}{\sqrt 2}$. Let
$F_{\overline{0}}$ and $F_{\overline{1}}$ be the corresponding eigenspaces.
Then $F_{\overline{\nu}}=\bigoplus_{k\in\nu+2\mathbb{Z}} F_k$.
Let $\mbox{Spin}\; V$ (resp $\mbox{Spin}\; V'$) be the spin group
generated by all invertible elements of the form $1+auv$, $a\in\mathbb{C}$,
$u,v\in V$ (resp. $u,v\in V'$) (see Section 2 for more details). Then clearly
$F$ (resp. $F_{\overline{\nu}}$) is a $\mbox{Spin}\; V$-
(resp $\mbox{Spin}\; V'$-)module. Let
$|1\rangle=\psi_{-\frac{1}{2}}|0\rangle$ and let
$O=\mbox{Spin}\; V|0\rangle$, (resp.
 $O_{\overline{\nu}}=\mbox{Spin}\; V'|\nu\rangle$)
be
the $\mbox{Spin}\; V$-
(resp $\mbox{Spin}\; V'$-) orbit of $ |0\rangle$ (resp.
$|\nu\rangle$). Then
$O_{\overline{\nu}} \subset O\cap F_{\overline{\nu}}$.

To a $\tau\in O$ there corresponds a point
$W_\tau\in
Gr$ given by
\[
W_\tau=\{v\in V|v\tau=0\},
\]
called the annihilator space of $\tau$.
Let
\[
S'=S^++S^-\quad\mbox{and}\quad S=\psi_0\otimes\psi_0+S'.
\]
One of the main observations of this paper is the following theorem,
proved in \S 2 using some simple arguments on geometry of spinors:
\begin{theorem}
\label{t1.8}
  \begin{enumerate}
  \item[(a)] If $\tau \in F$ and $\tau\ne 0$, then $\tau \in O$ if only if
$\tau$
    satisfies the equation
    \begin{equation}
      \label{1.9}
      S(\tau \otimes \tau) = \frac{1}{2} \tau \otimes \tau.
    \end{equation}
   \item[(b)] If $\tau \in F_{\overline{\nu}}$
($\nu\in\mathbb{Z}/2\mathbb{Z}$) and $\tau\ne 0$, then $\tau\in
     O_{\overline{\nu}}$ if and only if $\tau$ satisfies the equation
     \begin{equation}
       \label{1.10}
       S' (\tau \otimes \tau )=0.
     \end{equation}
  \end{enumerate}
\end{theorem}
Equation (\ref{1.9}) (resp. (\ref{1.10})) is called the fermionic BKP (resp.
DKP)
hierarchy.
As a consequence we obtain
\begin{corollary}
\label{c0.1}
Let $\tau\in F_k$ ($k\in\mathbb{Z}$), then $\tau$ satisfies the KP
hierarchy,i.e. equation
$(\ref{0.1})_\pm$,
if and only if $\tau$ satisfies the DKP (or BKP) hierarchy.
\end{corollary}

If we split up $\tau\in O$ into eigenspaces for $\psi_0$,
i.e., $\tau=\tau_0+\tau_1$ with $-\sqrt 2\psi_0\tau_\nu=(-)^\nu
\tau_\nu$, then one has
\begin{theorem}
  \label{t1.10}
Let $\tau\in F,\ \tau = \tau_0 + \tau_1$, where $\tau_{\nu} \in
F_{\overline{\nu}}$. Then $\tau$ satisfies the fermionic BKP hierarchy if and
only if both
$\tau_0$ and
$\tau_1 $ satisfy the DKP hierarchy (\ref{1.10}) and the Modified DKP
hierarchy:
\begin{equation}
\label{1.20}
S'(\tau_0\otimes\tau_1)=\tau_1\otimes\tau_0.
\end{equation}
\end{theorem}

As in the case of the KP hierarchy, we will again use the boson-fermion
correspondence to rewrite the BKP hierarchy in the bosonic picture.
Notice first that (\ref{1.9}) is equivalent to
\begin{equation}
\label{0.10}
\mbox{Res}_{z=0}dz\;\frac{1}{z}\psi_0\tau\otimes\psi_0\tau +
\psi^+(z)\tau\otimes\psi^-(z)\tau
-\psi^-(-z)\tau\otimes\psi^+(-z)\tau=\frac{1}{2}\tau\otimes\tau.
\end{equation}
Now apply $\sigma$ to (\ref{0.10}). Writing
$\sigma(\tau)=\sum_{n\in\mathbb{Z}}\tau_n(t)q^n$, we obtain
\begin{equation}
\label{0.11}
\begin{split}
\mbox{Res}&_{z=0}dz\; \sum_{n,m\in\mathbb{Z}}\frac{1}{2z}(-)^n\tau_n(t')(q')^n
(-)^m\tau_m(t'')(q'')^m\\[2mm]
&
+z^n e^{\xi(t',z)}e^{-\eta(t',z)}\tau_n(t')(q')^{n+1}
z^{-m} e^{-\xi(t'',z)}e^{\eta(t'',z)}\tau_m(t'')(q'')^{m-1}\\[2mm]
&
-(-z)^{-n} e^{-\xi(t',-z)}e^{\eta(t',-z)}\tau_n(t')(q')^{n-1}
(-z)^{m} e^{\xi(t'',-z)}e^{-\eta(t'',-z)}\tau_m(t'')(q'')^{m+1}\\[2mm]
&=\frac{1}{2}\sum_{n,m\in\mathbb{Z}}\tau_n(t')(q')^n\tau_m(t'')(q'')^m.
\end{split}
\end{equation}
Taking the coefficient of $(q')^n(q'')^m$, we find for all
$n,m\in\mathbb{Z}$ the following equation:
\begin{equation}
\label{0.12}
\begin{split}
\mbox{Res}&_{z=0}dz\;\frac{1}{2z}(-)^n\tau_n(t')
(-)^m\tau_m(t'')\\[2mm]
&+z^{n-1} e^{\xi(t',z)}e^{-\eta(t',z)}\tau_{n-1}(t')
z^{-m-1} e^{-\xi(t'',z)}e^{\eta(t'',z)}\tau_{m+1}(t'')\\[2mm]
&-(-z)^{-n-1} e^{-\xi(t',-z)}e^{\eta(t',-z)}\tau_{n+1}(t')
(-z)^{m-1} e^{\xi(t'',-z)}e^{-\eta(t'',-z)}\tau_{m-1}(t'')\\[2mm]
&=\frac{1}{2}\tau_n(t')\tau_m(t'').
\end{split}
\end{equation}
Now divide by $\tau_p(t')\tau_q(t'')$ for $p=n-1,n,n+1,\ q=m-1,m,m+1$
and view the obtained result as products of rows and columns of
$3\times 3$-matrices. In other words, let
\begin{equation}
\label{0.13}
\begin{array}[h]{lcl}
S(z)&=&\mbox{diag }(-1,\frac{1}{2z},1),\\[2mm]
Q^\pm(t,z)&=&\mbox{diag }(e^{\mp\xi(t,z)},1,e^{\pm\xi(t,z)}),\\[2mm]
R^\pm(n,z)&=&\mbox{diag }((-z)^{\mp n}, (-)^n,z^{\pm n}),\\[2mm]
\end{array}
\end{equation}
\begin{equation}
\label{0.14}P^\pm(n,t,z)=
\frac{1}{\tau_n(t)}
\left(
  \begin{array}[h]{ccc}
e^{\pm\eta(t,-z)}\tau_n(t)&
-\tau_{n\mp 1}(t)&
z^{-2}e^{\mp\eta(t,z)}\tau_{n\mp 2}(t)\\[2mm]
(-z)^{-1}e^{\pm\eta(t,-z)}\tau_{n\pm 1}(t)&
\tau_{n}(t)&
z^{-1}e^{\mp\eta(t,z)}\tau_{n\mp 1}(t)\\[2mm]
z^{-2}e^{\pm\eta(t,-z)}\tau_{n\pm 2}(t)&
-\tau_{n\pm 1}(t)&
e^{\mp\eta(t,z)}\tau_{n}(t)\\[2mm]
   \end{array}
\right),
\end{equation}
\begin{equation}
\label{0.15}
T^\pm(n,t,z)=
\left(
  \begin{array}[h]{ccc}
1&\tau_{n\mp 1}/\tau_n(t)&0\\[2mm]
0&1&0\\[2mm]
0&\tau_{n\pm 1}/\tau_n(t)&1\\[2mm]
\end{array}
\right)
\end{equation}
and put
\begin{equation}
\label{0.16}
\Psi^\pm(n,t,z)=P^\pm(n,t,z)R^\pm(n,z)Q^\pm(t,z),
\end{equation}
then (\ref{0.12}) is equivalent to
\begin{equation}
\label{0.17}
\mbox{Res}_{z=0}dz\;\Psi^+(n,t',z)S(z)\;^t\Psi^-(m,t'',z)=
\mbox{Res}_{z=0}dz\;T^+(n,t',z)S(z)\;^tT^-(m,t'',z).
\end{equation}
We can rewrite this once more to get a nicer formula. Define BKP
wave functions by
\begin{equation}
\label{0.18}
\Phi^\pm(n,t,z)=T^\pm(n,t,z)^{-1}\Psi^\pm(n,t,z),
\end{equation}
then (\ref{0.17}) turns into
\begin{equation}
\label{0.19}
\mbox{Res}_{z=0}dz\;\Phi^+(n,t',z)S(z)\;^t\Phi^-(m,t'',z)=
\mbox{diag }(0,\frac{1}{2},0).
\end{equation}
{}From the formulas (\ref{0.14}), (\ref{0.15}) and (\ref{0.18}), it is
easy to determine that $\Phi^-(n,t,z)$ is completely determined by
$\Phi^+(n,t,z)$. Using the bilinear identity (\ref{0.19}) one can in
fact show that $\Phi^+(n,t,z)$ determines its neighbors $\Phi^+(n\pm
1,t,z)$. The DKP bilinear identity on the wave functions is obtained
from (\ref{0.19}) by removing the second row and column from all
matrices (\ref{0.13})-(\ref{0.15}).

Differentiate (\ref{0.19}) by $t_\ell'$, then using
Lemma \ref{l4.1} of this paper, we obtain the Sato
equation of this BKP hierarchy. Fix $n\in\mathbb{Z}$, let
$\partial=\frac{\partial}{\partial t_1}$ and let

\[
\begin{array}[h]{lcl}
N(\ell,\partial)&=&\mbox{diag }(-(-\partial)^\ell,0,\partial^\ell),\\
K(\partial)&=&T^+(n,t,\partial)^{-1}P^+(n,t,\partial)=\sum_{j=0}^\infty
K^{(j)}\partial^{-j},
\end{array}
\]
then $K^{(j)}_{i2}=0$ for $i=1,2,3$ and $j=1,2,\ldots$.
One finds the following Sato equation:
\begin{equation}
\label{0.19a}
\frac{\partial K(\partial)}{\partial t_\ell}=
-(K(\partial)N(\ell,\partial)K(\partial)^{-1}S(\partial))_+
S(\partial)^{-1}K(\partial).
\end{equation}

We now return to equation (\ref{0.12}) and rewrite it into a
generating series of Hirota bilinear equations. Make the same change
of variables as before, viz. (\ref{0.3}) and use the elementary Schur
functions defined by (\ref{0.5}). Then (\ref{0.12}) is equivalent to
\begin{equation}
\label{0.20}
\begin{split}
\frac{1}{2}((-)^{n+m}-1)&\tau_n(s+t)\tau_m(t-s)+
\sum_{j=0}^\infty P_j(2s)P_{j+n-m-1}(-\tilde{\frac{\partial}{\partial s}})
\tau_{n-1}(s+t)\tau_{m+1}(t-s)+\\
\ &P_j(-2s)P_{j+m-n-1}(\tilde{\frac{\partial}{\partial s}})
\tau_{n+1}(s+t)\tau_{m-1}(t-s)
=0.
\end{split}
\end{equation}
Using Taylor's formula this turns into the following generating series
of Hirota bilinear equations:
\begin{equation}
\label{0.21}
\begin{split}
\frac{1}{2}((-)^{n+m}-1)&
e^{\sum_{r=1}^\infty s_rD_r}\tau_n\cdot\tau_m+
\sum_{j=0}^\infty P_j(2s)P_{j+n-m-1}(-\tilde D)e^{\sum_{r=1}^\infty s_rD_r}
\tau_{n-1}\cdot\tau_{m+1}+\\
\ &P_j(-2s)P_{j+m-n-1}(\tilde D)e^{\sum_{r=1}^\infty s_rD_r}
\tau_{n+1}\cdot\tau_{m-1}
=0.
\end{split}
\end{equation}
The simplest Hirota bilinear equations of the charged BKP
hierarchy, appear in the constant term when we take $m-n=3$ and $4$
($m-n=0,1,2$ give trivial equations):
\begin{equation}
\begin{split}
(D_1^2+D_2)\tau_{n+1}\cdot\tau_{n+2}=2\tau_n\tau_{n+3},\\[2mm]
(D_1^3+3D_1D_2+2D_3)\tau_n\cdot\tau_{n+2}=0.
\end{split}
\end{equation}
The coefficient of $s_3$ with $n-m=2$ gives:
\begin{equation}
(D_1^4-4D_1D_3+3D_2^2)\tau_n\cdot \tau_n=24\tau_{n+2}\tau_{n-2}.
\end{equation}

Another BKP hierarchy, which is natural to call the neutral BKP
hierarchy, was introduced in ~\cite{DJKM2}. It uses neutral fermions
and the neutral boson-fermion correspondence (cf. \S 3.2 of the
present paper).

In this paper we consider the general situation of $n$ charged
fermions and $n'$ neutral fermions. Using the multicomponent
boson-fermion correspondence we derive the general multicomponent BKP
and DKP hierarchies in the form of a system of Hirota bilinear
equations and in the form of bilinear equations on the wave functions
(\S\S 3 and 4). Note that though the BKP hierarchy in the
fermionic picture is always the same (eq. (\ref{1.9})), different
bosonizations (depending of the choice of the multicomcponent
fermions) produce quite different hierarchies of partial differential
equations.

As in the case of the $n$-component KP hierarchy (see ~\cite{KV}) we
go on to develop the formalism of Sato and Lax equations for the
multicomponent BKP hierarchies (\S 5). In conclusion we derive a
OSASM-type formula (\S 6), using the method of ~\cite{V} (see also ~\cite{KV}).

\section{Geometry of spinors}

Let $V$ be a vector space with a symmetric bilinear form $(\cdot ,
\cdot )$. We shall assume for simplicity that $V$ is at most countable
dimensional. Given a non-isotropic vector $\alpha$ (i.e. $(\alpha ,
\alpha) \neq 0)$, the associated reflection $r_{\alpha}$ is defined by
\[
r_{\alpha} (v) = v - \frac{2(\alpha, v)}{(\alpha, \alpha)} \alpha .
\]

We denote by $O(V)$ the group generated by all $-r_{\alpha}$ and by
$SO(V)$ its subgroup consisting of products of an even number of
reflections.
\ \\
Let $C\ell\; V$ be the associated Clifford algebra, that is the quotient
of the tensor algebra over $V$ by the ideal generated by relations
\begin{equation}
  \label{1.1}
  uv + vu = (u,v)1,\quad\mbox{where } u,v \in V.
\end{equation}
If $\{v_i\}_{i\in J}$ is a basis of $V$ indexed by an ordered set,
then the elements
\begin{equation}
  \label{1.2}
  v_{i_1}  v_{i_2} \ldots v_{i_s} \;\;\mbox{with}\;\; i_1 < i_2 < \ldots <
  i_s, \quad
s\in \mathbb{Z}_{+},
\end{equation}
form a basis of $C\ell\; V$. We have a $\mathbb{Z} /
2\mathbb{Z}$-gradation
\[
C\ell\; V = C\ell_{\overline 0}V\oplus C\ell_{\overline{1}}V
\]
where $C\ell_\nu V$ $ (\nu \in \mathbb{Z} / 2\mathbb{Z})$ is
spanned by all products of the form (\ref{1.2}) with $s\equiv \nu$
mod $2$. Here and further we identify $V$ with its image in $C\ell\;
V$.
\ \\
Let $(C\ell\; V)^\times$ denote the multiplicative group of invertible
elements of the algebra $C\ell\; V$. We denote by Pin $V$ the subgroup
of $(C\ell\; V)^\times$ generated by all the elements $a$ such that
$aVa^{-1}=V$ and let Spin $V=$ Pin $V\cap C\ell_{\overline{0}}V$.

If $\alpha \in V$ is a non-isotropic vector, then, by (\ref{1.1}):
\begin{equation}
  \label{1.3}
  \alpha^{-1} = \frac{2\alpha}{(\alpha , \alpha)} ,
\end{equation}
hence $\alpha \in (C\ell\; V)^\times$. From (\ref{1.1}) and (\ref{1.3}) we
obtain
\begin{equation}
  \label{1.4}
  \alpha v \alpha^{-1} = -r_{\alpha} (v) ,
\end{equation}
hence $\alpha \in \;\mbox{Pin}\; V$. We have a homomorphism $T:$ Pin
$V\rightarrow O(V), g\mapsto T_g$ defined by $(v\in V)$:
\[
T_g (v)= g v g^{-1}  \in V.
\]
\ \\
Let $U$ be a subspace of $\mathbb{C} 1 + V$. Let
\[
F(V,U) = C\ell\; V/ (C\ell\; V)U
\]
and let $|0\rangle $ denote the image of $1$ in $F(V,U)$. The space $F(V,U)$
caries a structure of a $C\ell\; V$-module induced by left
multiplication. This module restricted to Pin $V$ is called a {\it
  spin module} of the group Pin $V$.
\ \\
Consider the special case when $V$ has a basis
$\{v_i\}_{i\in \mathbb{Z}}$ such that
\begin{equation}
  \label{1.5}
  (v_i , v_j)= \delta_{i, - j} \quad i,j \in \mathbb{Z}.
\end{equation}
Then $U_0 : = \sum_{j<0}
\mathbb{C} v_j$ is a maximal isotropic subspace of the space
$V$. Let
\[U=
U_0 + \mathbb{C} (1+\sqrt{2} v_0)
\]
and consider the $C\ell\; V$-module $F:= F(V, U)$.
Note that
\[
v_0|0\rangle=-{1\over\sqrt 2}|0\rangle\quad\mbox{and }
v_j|0\rangle=0\quad\mbox{if }j<0.
\]
\ \\
Let
\begin{equation}
  \label{1.6}
  V' = \{ w \in V^0 \mid (w,v_0) = 0\}.
\end{equation}
Then $C\ell\; V'$ is a subalgebra of $C\ell\; V$
(it is actually the centralizer of $v_0$)
and Pin $V'$,
Spin $V'$ are subgroups of Pin $V$ and Spin $V$. The Spin
$V'$-module $F$ decomposes in the direct sum of
submodules  $F_{\overline{0}}$ and $F_{\overline{1}}$, called
{\it half spin modules}, defined by
\[
F_{\overline{\nu}} = \{ f \in F \mid -\sqrt{2} v_0 f = (-)^\nu f\}.
\]
The $C\ell\; V$- (resp. $C\ell\; V'$-)module $F$ is called the spin
module of type B (resp. D).

Given $f\in F$, let
\[
Ann\;  f = \{ v\in V \mid v f = 0\}.
\]

If moreover $f\in F_{\overline{\nu}}$, then
\[
Ann\;  f = \{ v\in V' \mid v f = 0\}.
\]

Recall that we have the vacuum vector $|0\rangle  \in F$ characterized (up
to a constant factor) among the vectors of $F$ by the property that
$Ann\;  |0\rangle  = U_0$. Let $|1\rangle  =  v_1 |0\rangle  \in
F_{\overline{1}}$. Note that $|1\rangle $ is characterized by the property
that $Ann\;  |1\rangle  = r_{v_1 + v_{-1}} (U_0)$.

\begin{remark}
It is clear that the $C\ell_{\overline{0}}V$-module $F$ is
irreducible. The $C\ell_{\overline{0}}V'$-modules $F_{\overline{0}}$
and $F_{\overline{1}}$ are also irreducible and moreover
inequivalent. The same holds with $C\ell_{\overline{0}}V$
(resp. $C\ell_{\overline{0}}V'$) replaced by Spin $V$
(resp. Spin $V'$).
\end{remark}

Introduce the following operators on the space $F\otimes F$:
\begin{equation}
\label{1.6a}
S' = \sum_{0\neq j \in  \mathbb{Z}} v_j \otimes v_{-j}
\;\;\mbox{and}\;\; S=v_0 \otimes v_0 + S' .
\end{equation}
Since $v_j |0\rangle  = v_j |1\rangle  = 0$ for $j<-1$, $v_{-1} |0\rangle  =
v_1 |1\rangle  =0$ and
$v_0 |0\rangle  = -\frac{1}{ \sqrt{2}} |0\rangle $, $v_0 |1\rangle  = \frac{1}{
\sqrt{2}} |1\rangle $ we find ($\epsilon = 0$ or $1$):

\begin{equation}
  \label{1.7}
  \begin{array}[h]{lcl}
S' (|\epsilon\rangle \otimes |\epsilon\rangle ) &=& 0,\\[2mm]
S (|0\rangle  \otimes |0\rangle ) &=& \frac{1}{2} |0\rangle  \otimes |0\rangle
,\\[2mm]
S' (|0\rangle  \otimes |1\rangle ) &=& |1\rangle  \otimes |0\rangle ,\\[2mm]
S' (|1\rangle  \otimes |0\rangle ) &=& |0\rangle  \otimes |1\rangle .
  \end{array}
\end{equation}

\begin{lemma}
\label{l1.2}
The operator $S$ (resp. $S'$) commutes with the diagonal action of the
group Pin $V$ (resp. Pin $V'$) on $F\otimes F$.
\end{lemma}

{\bf Proof.}
For $a \in$ Pin $V$ we have
\[
\begin{array}[h]{lcl}\displaystyle
(a\otimes a ) S(a^{-1} \otimes a^{-1}) &=&\displaystyle \sum_{j\in \mathbb{Z}}
T_a
(v_j) \otimes T_a (v_{-j})\\[3mm]
&=&\displaystyle \sum_{i\in \mathbb{Z}} \sum_{j,k\in \mathbb{Z}} a_{ij} v_j
\otimes
a_{-ik} v_k
=\displaystyle \sum_{i,j,k\in \mathbb{Z}} a_{ij} a_{-ik} v_j \otimes v_k
=\displaystyle \sum_{j\in \mathbb{Z}} v_j \otimes v_{-j},
\end{array}
\]

since $\sum_i a_{ij} a_{-ik} = \delta_{j,-k}$ (because $T_a \in
O(V)$). The proof for $S'$ is analogous.
\hfill{$\Box$}
\ \\

Let $(\nu =0,1)$:
\[
O = \mbox{Spin}\; V |0\rangle  \;\;\mbox{and}\;\; O_{\overline{\nu}} =
\mbox{Spin}\; V' |\nu\rangle
\]

be the orbit of the vector $|0\rangle $ (resp. $|\nu\rangle$) under the action
of the
group Spin $V$ (resp. Spin $V'$). Then clearly Pin $V |0\rangle  = O$ and Pin
$V' |\nu\rangle  = O_{\overline{0}} \cup O_{\overline{1}}$.

\begin{lemma}
\label{l1.3}
If $\tau \in O$ (resp. $\tau \in O_{\overline{\nu}}$) and $\alpha \in V$
(resp. $\alpha \in V'$), then $\alpha \tau \in O \cup \{ 0\}$
(resp. $\alpha \tau \in O_{\overline{\nu + 1}}\cup \{ 0\}$).
\end{lemma}

{\bf Proof.}
For simplicity we consider only the case $\tau\in
O_{\overline{\nu}}$. If $\alpha$ is non-isotropic then $\alpha \tau \in
O_{\overline{\nu + 1}}$ by definition. If $(\alpha , \alpha) = 0$ and
$\alpha \in$ (resp. $\not\in$) $Ann\;  |\nu\rangle$, then $\alpha |\nu\rangle
= 0$
(resp. there exists a $u \in Ann\;  |\nu\rangle $ such that $(\alpha , u)\neq
0$, so that $\alpha + u$ is non-isotropic, hence $\alpha | \nu > =
(\alpha + u) |\nu\rangle \in O_{\overline{\nu +1}}$). Since for some $g\in$
Spin
$V', \tau = g |\nu\rangle $ we have $\alpha \tau = g (g^{-1} \alpha g)
|\nu\rangle  = g
(T_{g^{-1}} (\alpha) ) |\nu\rangle  \in g(O_{\overline{\nu +1}} \cup\{ 0\})
\subset
O_{\overline{\nu +1}} \cup \{ 0 \}$.
\hfill{$\Box$}
\ \\

For $k\in \mathbb{Z}$, let $U_k = \sum_{j<-k} \mathbb{C}
v_j$. Introduce the Grassmannians $Gr$ and $Gr'$ as the sets of all
maximal isotropic subspaces $W$ of $V$, respectively $V'$, such that
$W\supset U_k$ for $k>>0$.

\begin{remark}
\label{r1.4}
For any $\tau \in O$ (resp. $\tau \in O_{\overline{\nu}}$) we have
\[
Ann\;  \tau \in Gr \;\;(\mbox{resp.}\;\; Ann\;  \tau \in Gr').
\]
\end{remark}
Indeed this is clear if $\tau = |0\rangle $ (resp. $\tau = |\nu\rangle $), and
for
$g\in$ Spin $V$ ($g\in$ Spin $V'$) we have
\begin{equation}
  \label{1.8}
  Ann\;  g\tau = T_{g} (Ann\;  \tau).
\end{equation}

\begin{lemma}
\label{l1.5}
  \begin{enumerate}
  \item[(a)] The group $SO(V)$ (resp. $SO(V')$) is the group of all
    orthogonal transformations $A$ of the space $V$ (resp. $V'$) such
    that the matrix $A-I$ has only finitely many non-zero entries and
    $\det A=1$.
  \item[(b)] The homomorphisms $T:{\rm Pin}\; V\rightarrow{  O(V)}$, $T:{\rm
Pin} V'
    \rightarrow { O(V')}$, $ T:{\rm Spin}\; V\rightarrow {
SO(V)}$ and $T:{\rm Spin}\;
    V' \rightarrow { SO(V')}$ are surjective.
  \item[(c)] The group $O(V)$ (resp. $O(V')$) acts transitively on $Gr$
    (resp. $Gr'$).
  \end{enumerate}
\end{lemma}

{\bf Proof.}
Let $U^k = \sum\limits_{0\neq \mid j\mid \leq k} \mathbb{C} v_j$ and
$V^k = U^k + \mathbb{C} v_0$. It is well-known that the group of all
orthogonal transformations of $U^k$ and $V^k$ is generated by
reflections. This proves (a).
(b) follows from (\ref{1.4}) and the
fact that $O(U^k)$ and $O(V^k)$ is generated by reflections.

(c): If
$W\in Gr$ (resp. $Gr'$), then $W=W' + U^k$ for $k>>0$, where $W'
\subset V^k$ (resp. $\subset U^k$). Now (c) follows from the
well-known fact that $O(V^k)$ resp. $O(U^k)$ acts transitively on the
set of maximal isotropic subspaces.\hfill{$\Box$}

\begin{corollary}
\label{c1.6}
The group $SO(V)$ acts transitively on $Gr$. The group $SO(V')$ has
two orbits on $Gr'$, that of $U_0$ and of $U'_0 =r_{{v_1 + v_{-1}}}( U_0)$.
\end{corollary}


\begin{corollary}
\label{c1.7}
Let $\tau \in F$ (resp. $\tau \in F_{\overline{\nu}}$). Then $Ann\;
\tau$ is a maximal isotropic subspace in $V$ (resp. $V'$) if and only
if $\tau \in O$ (resp. $\tau \in O_{\overline{\nu}}$).
\end{corollary}

{\bf Proof.}
Suppose that $Ann\;  \tau$ is maximal isotropic, then it lies in $Gr$
(resp. $Gr'$). Using corollary \ref{c1.6}, we may find a $g\in
SO(V)$ (resp. $SO(V')$) such that $g Ann\;  \tau = U_0$ (resp. $=U_0$
or $U'_0$). Taking $\tilde{g} \in$ Spin $V$ (resp. $\in$ Spin
$V'$), a preimage of $g$, we see that $Ann\; \tilde{g} \tau = U_0$
(resp. $=U_0$ or $U'_0$) by (\ref{1.8}). It follows that, up to a
constant, $\tilde{g} \tau = |0\rangle $ (resp. $|0\rangle $ or $|1\rangle $).
\hfill{$\Box$}

\ \\
We can now prove Theorem \ref{t1.8} stated in the Introduction, where $S$ and
$S'$
are given by  (\ref{1.6a}).

\ \\
{\bf Proof of Theorem \ref{t1.8}.}
We only give the proof of (a). The proof of (b) is completely
analogous.

Since $S$ commutes with the diagonal action of the group Spin$V$ (see
Lemma \ref{l1.2}) and since the second equation of (\ref{1.7}) holds
for $\tau = |0\rangle $, any $\tau \in O$ satisfies (\ref{1.9}).

For the converse, we introduce a gradation on $F$ by letting
\[
\deg |0\rangle  = 0 \qquad \deg v_i = i.
\]
Write $\tau = \sum\limits_{k=1}^N c_k \tau_k$, where any $\tau_k$ is a
simple vector
\[
\tau_k = v_{j_1} v_{j_2} \ldots v_{j_p} |0\rangle  \quad j_1 > j_2 > \ldots >
j_p > 0 .
\]
We can of course reorder and rescale in such a way that $\tau_1$ will
be of maximal degree and $c_1 = 1$. If among the $\tau_k$ with $k>1$
there exists a $\tau_{\ell}$ of the form
\begin{equation}
  \label{1.11}
  \tau_{\ell} = v_i v_j \tau_1 \quad \mbox{with}\quad i+j<0,
\end{equation}

we can remove this term by replacing $\tau$ by $(1-c_\ell v_i
v_j)\tau$, which again satisfies (\ref{1.9}). Such $1-c v_i v_j \in$
Spin $V$ for any $c\in \mathbb{C}$ and $i+j \neq 0$. Repeating this
procedure a finite number of times we obtain an element of the form
$\tau_1 + \varphi\ (\varphi = \sum\limits^N_{k=2} c_k \tau_k)$ where
none of the simple vectors $\tau_k$ appearing in $\varphi$ is equal for
$\tau_1$, or is of the form (\ref{1.11}). Since $\tau_1$ is a simple
vector, which by Lemma \ref{l1.5} lies in $O$, $\tau_1$ must satisfy
(\ref{1.9}). Hence $\tau_1$ and $\varphi$ satisfy:
\[
\sum_{j\in \mathbb{Z}} v_j \varphi \otimes v_{-j} \tau_1 +
  v_j \tau_1 \otimes v_{-j} \varphi + v_j \varphi \otimes v_{-j}
\varphi = \frac{1}{2} (\varphi \otimes \tau_1 + \tau_1 \otimes \varphi
+ \varphi \otimes \varphi).
\]

If $\varphi$ is non-zero, then (since $\varphi$ does not contain a
simple vector proportional to $\tau_1$) there exists a $j\neq 0$ such
that $v_j \tau_1 \otimes v_{-j} \varphi \neq 0$. Then $v_j \tau_1$
must be equal to either
(1) $\lambda v_{-i} \tau_{\ell}$ for some $-i>j$,
(2) $\lambda \tau_{\ell}$ for some $\ell > 1$ or to
(3) $\lambda \tau_1$, where for all 3 cases $\lambda \in \mathbb{C}^\times$.
Clearly (3) is impossible and (1) and (2) are  also not possible since
$\tau_{\ell}$ is not of the form (\ref{1.11}) (where $i=0$ for case
(2)). Hence $\varphi = 0$ and $\tau = \tau_1$ is a simple vector,
which by Lemma \ref{l1.5} lies in $O$.
\hfill{$\Box$}
\ \\

Equation (\ref{1.9}) (resp. (\ref{1.10}) is called the BKP (resp. DKP)
hierarchy because the group $SO(V)$ (resp. $SO(V')$) is of $B$
(resp. $D$) type and because after bosonization this equation becomes
a hierarchy of PDE's similar to the $n$-component KP hierarchy (cf.
Introduction).

Let, for $g\in$ Spin$V'$, $\tau_\nu = g | \nu \rangle$ and $\tau_{\mu} = g |
\mu\rangle$, where
$\nu, \mu = 0$ or $1$ and $\mu \neq \nu$. Then by Lemma \ref{l1.3}
and the third and fourth equation of (\ref{1.7}), $\tau_{\nu}$ and
$\tau_{\mu}$ satisfy the Modified DKP hierarchy (see \ref{1.20})
\[
  S'(\tau_{\nu} \otimes \tau_{\mu}) = \tau_{\mu} \otimes \tau_{\nu} .
\]
It has the following
geometrical interpretation.

\begin{theorem}
\label{t1.9}
Elements $\tau_{\nu} \in O_{\overline{\nu}}, \tau_{\mu} \in
O_{\overline{\mu}} ,\ \mu\neq \nu$, satisfy the Modified DKP hierarchy
if and only if the space
\[
(Ann\;  \tau_0 + Ann\;  \tau_1 )/ (Ann\;  \tau_0 \cap Ann\;  \tau_1)
\]

is 2-dimensional and the induced bilinear form on it is non-degenerate.
\end{theorem}

{\bf Proof.}
Without loss of generality we may assume that $\nu = 1$ and $\mu =0$
and that $\tau_{\mu} = |0\rangle $. Suppose first that (\ref{1.20}) holds,
then
\[
\sum_{j<0} v_j \tau_1 \otimes v_{-j} |0\rangle  = |0\rangle  \otimes \tau_1 .
\]

Since all $v_{-j} |0\rangle $ for $j<0$ are linearly independent, we conclude
that $(\sum\limits_{j<0} a_j v_j)\tau_1 = |0\rangle $. Hence there exists an
isotropic vector $w_0 \in V'$ such that $|0\rangle  = w_0 \tau_1$. Now notice
that $w_0 |0\rangle  = w_0^2 \tau_1 = \frac{1}{2} (w_0 , w_0)\tau_1 =
0$. Because of the symmetry of the equation (\ref{1.20}), there also
exists an isotropic vector $w_1 \in V'$ such that $\tau_1 = w_1
|0\rangle $. Hence
\[
|0\rangle  = w_0 w_1 |0\rangle  = ((w_0 , w_1 ) -w_1 w_0) |0\rangle
= (w_0 , w_1) |0\rangle
\]

and therefore $(w_0 , w_1) = 1$. This leads to the statement of the
theorem about the annihilator spaces of $\tau_0$ and $\tau_1$.

Conversely,assume that $\tau_1 = w_1 \tau_0$ for some isotropic vector $w_1
\in V'$. Then
\[
\begin{array}[h]{lcl}\displaystyle
\sum_{0\neq i\in\mathbb{Z}} v_i \tau_1 \otimes v_{-i} \tau_0 &=&\displaystyle
\sum_{0 \neq i \in \mathbb{Z}} v_i w_1 \tau_0 \otimes v_{-i} \tau_0\\[3mm]
&=&\displaystyle \sum_{0\neq i\in \mathbb{Z}} (v_i , w_1)-w_1 v_i \tau_0
\otimes
v_{-i} \tau_0\\[3mm]
&=&\displaystyle \tau_0 \otimes w_1 \tau_0 - (w_1 \otimes 1) S' (\tau_0 \otimes
\tau_0)\\[3mm]
&=&\displaystyle \tau_0 \otimes \tau_1 ,
\end{array}
\]
since $\tau_0$ satisfies (\ref{1.10}).\hfill{$\Box$}
\ \\

Let $\tau\in O$, we split up $\tau$ as $\tau = \tau_0 + \tau_1$, where
$\tau_{\nu} \in F_{\overline{\nu}}$ (recall that
$F_{\overline{\nu}}$ are eigenspaces for $-\sqrt{2} v_0$). We now
rewrite (\ref{1.9}):
\begin{equation}
  \label{1.13}
  S' ((\tau_0 + \tau_1 ) \otimes (\tau_0 + \tau_1))+v_0 (\tau_0 +
  \tau_1)\otimes v_0 (\tau_0 + \tau_1) = \frac{1}{2} (\tau_0 + \tau_1)
  \otimes (\tau_0 + \tau_1).
\end{equation}

The left-hand side of this equation is equal to
\[
S' ((\tau_0 + \tau_1) \otimes (\tau_0 + \tau_1)) + \frac{1}{2} (\tau_0
- \tau_1) \otimes (\tau_0 - \tau_1).
\]

Hence, (\ref{1.13}) splits up into 4 separate equations:
\begin{equation}
  \label{1.14}
  \begin{array}[h]{lcl}
S' (\tau_{\nu} \otimes \tau_{\nu}) &=& 0 \qquad\qquad (\nu = 0
\;\;\mbox{or}\;\; 1),\\[2mm]
S' (\tau_0 \otimes \tau_1 ) &=& \tau_1 \otimes \tau_0,\\[2mm]
S' (\tau_1 \otimes \tau_0) &=& \tau_0 \otimes \tau_1 ,
  \end{array}
\end{equation}

which are exactly the equations of the DKP and Modified DKP-hierarchy
(\ref{1.10}) and (\ref{1.20}). Hence we have proved Theorem
\ref{t1.10}, stated in the Introduction (where $S$ and $S'$ are given by
(\ref{1.6a})).

So we  conclude that the fermionic BKP hierarchy contains the DKP hierarchy
and the modified DKP hierarchy. We will now show that it also contains
the KP hierarchy and modified KP hierarchy as well.
We split up in some way the maximal isotropic subspace $U_0$ into two infinite
dimensional subspaces $U^{\pm}_0$:
\[
U_0 = U^+_0 \oplus U^-_0 .
\]

Choose bases $\{ v^{\pm}_{-i} \}_{ i\in \mathbb{Z}_++\frac{1}{2}}$
of $U^{\pm}_0$ and include these bases in a basis of $V$ by adding
vectors
$\{ v^{\mp}_i \}_{i\in \mathbb{Z}_+ + \frac{1}{2}}$ and $v_0$
such that
\[
(v_i^{\pm} , v_j^{\pm} ) =0 , (v_i^{\pm} ,\ v_j^{\mp}) = \delta_{i,
-j},\
(v^{\pm}_i , v_0) = 0,\ (v_0 , v_0) = 1.
\]
We will assume from now on that we have chosen these bases in such a
way that
\[
v_{\frac{1}{2}}^+ |0\rangle  = v_1 |0\rangle .
\]

The Grassmannian $Gr$ (resp. $Gr'$) translates in this picture to
the set of all maximal isotropic
subspaces $W\subset V$ (resp. $\subset V'$) such that $U_k \subset W$ for
$k>>0$ , where $U_k = U^+_k \oplus U^-_k$ and $U^{\pm}_{k} = \sum_{j <
  -k} \mathbb{C} v_j^{\pm}$.

Define
\[
V^{\pm} = \sum_{j\in \mathbb{Z}+\frac{1}{2}} \mathbb{C} v_j^{\pm} ,
\]
so that $V=V^+ +\mathbb{C}v_0+V^-$.
Consider the subgroup $\overline{\mbox{Spin}\; V}$ of Spin $V$ generated
by all elements of the form $1+a v_i^{+} v^-_j$ , $a\in
\mathbb{C}$. It preserves each of the subspaces $V^+$ and $V^-$ and we
have the induced isomorphisms $\overline{\mbox{Spin}\; V}
\tilde{\rightarrow} GL$, where $GL$ is the subgroup of
$SO(V)$, consisting of all elements $g\in SO(V)$ such that $g(V^{\pm})
= V^{\pm}$.

\begin{lemma}
  \label{l1.11}
Let $g\in GL$ be such that
\[
g v_j^{+} = \sum_{i\in\mathbb{Z}+\frac{1}{2}} A_{ij} v_i^+
\]
and let $A=(A_{ij})_{i,j \in
\mathbb{Z}+\frac{1}{2}
}$ and
$A^{-1} = (B_{ij})_{i,j\in\mathbb{Z}+\frac{1}{2}
}$. Then
\[
g v^-_j = \sum_{i\in\mathbb{Z}+\frac{1}{2}
} B_{-ji} v^-_{-i}
\]
and
\[
g v_0 = v_0 .
\]
\end{lemma}

{\bf Proof.}
Suppose that $g v_{\ell}^- = \sum_k B_{-\ell , k} v_{-k}^-$, then it
follows from
\[
\begin{array}[h]{lcl}
\delta_{j, -\ell} = (v^+_j , v_{\ell}^-) &=& (\sum_i A_{ij} v^+_i ,
B_{-\ell k} v_{-k}^-)\\
&=& \sum_i A_{ij} B_{-\ell , i}
\end{array}
\]

that $B=A^{-1}$. Since $g\in SO(V),\ gv_0 = v_0$.
\hfill{$\Box$}
\ \\
In order to describe the group orbits of $\overline{\mbox{Spin}\; V}$,
we introduce the restricted Grassmannian
\[
\overline{Gr} = \bigcup_{\ell \in \mathbb{Z}} \overline{Gr}^{(\ell)} ,
\]
where $\overline{Gr}^{(\ell)}$ is the set of all maximal isotropic subspaces
$W\subset V' \subset V$ such that $W=W^+ \oplus W^-$ with $U^{\pm}_k
\subset W^{\pm} \subset V^{\pm}$ for $k>>0$ and $\dim (W^+ / U^+_{k} )
-\dim (W^- / U^-_k)=2\ell$.
We will call $\ell$ the {\it charge} of
$\overline{Gr}^{(\ell)}$. Notice that, since $W$ is maximal isotropic,
\begin{equation}
  \label{1.15}
  W^-= (W^+)^{\perp} \cap V^- .
\end{equation}

Hence $W^+$ determines $W$.

We decompose the space $F$ into charge sectors as follows. Define
\[
\mbox{charge}\;\; |0\rangle  = 0 ,\quad \mbox{charge}\;\; v^{\pm}_j = \pm 1
,\quad
\mbox{charge}\;\; v_0=0,
\]
and let
\[
F = \bigoplus_{k\in \mathbb{Z}} F_k, \;\;\mbox{where}\;\; F_k = \{ f\in
F|\;\;\mbox{charge}\;\; f = k\} .
\]

Since charge $(1+a v_i^+ v_j^-)=0$, $\overline{\mbox{Spin}\; V}$ leaves
all spaces $F_k$ invariant.

Define the following vectors in $F_k$:
\[
|k\rangle = \left\{
\begin{array}[h]{lcl}
v^+_{k-\frac{1}{2}} v^+_{k-\frac{3}{2}} \ldots v^+_{\frac{1}{2}} | 0 \rangle
&\mbox{for}& k\geq 0 ,\\
v^-_{-k-\frac{1}{2}} v^-_{-k-\frac{3}{2}} \ldots v^-_{\frac{1}{2}} | 0 \rangle
&\mbox{for}& k< 0 .
\end{array}\right.
\]

One has
\begin{equation}
  \label{1.16}
  Ann\;  | k> = \sum_{j< k} \mathbb{C} v_j^+ + \sum_{j>k}\mathbb{C} v^-_{-j} ,
\end{equation}

and clearly $Ann\;  |k\rangle  \in \overline{Gr}^{(k)}$.

\begin{lemma}
  \label{l1.12}
The group $GL$ acts transitively on $\overline{Gr}^{(\ell)}$
for any $\ell \in \mathbb{Z}$.
\end{lemma}

{\bf Proof.}
Let $W\in \overline{Gr}^{(\ell)}$, then by (\ref{1.15}) $W=W^+ \oplus
((W^+)^{\perp} \cap V^-)$, hence $W^+$ determines $W$. Now choose
a basis $w_{\ell -\frac{1}{2}} , w_{\ell -\frac{3}{2}}, \ldots$ of
$W^+$ in such a way that $w_{\ell - p} = v^+_{\ell - p}$ for
  $p>>0$. Clearly there exists an invertible matrix $A=(A_{ij})_{i,j
    \in \mathbb{Z}_{\frac{1}{2}}}$ such that $w_j = \sum_i A_{ij}
  v_i^+$. Hence by lemma (\ref{l1.11}), this $A$ determines a unique
  $g\in GL$ such that $g(\sum_{j<\ell} \mathbb{C}
  v_j^+)=W^+$. Thus $g(Ann\;  |\ell \rangle)=W$.
\hfill{$\Box$}

We express the operator $S$ in terms of the new basis
\[
S = v_0 \otimes v_0 +\sum_{j\in \mathbb{Z}+{\frac{1}{2}}} v_j^+ \otimes
  v^-_{-j} + v_j^- \otimes v^+_{-j} .
\]

We split this operator as follows
\[
\begin{array}[h]{lcl}
S &=& S^0 + S^+ + S^- \quad \mbox{with}\\
S^0 &=& v_0 \otimes v_0 \;\;\mbox{and}\;\; S^{\pm} = \sum_{j\in
  \mathbb{Z}+{\frac{1}{2}}} v_j^{\pm} \otimes v_{-j}^{\mp} .
\end{array}
\]

It is straightforward to check that for $k, \ell \in \mathbb{Z}$, with
$\ell \geq 0$, one has
\begin{equation}
  \label{e1.16}
  \begin{array}[h]{lcl}\displaystyle
S^0 (|k + \ell \rangle \otimes |k\rangle ) &=&\displaystyle
  \frac{(-)^{\ell}}{2}
|k + \ell \rangle \otimes |k\rangle ,\\[3mm]
\displaystyle S^+ (|k + \ell \rangle \otimes |k\rangle ) &=&\displaystyle  S^-
(|k\rangle \otimes
|k + \ell\rangle )=0,\\[3mm]
\displaystyle
(S^+)^{\ell} (|k\rangle \otimes |k + \ell\rangle ) &=&\displaystyle
(-)^{\frac{\ell (\ell
    -1)}{2}} \ell ! |k+\ell \rangle \otimes |k\rangle ,\\[3mm]
\displaystyle
(S^-)^{\ell} (|k + \ell\rangle  \otimes |k\rangle ) &=&\displaystyle
(-)^{\frac{\ell (\ell
    -1)}{2}} \ell ! |k\rangle  \otimes |k+\ell \rangle ,\\[3mm]
\displaystyle
(S^+)^{\ell +1} (|k\rangle \otimes |k+\ell\rangle ) &=&\displaystyle
(S^-)^{\ell +1} (|k +
\ell \rangle \otimes |k\rangle )=0.
  \end{array}
\end{equation}

By Lemma \ref{l1.2}, $S$ commutes with the diagonal action of Pin
$V$. Since for any $g\in \overline{\mbox{Spin}\; V}, T_g (V^{\pm}) =
V^{\pm}$ and $T_g (v_0) = v_0$, we deduce that $S^0, S^+$ and $S^-$
separately commute with the diagonal action of $\overline{\mbox{Spin}
  V}$. Denote by $\overline{O_k}$ the $\overline{\mbox{Spin}\; V}$ group
orbit of $|k\rangle $. Let $g\in \overline{\mbox{Spin}\; V}$ and let $\tau_k =
g|k\rangle $ for all $k\in \mathbb{Z}$, then since (\ref{e1.16}) holds one has
for $k, \ell \in \mathbb{Z} \;\;\ell \geq 0$:
\begin{equation}
  \label{1.17}
  \begin{array}[h]{lcl}\displaystyle
S^0 (\tau_{k+\ell} \otimes \tau_k) &=&\displaystyle \frac{(-)^{\ell}}{2}
\tau_{k+\ell} \otimes \tau_k ,\\[3mm]
\displaystyle S^+ (\tau_{k+\ell} \otimes \tau_k) &=&
\displaystyle   S^- (\tau_k \otimes
\tau_{k+\ell})=0,\\[3mm]
\displaystyle (S^+)^{\ell} (\tau_k \otimes \tau_{k+\ell})
&=&\displaystyle  (-)^{\frac{\ell (\ell
    -1)}{2}} \ell ! (\tau_{k+\ell} \otimes \tau_k ),\\[3mm]
\displaystyle (S^-)^{\ell} (\tau_{k+\ell} \otimes \tau_k)
&=&\displaystyle  (-)^{\frac{\ell(\ell
    -1)}{2}} \ell ! \tau_k \otimes \tau_{k+\ell} ,\\[3mm]
\displaystyle (S^+)^{\ell + 1} (\tau_k \otimes \tau_{k+\ell})
&=& \displaystyle  (S^-)^{\ell +
  1} (\tau_{k+\ell} \otimes \tau_k)=0 .
  \end{array}
\end{equation}

The second equation of (\ref{1.17}) is called the $\ell$-th modified KP
hierarchy;
the $0$-th modified KP hierarchy is also known as the KP hierarchy.

\begin{theorem}
\label{t1.13}
Let $\tau_k \in F_k , k\in \mathbb{Z}$. Then $\tau_k \in
\overline{O_k}$ if and only if $\tau_k$ satisfies the KP hierarchy,
i.e., satisfies the second equation of (\ref{1.17}) for $\ell = 0$.
\end{theorem}

The proof of this Theorem is similar to the proof of theorem
\ref{t1.8}. For a proof see~\cite{KacRaina}.

\begin{theorem}
  \label{t1.14}
Let $k\in \mathbb{Z}, \ell \in \mathbb{Z}_+ , \tau_k \in
\overline{O_k}$ and $\tau_{k+\ell} \in \overline{O_{k+\ell}}$, then
one has the following 6 equivalent formulations of the $\ell$-th
modified KP hierarchy.
\begin{enumerate}
\item[(a)] $S^+ (\tau_{k+\ell} \otimes \tau_k) = 0$,
\item[(b)] $S^- (\tau_k \otimes \tau_{k+\ell} ) =0$,
\item[(c)] $(S^+)^{\ell} (\tau_k \otimes \tau_{k+\ell}
  )=(-)^{\frac{\ell (\ell -1)}{2}} \ell !\; \tau_{k+\ell} \otimes
  \tau_k$,
\item[(d)] $(S^-)^{\ell} (\tau_{k+\ell} \otimes \tau_{k}
  )=(-)^{\frac{\ell (\ell -1)}{2}} \ell !\; \tau_{k} \otimes
  \tau_{k+\ell}$,
\item[(e)] $Ann\;  \tau_k \cap V^+ \subset Ann\;  \tau_{k+\ell} \cap V^+$,
\item[(f)] $Ann\;  \tau_{k+\ell} \cap V^- \subset Ann\;  \tau_k \cap V^-$.
\end{enumerate}
\end{theorem}

{\bf Proof.}
The equivalences (a) $\Leftrightarrow$ (b), (c) $\Leftrightarrow$ (d)
and (e) $\Leftrightarrow$ (f) are obvious. Without loss of generality
we may assume from now on in this proof that $\tau_k = |k\rangle $.

First assume (a) holds; since $\tau_k = |k\rangle $ one has
\[
\sum_{i<k} v_i^+ \tau_{k+\ell} \otimes v^-_{-i} |k\rangle  = 0.
\]

Hence, $v_i^+ \tau_{k+\ell} = 0$ for all $i<k$, so (e) holds:
\[
Ann\;  |k\rangle  \cap V^+ \subset Ann\;  \tau_{k+\ell} \cap V^+ .
\]

Now suppose (e) holds, then
\[
\sum_{i\in \frac{1}{2} + \mathbb{Z}} v_i^+ \tau_{k+\ell} \otimes
|k\rangle = \sum_{i<k} v_i^+ \tau_{k+\ell} \otimes v_{-i}^- |k\rangle  ,
\]

but this is equal to $0$ since all $v_i^+ \in Ann\;  |k\rangle $ for $i<k$
and by (e) also $v_i^+ \in Ann\;  \tau_{k+\ell}$ for $i<k$. This
proves the equivalence (a) $\Leftrightarrow$ (e).

Next assume that (c) holds, which is equivalent to
\[
\sum_{i_1 > i_2 > \ldots > i_{\ell} > k} v_{i_1}^+ \ldots
v^+_{i_{\ell}} |k\rangle  \otimes v_{-i_1}^- \ldots v^-_{-i_{\ell}}
\tau_{k+\ell} = (-)^{\frac{\ell (\ell -1)}{2}} \tau_{k+\ell} \otimes
|k\rangle  .
\]

Since all $v_{i_1}^+ \ldots v^+_{i_{\ell}} |k\rangle $ are linearly
independent we find that
\[
\tau_{k+\ell} = \sum_{i_1 > i_2 > \ldots i_{\ell} > k} \mu_{{i_1} ,
\ldots i_{\ell}} v_{i_1}^+ v_{i_2}^+ \ldots v_{i_{\ell}}^+ |k\rangle  .
\]

{}From which one deduces that
\[
Ann\;  |k\rangle  \cap V^+ \subset Ann\;  \tau_{k+\ell} \cap V^+ .
\]

Hence (c) implies (e).

Now assume that (e) holds; one has
\[
Ann\;  \tau_{k+\ell} \cap V^+ = (Ann\;  |k\rangle \cap V^+) \cup <w^+_1 ,
w^+_2 , \ldots , w^+_{\ell}>
\]

for certain $w^+_i \in V^+ \;\; w^+_i \not\in Ann\;  |k\rangle  , 1\leq i
\leq \ell $.

Without loss of generality we may assume that $\tau_{k+\ell} = w^+_1
w^+_2 \ldots w^+_{\ell} |k\rangle $, so
\[
\begin{array}[h]{lcl}
(S^+)^{\ell} (|k\rangle  \otimes \tau_{k+\ell}) &=&\displaystyle \sum_{i_1 ,
\ldots ,
  i_{\ell} \in \mathbb{Z} + \frac{1}{2}} v_{i_1}^+ \ldots
v^+_{i_\ell} |k\rangle  \otimes v^-_{i_1}\ldots v^-_{-i_{\ell}} w^+_1 \ldots
w^+_{\ell}
|k\rangle \\[3mm]
&=&\displaystyle \sum_{i_1 , \ldots ,
  i_{\ell} \in \mathbb{Z} + \frac{1}{2}} \sum_{\sigma\in S_{\ell}}
(-)^{\frac{\ell(\ell -1)}{2}} (-)^{\mbox{sign} (\sigma)} (v^-_{-i_1} ,
w^+_{\sigma(i_1)})\times \\[3mm]
&&\displaystyle (v^-_{-i_2} , w^+_{\sigma(i_2)}) \ldots (v^-_{-i_{\ell}} ,
w^+_{\sigma(i_{\ell})}) v^+_{i_1} v^+_{i_2} \ldots v^+_{i_{\ell}} |k\rangle
\otimes |k\rangle \\[3mm]
&=&\displaystyle (-)^{\frac{\ell(\ell -1)}{2}} \ell ! \; w^+_1 w^+_2 \ldots
w^+_{\ell} |k\rangle  \otimes |k\rangle \\[3mm]
&=&\displaystyle (-)^{\frac{\ell(\ell -1)}{2}} \ell ! \; \tau_{k+\ell} \otimes
|k\rangle .
\end{array}
\]

Hence (e) implies (c).
\hfill{$\Box$}
\\

Since $|1\rangle =v_1 |0\rangle  = v^+_{\frac{1}{2}} |0\rangle $, we
find that $|k\rangle  \in F_{\overline{k}}$ (here, as
before, $\overline k=k\mod 2$). Now $Ann\;  |k\rangle $ is an element
of $Gr$ and $Gr'$ hence $|k\rangle \in O$ and $|k\rangle  \in
O_{\overline{k}}$. Therefore one has the following

\begin{proposition}
  \label{p1.15}
Let $\tau_k \in F_k , \tau_{k+1} \in F_{k+1}$ (so that $\tau_k \in
F_{\overline{k}}$ and $\tau_{k+1} \in F_{\overline{k+1}}$). Then:
\begin{enumerate}
\item[(a)] $\tau_k $ satisfies the KP hierarchy
           if and only if it satisfies the DKP hierarchy.
\item[(b)] $\tau_k$ and $\tau_{k+1}$ satisfy the first modified KP
           hierarchy if and only if they satisfy
           the modified DKP hierarchy.
\item[(c)] $\tau_k$ and $\tau_{k+1}$ satisfy the KP hierarchy and the
           first modified KP
           hierarchy, if and only if $\tau_k + \tau_{k+1}$ satisfies
           the BKP hierarchy.
\end{enumerate}
\end{proposition}
The proof is straightforward.

\section{Orthogonal multicomponent boson-fermion correspondence}

Using a bosonization one can rewrite (\ref{1.9}), (\ref{1.20}) and
(\ref{1.17}) as a system of partial differential equations. There are
however many different bosonizations. In this paper we will consider
several possibilities. The simplest one is given in the introduction.

\subsection{$n$-component charged fermions and $D^{(1)}_{n}$}
\label{subs2.2}
In this section we will explain the $n$-component generalization of
the classical boson-fermion correspondence, which was described in
detail in the Introduction.

Choose bases

$ \{
\psi^a_i \}_{i\in \mathbb{Z} +\frac{1}{2}}$, where  $a=1,2,\dots,n$,
of $V^\pm$ such that
$\psi^{\pm a}_{i}
\in U_0$for $i>0$ and
\begin{equation}
  \label{2.7}
  (\psi^a_i , \psi^b_j) = \delta_{a, -b} \delta_{i,-j} .
\end{equation}

Define the fermionic fields ($a\in\mathbb{Z},\ 0<|a|\le n$):
\begin{equation}
\label{2.7a}
\psi^{a} (z) = \sum_{i\in \frac{1}{2} + \mathbb{Z}} \psi^a_i
z^{-i-\frac{1}{2}} .
\end{equation}

Then one has the following commutation relations:
\begin{equation}
  \label{2.8}
  [\psi^a (y) ,\psi^b (z)]_+= \delta_{a, -b}\delta(y-z) .
\end{equation}

We let
\[
e^{ab} (z) = \sum_{m\in \mathbb{Z}} e^{ab}_m z^{-m-1} =: \psi^a (z)
\psi^b (z) :
\]

where the normal ordering is defined as in the introduction.

It is immedate that
\[
e^{ab}_m = -e^{ba}_m \quad \mbox{hence}\quad e^{aa}_m = 0.
\]

Using Wick's formula, it is straightforward to check that the
operators $e^{ab}_m$ form a representation in $F$ of the affine
Kac-Moody algbera $D^{(1)}_n$ of level 1~\cite{K}:
\[
\begin{array}[h]{lcl}
[e^{ab}(y) , e^{cd}(z) ] &=& (\delta_{b,-c} e^{ad}(z) + \delta_{b, -d}
e^{cd}(z) -\delta_{a,-c} e^{bd}(z)
-\delta_{a,-d} e^{cb}(z))\delta(y-z)\\[3mm]
&+& (\delta_{a, -d} \delta_{b, -c} -
 \delta_{a, -c} \delta_{b, -d}) \delta_z^\prime(y-z).
\end{array}
\]
The Lie algebra $D^{(1)}_n$ acts on the Clifford algebra via the
adjoint representation
\[
[e^{ab}(y) , \psi^c(z)]=(\delta_{b,-c} \psi^a(z) - \delta_{a, -c}
\psi^b(z))\delta(y-z) .
\]
Let $\alpha^a (z) = e^{a, -a} (z)$, then we have :

\begin{equation}
  \label{2.9}
  \begin{array}[h]{lcl}
[\psi^a (y), \alpha^b (z)] &=&-s(a) \delta_{|a| , b}
\psi^a (z)\delta(y-z), \\[3mm]
[\alpha^a (y), \alpha^b (z)] &=&\delta_{a,b}\delta_z^\prime(y-z),
  \end{array}
\end{equation}
in particular, the $\alpha_m^a$ form an oscillator algebra. Here and
further we let
$s(a)= a/|a|$.

The charge operator $\alpha_0$ of the previous section is now
$\alpha_0 = \sum\limits_{a=1}^n \alpha^a_0$. Again it is possible to
express the fermionic fields in terms of the oscillator
algebra. However in the $n$-component case we need $n$ additional
operators $Q_a (a=1,2,\ldots , n)$ on $F$ defined (uniquely) by
\[
\begin{array}[h]{lcl}
Q_a |0\rangle  &=& \psi^a_{-\frac{1}{2}} |0\rangle , \\[3mm]
Q_a \psi^b_k &=& (-)^{1-\delta_{a,|b|}} \psi^b_{k-s(b)
  \delta_{a,|b|}} Q_a.
\end{array}
\]
They satisfy the following commutation relations
\[
\begin{array}[h]{lcl}
Q_a Q_b &=& -Q_b Q_a \quad \mbox{if}\quad a\neq b,\\[3mm]
{[}\alpha_k^a , Q_b {]} &=& \delta_{ab} \delta_{k,0} Q_b,\\[3mm]
Q_av_0&=&-v_0Q_a. .
\end{array}
\]

As in the $1$-component case one has

\begin{theorem}
  \label{t2.2}(~\cite{TV})
  \begin{equation}
    \label{2.10}
    \psi^a (z) = Q_a^{s(a)} z^{s(a) \alpha_0^a} \exp
    (-s(a)\sum_{k<0} \frac{\alpha^a_k}{k} z^{-k}) \exp
    (-s(a) \sum_{k>0} \frac{\alpha^a_k}{k} z^{-k}) .
  \end{equation}
\end{theorem}

We can now describe the $n$-component  charged (or untwisted) boson-fermion
correspondence. Let $\frak{L}$ be the root lattice of type $B_n$, i.e.,
$\frak{L}$ is a
lattice with basis $\delta_a (1\leq q\leq n)$ over $\mathbb{Z}$ and
symmetric bilinear form $(\delta_i | \delta_j)=\delta_{ij}$. Let
\begin{equation}
  \label{2.11}
  \epsilon_{ij} = \left\{
  \begin{array}[h]{lcl}
-1 &\mbox{if} & i>j,\\
1 & \mbox{if} & i\leq j.
  \end{array}\right.
\end{equation}

Define a bimultiplicative function $\epsilon : \frak{L} \times \frak{L}
\rightarrow
\{\pm 1\}$ by letting
\begin{equation}
  \label{2.12}
  \epsilon (\delta_i , \delta_j) = \epsilon_{ij} .
\end{equation}

Let $\delta = \delta_1 + \delta_2 + \ldots + \delta_n$ and set $\frak{M}=\{
\gamma
\in\frak{L} | (\delta | \gamma)=0\}$  and $\frak{N}=\{ \gamma \in \frak{L} |
(\delta |
\gamma)\in 2\mathbb{Z}\}$. Then $\frak{M}$ (resp. $\frak{N}$) is the root
lattice of
$A_{n-1}$ (resp. $D_n$). Consider the vector space
$\mathbb{C}_\epsilon [\frak{L}]$ with basis
$e^{\gamma} , \gamma \in\frak{L} $, and the following twisted group algebra
product
\[
e^{\alpha} e^{\beta} = \epsilon (\alpha , \beta) e^{\alpha + \beta} .
\]

In the $1$-component case $q=e^{\delta_{\frac{1}{2}}} $. Let $\mathbb{C}[x]$
be the space of polynomials in the indeterminates $t=\{ t^{(a)}_k \},
k=1,2,\ldots , a=\frac{1}{2}, \frac{3}{2}, \ldots ,
n-\frac{1}{2}$. Define $B=\mathbb{C} [t] \otimes_ \mathbb{C}
\mathbb{C}_\epsilon [\frak{L}]$, the tensor product of algebras. Then the
$n$-component boson-fermion correspondence is the vector space
isomorphism
\[
\sigma:F\tilde{\rightarrow} B
\]

given by
\begin{equation}
 \label{2.13}
  \begin{array}[h]{l}
\sigma (\alpha^{a_1}_{-m_1} \alpha^{a_2}_{-m_2} \ldots
\alpha^{a_s}_{-m_s} Q_{\frac{1}{2}}^{k_{\frac{1}{2}}}
Q_{\frac{3}{2}}^{k_{\frac{3}{2}}} \ldots
Q_{n-\frac{1}{2}}^{k_{n-\frac{1}{2}}}|0\rangle )=\\[2mm]
m_1 m_2 \ldots m_s
t_{m_1}^{(a_1)} t_{m_2}^{(a_2)} \ldots t_{m_s}^{(a_s)} \otimes
e^{k_{\frac{1}{2}}\delta_{\frac{1}{2}} + k_{\frac{3}{2}} \delta_{\frac{3}{2}} +
\ldots
+k_{n-\frac{1}{2}} \delta_{n-\frac{1}{2}}}
  \end{array}
\end{equation}

The transported charge is then as follows
\[
\mbox{charge} \;(p(t) \otimes e^{\gamma}) = (\delta \mid \gamma).
\]

We denote the transported charge decomposition by
\[
B = \bigoplus_{m\in \mathbb{Z}} B_m .
\]

Notice that one has the following isomorphism
\[
\displaystyle \sigma(F_{\overline{\nu}}) = \bigoplus_{m\in \nu + 2 \mathbb{Z}}
B_m.
\]

The transported action of the operators is as follows
\begin{equation}
  \label{2.14}
  \begin{array}[h]{lcl}
\sigma \alpha^a_{-m}\sigma^{-1} (p(t) \otimes e^{\gamma})
&=& mt^{(a)}_m p(t) \otimes
e^{\gamma},\\[3mm]
\sigma \alpha^a_m \sigma^{-1} (p(t) \otimes e^{\gamma})
&=&\displaystyle
 \frac{\partial
  p(t)}{\partial t^{(a)}_m} \otimes e^{\gamma},\\[3mm]
 \sigma\alpha^a_0 \sigma^{-1} (p(t) \otimes e^{\gamma}) &=& (\delta_a \mid
\gamma) p(t) \otimes e^{\gamma},\\[3mm]
\sigma Q_a \sigma^{-1} (p(t) \otimes e^{\gamma}) &=& \epsilon (\delta_a ,
\gamma) p(t) \otimes e^{\gamma + \delta_a},
  \end{array}
\end{equation}

so $\sigma Q_a \sigma^{-1} = e^{s(a) \delta_a}$. Introduce for
$\alpha\in \frak{L}$ the operator $Z^\alpha$ on $B$:
\begin{equation}
  \label{2.15}
  z^\alpha ((p(t)\otimes e^{\gamma}) = z^{(\alpha \mid \gamma)}p(t)\otimes
e^{\gamma} .
\end{equation}

Then
\begin{equation}
  \label{2.16}
  \sigma \psi^a (z) \sigma^{-1} = e^{s(a) \delta_a} z^{s(a)
    \delta_a} e^{s(a) \xi_a (t,z)} e^{-s(a) \eta_a (t,z)},
\end{equation}

where
\begin{equation}
  \label{2.17}
  \xi_a (t,x) = \sum^{\infty}_{i=1} t^{(a)}_i z^i \quad
  \mbox{and}\quad \eta_a (t,z) = \sum^{\infty}_{i=1} \frac{1}{i}
  \frac{\partial}{\partial t^{(a)}_i} z^{-i} .
\end{equation}

\subsection{$1$-component neutral fermions}
\label{subs2.3}

Here we recall the classical neutral boson-fermion correspondence. Let
Let $\{ \psi_i\}_{i\in \mathbb{Z}}$ be a basis of $V$ such that
\[
(\psi_i,\psi_j)=(-)^i\delta_{i,-j}
\]
and $\psi_i$ with $i>0$ lie in $U_0$. Define the
generating series
\[
\psi (z) = \sum\limits_{j\in \mathbb{Z}} \psi_j z^{-j} .
\]
Then we have:
\begin{equation}
\label{2.18}
[\psi(y),\psi(-z)]_+=z\delta(y+z).
\end{equation}
Let $\psi (z)_+ = \sum^{\infty}_{i=0} \psi_{-i} z^i$ and $\psi (z)_- =
\psi (z) - \psi(z)_+$. Define
\[
\alpha(z) = \sum\limits_{k\in 2 \mathbb{Z}+1} \alpha_k
 z^{-k-1} = :
\psi(z)  \frac{\psi(-z)}{z}:=\psi (z)_+  \frac{\psi(-z)}{z} -
\frac{\psi(-z)}{z} \psi(z)_- .
\]
Then it is straightforward to check (using Wick's formula) that:
\begin{equation}
  \label{2.20}
  \begin{array}[h]{lcl}
[\psi (y), \alpha(z)] &=&\psi(-z)\delta(y+z) -\psi(z)\delta(y-z),\\[3mm]
[\alpha(y), \alpha(z)] &=&\delta_z^\prime (y-z) - \delta_z^\prime (y+z) ,
  \end{array}
\end{equation}

which means that ($m\in2\mathbb{Z}+1,n\in\mathbb{Z}$):
\[
\begin{array}[h]{lcl}
[\alpha_m , \psi_n ] &=& 2 \psi_{n-m},\\[2mm]
{[}\alpha_m , \alpha_n{]} &=& 2m \delta_{m, - n} .
\end{array}
\]

The $\alpha_m$ form a {\it twisted oscillator algebra}. There
is no charge operator $\alpha_0$, therefore the elements $\psi_i$ from
the Clifford algebra are called neutral fermions. In this case it is
possible to express the fermions completely in terms of the oscillator
algebra.

\begin{theorem}(~\cite{DJKM2})
  \label{t2.3}
\[
\psi(z) = -\frac{1}{\sqrt{2}} \exp (-\sum\limits_{k<0, \mbox{odd}}
\frac{\alpha_k}{k} z^{-k} ) \exp ( -\sum\limits_{k>0,\mbox{odd}}
\frac{\alpha_k}{k} z^{-k}) .
\]
\end{theorem}

The neutral or twisted boson-fermion correspondece consists of
identifying the space $F$ with the space $B=\mathbb{C} [ t_1 , t_3 ,
t_5 , \ldots ]$ via the vector space isomorphism
$
\sigma : F \rightarrow B
$
given by
\[
\sigma (\alpha_{-m_1} \alpha_{-m_2} \ldots \alpha_{-m_s} |0\rangle ) = m_1
m_2 \ldots m_s t_{m_1} t_{m_2} \ldots t_{m_s}.
\]

The transported action of the operators $\alpha_m$ is as follows
\[
\begin{array}[h]{lcl}
\sigma \alpha_{-m} \sigma^{-1} (p(t)) &=& m t_m p(t),\\[3mm]
\sigma \alpha_m \sigma^{-1} (p(t)) &=&\displaystyle  2 \frac{\partial
p(t)}{\partial t_m}.
\end{array}
\]

Then
\[
\sigma \psi(z) \sigma^{-1} =-\frac{1}{2} \sqrt{2} e^{\xi (t,z)} e^{-\eta
(t,z)}\qquad \mbox{where}
\]

\[
\xi (t,z) = \sum\limits_{i=0}^{\infty} t_{2i + 1} z^{2i + 1} \quad
\eta (t,z) = \sum\limits^{\infty}_{i=0} \frac{2}{2i+1}
\frac{\partial}{\partial t_{2i + 1}} z^{-2i -1}.
\]

\subsection{$n$-component neutral fermions}
\label{subs2.4}
In the {\it neutral} (or {\it twisted}) case we make a small distinction
between $n$ even and $n$ odd. So let $n=2m+1$ if $n$ is odd and $n=2m$
if $n$ is even. We choose the following basis of $V$.

\begin{equation}
  \label{2.21}
  \begin{array}[h]{lll}
\{\psi^a_i\}_{i\in \mathbb{Z} , a=0, \pm 1 , \pm 2 , \ldots \pm m}
&\mbox{if} & n \;\;\mbox{is odd},\\[2mm]
\{\psi^a_i\}_{i\in \mathbb{Z} , a=\pm 1 , \pm 2 , \ldots \pm m} \cup \{v_0\}
&\mbox{if} & n \;\;\mbox{is even},
  \end{array}
\end{equation}

such that they satisfy the following conditions
\begin{equation}
  \label{2.22}
  \begin{array}[h]{lll}
(\psi^a_i , \psi^b_j ) = (-)^i \delta_{a,b} \delta_{i,-j}, & & \\[2mm]
(\psi^a_i , v_0) = 0 & \mbox{if} & n \;\;\mbox{is even},\\[2mm]
\psi^a_i \in U_0 &\mbox{if}& i>0.
  \end{array}
\end{equation}
Then
\begin{equation}
\label{2.19}
[\psi^a(y),\psi^b(-z)]_+=\delta_{ab}z\delta(y+z).
\end{equation}
For simplicity we assume that $\psi^0_0 = v_0$ (if  $n$ is odd). The
elements $\psi^{\pm a}_{0} (1\leq a \leq m)$ are combinations of both
creation and annihilation operators. To be more precise, there exist
elements $u_{-1} , u_{-2}, \ldots , u_{-m} \in U_0$ and $u_1, u_2,
\ldots , u_m \in (U_0 + \mathbb{C} v_0)^{\perp}$ such that
\begin{equation}
  \label{2.23}
  \begin{array}[h]{llllcll}
(u_j , \psi^a_i) &=& 0 & \mbox{for all}\;\; a \;\;\mbox{and}\;\; i\neq
0,\\[2mm]
(u_i , u_j) &=& \delta_{i, -j} & &
  \end{array}
\end{equation}
and
\begin{equation}
  \label{2.24}
  \psi_0^j = \left\{
  \begin{array}[h]{ll}
\frac{1}{2} \sqrt{2} (u_j + u_{-j}) &j>0,\\[2mm]
\frac{i}{2} \sqrt{2} (u_{-j} - u_j ) & j<0.
  \end{array}\right.
\end{equation}
Then
\begin{equation}
  \label{2.25}
  \psi^j_0 | 0 \rangle = \left\{
  \begin{array}[h]{rl}
\frac{1}{\sqrt{2}} u_j | 0\rangle & j>0,\\[2mm]
\frac{i}{\sqrt{2}} u_j | 0\rangle & j<0,\\[2mm]
-\frac{1}{ \sqrt{2}} | 0 \rangle & j=0.
  \end{array}\right.
\end{equation}

As in the previous section we introduce
\begin{equation}
  \label{2.26}
  \alpha^a (z) = \sum\limits_{k\in 2\mathbb{Z} +1} \alpha_k^a z^{-k-1} =
  : \psi(z) \frac{\psi (-z)}{z} :.
\end{equation}

Using (\ref{2.20}) we find
\begin{equation}
  \label{2.27}
  \begin{array}[h]{lcl}
[\psi^a (y), \alpha^b (z)] &=&\delta_{ab} (\psi(-z)\delta(y+z) -
\psi (z)\delta(y-z)),\\[3mm]
[\alpha^a (y), \alpha^b (z)] &=&\delta_{ab} (\delta_z^\prime (y-z) -
\delta_z^\prime (y+z) ).
  \end{array}
\end{equation}

The fermionic fields can be expressed in terms of the oscillators
$\alpha^a_k$. However, in the multicomponent case one needs some extra
operators $Q_a, 0\leq | a| \leq m$, which are uniquely defined by
\begin{equation}
  \label{2.28}
  \begin{array}[h]{lcl}
Q_a\psi^b_n        &=& -(-)^{\delta_{ab}} \psi^b_n Q_a ,\\[2mm]
Q_a |0\rangle  &=& \sqrt{2} \psi^a_0 |0\rangle .
  \end{array}
\end{equation}

It is straightforward to check that
\[
\begin{array}[h]{lclcl}
Q_a Q_b        &+& Q_b Q_a        &=& 2\delta_{ab},\\[2mm]
Q_a \alpha_m^b &-& \alpha_m^b Q_a &=& 0 .
\end{array}
\]

Notice that the space of highest weight vectors for the oscillator
algebras, i.e., $f\in F$ such that $\alpha^a_k f=0$ for all $a
$ and $k>0$ is $2^m$-dimensional. It has as basis elements the set
\[
\begin{array}[h]{l}
Q_1^{k_1} Q^{k_2}_2 \ldots Q^{k_m}_m | 0 \rangle =
u_1^{k_1} u_2^{k_2} \ldots u_m^{k_m}  | 0\rangle,\quad \hbox{where }k_i = 0
\;\;\mbox{or}\;\; 1 .
\end{array}
\]

One now has
\begin{theorem}(~\cite{TV2})
  \label{t2.4}
  \begin{equation}
    \label{2.29}
    \psi^a (z) = \frac{Q_a}{\sqrt{2}} \exp
    (-\sum\limits_{k<0,\mbox{odd}} \frac{\alpha_k^a}{k} z^{-k})\exp
    (-\sum\limits_{k>0,\mbox{odd}} \frac{\alpha_k^a}{k} z^{-k}).
  \end{equation}
\end{theorem}
We will now describe the $n$-component neutral (or twisted)
boson-fermion correspondence. It caries the name twisted, because the
expressions in Theorem \ref{t2.4} will be related to the twisted
vertex operators as described in~\cite{FLM}, ~\cite{KP}.

Let, as in section 2.2, $\frak{L}$ be the lattice over $\mathbb{Z}$ with
basis $\delta_a, \;\;1\leq a\leq m$ and symmetric bilinear form
$(\delta_a| \delta_b) = \delta_{ab}$. Let  $\delta=\delta_1 + \delta_2 +
\ldots + \delta_m$ and let $\epsilon :  \frak{L}\times \frak{L} \rightarrow
\{\pm 1\}$
be the bimultiplicative function defined by (\ref{2.10}),
(\ref{2.11}). Let $\frak{K}$ be the even sublattice with basis $2\delta_a,
\;\;1\leq a\leq m$. Then $\varepsilon$ induces a bimultiplicative function,
which we also denote by $\varepsilon$, on $\frak{L}/\frak{K}$. Consider the
vector space
$\mathbb{C}_\epsilon [\frak{L}/\frak{K}]$ with basis $e^{\gamma} , \gamma
\in\frak{L}/\frak{K} $ and the
following twisted group algebra product
\[
e^{\alpha} e^{\beta} = \epsilon (\alpha , \beta) e^{\alpha + \beta} .
\]

Let $B= \mathbb{C} [t] \otimes_{\mathbb{C}} \mathbb{C}_\epsilon
[\frak{L}/\frak{K}]$ where
$\mathbb{C} [t]$ is the space of polynomials in the indeterminates
$t_k^{(a)}$, $k=1,3,5,\ldots$, $1\leq a\leq n$. Then the neutral or
twisted boson-fermion correspondence is the vector space isomorphism
$\sigma : F\rightarrow B$

given by
\[
\begin{array}[h]{l}
\sigma (\alpha^{a_1}_{-m_1} \alpha^{a_2}_{-m_2} \ldots
\alpha^{a_s}_{-m_s} Q_1^{k_1} Q_2^{k_2} \ldots Q_m^{k_m} | 0\rangle) = \\[3mm]
m_1 m_2 \ldots m_s t_{m_1}^{(a_1)} t_{m_2}^{(a_2)} \ldots
t_{m_s}^{(a_s)} \otimes e^{k_1 \delta_1 + \kappa_2 \delta_2 + \ldots +
  k_m \delta_m}.
\end{array}
\]

One has the following isomorphism
\[
\sigma(F_{\overline{\nu}}) = B_{\overline{\nu}} ,
\]

where $B_{\overline{\nu}}$ are the eigenspaces of the operator $1\otimes
(-1)^{\delta}$, i.e. $B=B_{\overline{0}} + B_{\overline{1}}$ with
\[
B_{\overline{\nu}} = \{ b\in B | 1 \otimes (-1)^{\delta} b=
(-)^{\nu}b \}.
\]
The transported action of the operators is as
follows $(k>0, 1\leq a < m,\  0\leq |b| \leq m)$:
\begin{equation}
  \label{2.30}
  \begin{array}[h]{lcl}
\sigma \alpha^{b}_{-k} \sigma^{-1} (p(t) \otimes e^{\gamma})&=&
kt_k^{(b)} p(t) \otimes e^{\gamma}, \\[3mm]
\sigma \alpha^b_{k} \sigma^{-1}  (p(t) \otimes e^{\gamma}) &=&\displaystyle
2\frac{\partial p(t)}{\partial t^{(b)}_{k}} \otimes e^{\gamma},\\[4mm]
\sigma Q_a \sigma^{-1} (p(t) \otimes e^{\gamma}) &=& \epsilon(\delta_a
, \gamma) p(t) \otimes e^{\gamma + \delta_a},\\[3mm]
\sigma Q_{-a} \sigma^{-1} (p(t) \otimes e^{\gamma}) &=&
i(-)^{(\delta_a | \gamma)} \epsilon (\delta_a , \gamma ) p(t) \otimes
e^{\gamma + \delta_a}, \\[3mm]
\sigma Q_0 \sigma^{-1} (p(t) \otimes e^{\gamma}) &=& -(-)^{(\delta |
  \gamma)} p(t) \otimes e^{\gamma}.
  \end{array}
\end{equation}

{}From now on we omit the tensor symbol $\otimes$ and write as before
$\sigma Q_a \sigma^{-1} =e^{\delta_a}$ for $0\le a\leq m$. Then one has
for $0\leq a\leq m$:
\begin{equation}
  \label{2.31}
  \begin{array}[h]{lcll}
\sigma \psi^a (z) \sigma^{-1} &=& \displaystyle\frac{e^{\delta_a}}{\sqrt{2}}
\exp (\xi_a (t,z) \exp (-\eta_a (t,z)), &\\[3mm]
\sigma \psi^0 (z) \sigma^{-1} &=&\displaystyle \frac{-(-)^{\delta}}{\sqrt{2}}
\exp (\xi_0 (t,z) \exp (-\eta_0 (t,z)) &\mbox{if $n$ is odd},\\[3mm]
\sigma \psi^{-a} (z) \sigma^{-1} &=&\displaystyle
\frac{ie^{\delta_a}}{\sqrt{2}}
(-)^{\delta_a} \exp (\xi_{-a} (t,z) \exp (-\eta_{-a} (t,z)), &
  \end{array}
\end{equation}

where
\begin{equation}
  \label{2.32}
\xi_b (t,z) = \sum\limits^{\infty}_{k=1} t^{(b)}_{2k-1} z^{2k-1}\quad
\mbox{and}\quad
\eta_b (t,z) = \sum\limits^{\infty}_{k=1} \frac{2}{2k-1}
\frac{\partial}{\partial t_{2k-1}^{(b)}} z^{1-2k} .
\end{equation}

\subsection{General case}
\label{subs2.5}

The general case will be a combination of $n$ charged fermions and
$n'$ neutral fermions, hence it combines the sections \ref{subs2.2}
and \ref{subs2.4}. As before we assume that $n' = 2m + 1$ if $n'$ is
odd and $n'=2m$ if $n'$ is even. For notational convenience, we
introduce three sets $CF, NF$ and $SF$ (Charged, Neutral and Simple
Fermions):
\[
\begin{array}[h]{lcl}
NF &=& \left\{
      \begin{array}[h]{ll}
        \{0,\pm 1, \pm 2, \ldots , \pm m\} & \mbox{if $n'$ is odd},\\[2mm]
        \{ \pm 1, \pm 2, \ldots ,\pm m\}   & \mbox{if $n'$ is even},\\[2mm]
      \end{array}\right.\\[2mm]
CF &=& \{ \pm (m+1) , \pm (m+2) , \ldots , \pm (m+n)\},\\[2mm]
SF &=& \left\{
       \begin{array}[h]{ll}
         \{ 0\} &\mbox{if $n'$ is even},\\[2mm]
         \emptyset   &\mbox{ otherwise}.
       \end{array}\right.
\end{array}
\]

We choose the basis elements $\psi^a_i$ of $V$, where $i\in
\frac{1}{2} +\mathbb{Z}, i\in \mathbb{Z}, i=0$ for $a\in CF, NF, SF$,
respectively. These $\psi^a_i$ satisfy
\[
(\psi^a_i , \psi^b_j) = \left\{
\begin{array}[h]{ll}
(-)^i \delta_{a,b} \delta_{i, -j} & \mbox{if}\;\; a\in NF,\\[2mm]
\delta_{a,-b} \delta_{i,-j}       & \mbox{if}\;\; a\in CF,\\[2mm]
\delta_{ab} \delta_{0,j}          & \mbox{if}\;\; a\in SF
\;\;\mbox{with}\;\;i=0.
\end{array}\right.
\]

Moreover we assume that
\[
\psi^a_i \in U_0\;\;\mbox{if}\;\; i>0\quad \psi^0_0 = v_0 ,
\]

and that the $\psi^a_0$, $ a\neq 0$, are given by (\ref{2.24}), where
$u_{-1}, u_{-2} , \ldots , u_{-m} \in U_0$ and $u_1, u_2 , \ldots , u_m
\in (U_0 \oplus \mathbb{C} v_0)^{\perp}$ are linearly independent andsatisfy
(\ref{2.23}). Introduce the fermionic fields $\psi^a (z)$ by

\[
\psi^a (z) = \left\{
\begin{array}[h]{ll}
\sum\limits_{i\in \frac{1}{2} + \mathbb{Z}} \psi^a_i z^{-i
  -\frac{1}{2}} & \mbox{for}\;\; a\in CF,\\[2mm]
\sum\limits_{i\in \mathbb{Z}} \psi^a_i z^{-i} &\mbox{for}\;\; a\in NF.
\end{array}\right .
\]

The commutators $[\psi^a (y), \psi^b (z)]$ (resp. $\psi^a (y) \frac{\psi^b
  (-z)}{z}$) for $a\in CF$ (resp. $a\in NF$) are given by (\ref{2.8})
(resp. (\ref{2.19})). Introduce as before the oscillator algebra
\[
\alpha^a (z) = \left\{
\begin{array}[h]{lcl}
\sum\limits_{i\in \mathbb{Z}} \alpha^a_i z^{-i-1} &=& :\psi^a (z)
\psi^{-a} (z) :\quad \mbox{for }a\in CF,\\[2mm]
\sum\limits_{i\in 2\mathbb{Z} +1} \alpha^a_i z^{-i-1} &=& : \psi^a (z)
\frac{\psi^a(-1)}{z}: \quad\quad \mbox{for }a\in NF.
\end{array}\right.
\]

Then the commutators $[\psi^a (y), \alpha^b (z)]$ and $ [\alpha^a(y,)
\alpha^b (z)]$
for
$a\in CF$ (resp. $a\in NF$) are given by (\ref{2.9})
(resp. (\ref{2.27})).

In order to express the fermionic fields in terms of the oscillator
algebra, we have to introduce,as before,  some new operators $Q_a , -m \leq a
\leq
m+n$, $a\ne 0$ and we let, for notational convenience, that $Q_0 =
\sqrt{2} v_{0}$ if $0\in SF$. These $Q_a$ are uniquely defined by
\[
\begin{array}[h]{lcr}
Q_a \psi^b_i &=& -\psi^b_i Q_a\quad \mbox{for} \;\; a\neq |b|,\quad\;\\[2mm]
Q_a \psi^{\pm a}_{i} &=& \left\{
     \begin{array}[h]{rl}
       \psi^{\pm a}_{i\mp s(a)} Q_a & \mbox{for}\;\; a\in CF,\\[2mm]
       \pm \psi^{\pm a}_i Q_a & \mbox{for}\;\; a\in NF,\\[2mm]
     \end{array}\right.\\[2mm]
Q_0 \psi^0_0 &=& \psi^0_0 Q_0, \qquad\qquad\qquad\quad\\[2mm]
Q_a | 0\rangle &=& \left\{
     \begin{array}[h]{rl}
       \psi^a_{-\frac{1}{2}} | 0 \rangle&\mbox{for}\;\; a\in CF,\\[2mm]
       \psi^a_0 | 0\rangle &\mbox{for} \;\; a\not\in CF.\\[2mm]
     \end{array}\right.
\end{array}
\]

So they satisfy the following commutation relations

\[
\begin{array}[h]{ll}
Q_a Q_b + Q_b Q_a = 0 & \mbox{for}\;\; a\neq b,\\[2mm]
Q^2_a = 1             & \mbox{if}\;\; a\not\in CF.\\[2mm]
{[} \alpha^a_k , Q_b{]} = \delta_{k,0} \delta_{ab} Q_b &
\end{array}
\]

\begin{theorem}
  \label{t2.5}
Let $a\in CF$ $a>0$, $b\in NF$, then
\[
\begin{array}[h]{lcl}
\psi^{\pm a} (z) &=&\displaystyle Q^{\pm 1}_a z^{\pm \alpha^a_0} \exp (\mp
\sum\limits_{k<0} \frac{\alpha_k}{k} z^{-k}) \exp (\mp
\sum\limits_{k>0} \frac{\alpha_k}{k} z^{-k}),\\[3mm]
\psi^b (z) &=&\displaystyle \frac{Q_b}{\sqrt{2}} \exp (- \sum\limits_{k<0,
  \mbox{odd}} \frac{\alpha^b_k}{k} z^{-k})  \exp (-
\sum\limits_{k>0,\mbox{odd}} \frac{\alpha^b_k}{k} z^{-k}).
\end{array}
\]
\end{theorem}

The space of highest weight vectors for the oscillator algebra,
i.e. $f\in F$ such that $\alpha^a_k f=0$ for all $k>0$, has as basis
the following vectors
\[
Q_1^{k_1} Q_2^{k_2} \ldots Q_{m+n}^{k_{m+n}} | 0 > ,
\]

where $k_i = \pm 1$ for $i=1,2,\ldots , m$ and $k_i \in \mathbb{Z}$ for
$i = m+1 , m+2, \ldots , m+n$.

In a similar way as in the previous sections we introduce the lattice
$\frak{L}$ over $\mathbb{Z}$ with basis $\delta_a$, $a=1,2,\ldots , m+n$ and
symmetric bilinear form $(\delta_a | \delta_b) = \delta_{ab}$. Let
$\delta = \delta_1 + \delta_2 + \ldots + \delta_{n+m}$ and let
$\epsilon : \frak{L} \times  \frak{L}\rightarrow \{ \pm 1 \}$ be the
bimultiplicative function defined by (\ref{2.11} - \ref{2.12}). Let
$\frak{K}$ be the even sublattice of $\frak{L}$ generated by $2\delta_a \;\;
1\leq
a\leq m$. Then $\epsilon$ induces a bimultiplicative function, which
we also denote by $\epsilon$, on $\frak{L}/\frak{K}$.

It will be convenient to introduce the sublattice
$\frak{N}=\frak{N}_{\overline{0}} =
\{ \alpha \in \frak{L} | (\alpha | \delta) \in 2\mathbb{Z} \}$ and let
$\frak{N}_{\overline{1}} = \{ \alpha \in\frak{L} | (\alpha | \delta)\in
2\mathbb{Z} +
  1\}$. Then $\frak{N}/\frak{K}$ is well defined. From now on we will use the
  notation $\overline{\frak{L}}, \overline{\frak{L}}_{\overline{0}}$ and
  $\overline{\frak{L}}_{\overline{1}}$ for $\frak{L}/\frak{K},
,\frak{N} /\frak{K},\frak{N}_{\overline{1}}/\frak{K}$
  respectively. If $n' = 0$ we also introduce
\[
\frak{M}_k = \{ \alpha \in \frak{L} | (\alpha | \delta) = k\},
\]

then notice that $\frak{M}_0$ is the root lattice of $s\ell_n$. The sets
$\overline{\frak{L}}, \overline{\frak{L}}_{\overline{0}}$ and
$\overline{\frak{L}}_{\overline{1}}$ and $\frak{M}_k$ will be related to $F,
F_{\overline{0}}, F_{\overline{1}}, F_k$, respectively.

Let $B=\mathbb{C} [t] \otimes_{\mathbb{C}}
\mathbb{C}_\epsilon[\overline{\frak{L}}]$
where $\mathbb{C} [t]$ is the space of polynomials in the
indeterminates $t= \{ t^{(a)}_{k}\},\ k = 1,2,\ldots$ for $0<a\in CF$
and $k=1,3,5$ for $a\in NF$. The {\it orthogonal boson-fermion
correspondence}
is the isomorphism
$\sigma : F\tilde{\rightarrow} B$
given by
\[
\begin{array}[h]{l}
\sigma (\alpha^{a_1}_{-m_1} \alpha^{a_2}_{-m_2} \ldots
\alpha^{a_s}_{-m_s} Q_1^{k_1} Q_2^{k_1} \ldots Q^{k_{n+m}}_{n+m} |
0\rangle)=\\[2mm]
m_1 m_2 \ldots m_s t_{m_1}^{(a_1)} t_{m_2}^{(a_2)} \ldots
t^{(a_s)}_{m_s} e^{k_1 \delta_1 + k_2 \delta_2 + \ldots + k_{n+m}
  \delta_{n+m}} .
\end{array}
\]

The transported operators are then given by (\ref{2.30}) for $|a|<m$
and (\ref{2.14}) for $a=m+1, \ldots , m+n$. Then the formulas for the
transported fields $\sigma \psi^a (z) \sigma^{-1}$ are given by
(\ref{2.31}) and (\ref{2.32}) for $|a| < m$ and (\ref{2.16}) and
(\ref{2.17}) for $a\in CF$.

\section{Multicomponent BKP and DKP hierarchies; bilinear equations
  for the $\tau$-functions and the wave functions}

Using the orthogonal boson-fermion correspondence, we can reformulate
the BKP (\ref{1.9}), DKP (\ref{1.10}), Modified DKP (\ref{1.20}), KP
and modified KP hierarchies (\ref{1.17}) as  systems of Hirota
bilinear equations. The BKP hierarchy contains all the equations
of all the other hierarchies.The DKP, KP and first modified KP
hierarchies can be obtained by a reduction of BKP when $n'$ is even,
$n'=0$ and $n'=0$, respectively.  The boson-fermion correspondece in
the $n'$ odd case does
not allow to express the operators $S'$ in terms of fermionic
fields. Reduction to the KP and modified KP hierarchies are only
possible for $n' = 0$, since only then it is possible to express $S^+$ and
$S^-$ in terms of the fermionic fields.

It will be convenient to make
a distinction between $n'$ even and $n'$ odd. For that purpose we
introduce
\[
 \varepsilon(n') = \left\{
 \begin{array}[h]{llll}
0 &\mbox{if} & n' &\mbox{is even},\\
1 &\mbox{if} & n'  &\mbox{is odd}.
 \end{array}\right.
\]

Observe that we can rewrite the BKP hierarchy (\ref{1.9}) in terms of the
fermionic
fields as follows:

\begin{equation}
  \label{3.1}
\begin{array}{rcl}
 \displaystyle
 &\mbox{Res}_{z=0} dz \{\displaystyle (1-\epsilon (n')) \frac{\psi^0_0 \otimes
    \psi^0_0}{z} + \epsilon (n') \psi^0 (z) \otimes \frac{\psi^0
  (-z)}{z}\\[3mm]
&\displaystyle
 +\sum\limits^m_{a=1}(\psi^a (z) \otimes \frac{\psi^a (z)}{-z} +
\psi^{-a} (z) \otimes \frac{\psi^{-a} (-z)}{z})\\[3mm]
&\displaystyle
 +\sum\limits^{m+n}_{a=m+1} (\psi^a (z) \otimes \psi^{-a} (z) -
\psi^{-a} (-z) \otimes \psi^a (-z))\} \tau
\otimes \tau\\[3mm]
& =\frac{1}{2} \tau\otimes \tau .
\end{array}
\end{equation}
We write $\sigma (\tau) =
\sum_{\gamma \in \overline{\frak L}} \tau_{\gamma} (t) e^{\gamma}$,
then using the orthogonal boson-fermion correspondence of section
\ref{subs2.5},
we rewrite (\ref{3.1}) and take the coefficient of $e^{\alpha} \otimes
e^{\beta}$, for every $\alpha, \beta \in \overline{\frak L}$. One thus
obtains the following equation:
\begin{equation}
  \label{3.2}
  \begin{array}[h]{cl}
& \mbox{Res}_{z=0} dz \{ \frac{1-\epsilon (n')}{2z} ((-)^{(\delta \mid
  \alpha + \beta)} - 1) \tau_{\alpha} \otimes \tau_{\beta} \\[3mm]
+ & \frac{\epsilon (n')}{2z} \left( (-)^{(\delta \mid \alpha)} E^0 (t,
z)\tau_\alpha\otimes (-)^{(\delta \mid \beta)} E^0 (t,-z)\tau_{\beta}
- \tau_{\alpha} \otimes \tau_{\beta}\right)\\[3mm]
+& \frac{1}{2z} \sum\limits^m_{a=1} \epsilon(\delta_a , \alpha) E^a
(t, z) \tau_{\alpha - \delta_{a}} \otimes \epsilon (\delta_{a},
\beta) E^a (t, -z)\tau_{\beta + \delta_a} \\[3mm]
-& \frac{1}{2z} \sum\limits^m_{a=1} \epsilon(\delta_{a} , \alpha)
(-)^{(\delta_a \mid \alpha)} E^{-a} (t, z) \tau_{\alpha + \delta_{a}}
\otimes \epsilon (\delta_a , \beta) (-)^{(\delta_a \mid \beta)}
E^{-a} (t_, -z) \tau_{\beta - \delta_a} \\[3mm]
+& \sum\limits_{a=m+1}^{m+n} \epsilon (\delta_a , \alpha) z^{(\delta_a
\mid \alpha)-1} E^a_+ (t, z) \tau_{\alpha - \delta_a} \otimes
\epsilon({\delta_a , \beta})z^{-(\delta_a \mid \beta)-1} E^a_- (t,- z)
\tau_{\beta + \delta_a} \\[3mm]
-& \sum\limits^{m+n}_{a=m+1} \epsilon(\delta_a , \alpha)
(- z)^{-(\delta_a \mid \alpha) -1} E^a_- (t,
z) \tau_{\alpha + \delta_a} \otimes \epsilon(\delta_a , \beta)
(-z)^{(\delta_a \mid \beta)-1} E^a_+ (t, -z)
\tau_{\beta - \delta_a} \}\\[3mm]&\qquad = 0.
  \end{array}
\end{equation}
Here
\begin{equation}
  \label{3.2b}
\begin{array}{lcl}
  E^a (t, z) &=& e^{\xi_a (t, z)} e^{-\eta_a (t, z)} \\[3mm]
              &=& e^{\sum\limits^{\infty}_{k=1} t^{(a)}_{2k-1}
                z^{2k-1}} e^{-\sum\limits^{\infty}_{k=1}
                \frac{\partial}{\partial t^{(a)}_{2k-1}}
\frac{z^{1-2k}}{2k-1}}\;\;
                \mbox{for}\;\;\mid a\mid \leq m,
\end{array}
\end{equation}

\begin{equation}
  \label{3.3}
  \begin{array}[h]{lcl}
E^a_{\pm} (t, z) &=& e^{\pm \xi_a (t,{\pm} z)} e^{\mp \eta_a (t,
  \pm z)}\\[3mm]
                  &=& e^{\pm \sum\limits^{\infty}_{k=1} t_k^{(a)} (\pm
                    z)^k} e^{\mp \sum\limits^{\infty}_{k=1}
                    \frac{\partial}{\partial t^{(a)}_k} \frac{(\pm
                      z)^{-k}}{k}} \;\;\mbox{for}\;\; a>m.
  \end{array}
\end{equation}

Further notice that
$\tau_{\alpha - \delta_a}=\tau_{\alpha + \delta_a}$
for $\mid a \mid \leq
m$. Identifying $\mathbb{C} [t] \otimes \mathbb{C} [t]$ with
$\mathbb{C} [t', t'']$, where we write $t'$ (resp. $t''$) for the
first (resp. second) component of the tensor product. Making the
change of variables
\begin{equation}
  t^{(a)}_k = \frac{1}{2} (t^{(a)'}_k + t^{(a)''}_k ) \quad s^{(a)}_k
  = \frac{1}{2} (t^{(a)'}_k - t^{(a)''}_k )
\end{equation}
(\ref{3.2}) becomes:

\begin{equation}
  \label{3.5}
  \begin{array}[h]{ll}
 & \mbox{Res}_{z=0} dz \{ ( \frac{(-)^{(\delta_a \mid \alpha + \beta)}
   -1}{2z} + \frac{\epsilon (n')}{2z} (-)^{(\delta_a \mid \alpha +
   \beta)} (e^{\xi_0 (2s,z)} e^{-\eta_0 (s,z)} -1))\times\\[3mm]
  & \tau_{\alpha}
 (t+s) \tau_{\beta} (t-s) +\\[3mm]
+ & \frac{1}{2z} \sum\limits^m_{a=1} \epsilon(\delta_a , \alpha +
\beta)(e^{\xi_a (2s,z)} e^{-\eta_a (s,z)} -(-)^{(\delta_a \mid \alpha
  + \beta)} e^{\xi_{-a}(2s,z)} e^{-\eta_{-a} (s,z)} ) \times\\[3mm]
 &\tau_{\alpha - \delta_a} (t+s) \tau_{\alpha + \delta_b}(t-s) +\\[3mm]
+ & \sum\limits^{m+n}_{a=m+1} \epsilon (\delta_a , \alpha + \beta)
z^{(\delta_a \mid \alpha - \beta)-2} e^{\xi_a (2s,z)} e^{-\eta_a
  (s,z)} \tau_{\alpha - \delta_a} (s+t) \tau_{\beta + \delta_a} (t-s)+
\\[3mm]
+ & \sum\limits^{m+n}_{a=m+1} \epsilon(\delta_a , \alpha + \beta)
z^{(\delta_a \mid \beta - \alpha)-2} e^{-\xi (2s,z)} e^{\eta_a (s,z)}
\tau_{\alpha + \delta_a} (s+t) \tau_{\beta -\delta_a} (t-s)\} =0.
  \end{array}
\end{equation}

Here we have used the fact that Res$_{z=0} dz f(-z)=-$ Res$_{z=0} dz
f(z)$. We can rewrite (\ref{3.5}) using elementary Schur polynomials,
which are defined by (\ref{0.5}):

\[
\begin{array}[h]{ll}
& (\frac{(-)^{(\delta_a \mid \alpha + \beta)-1}}{2} +
\frac{\epsilon(n')}{2} (-)^{(\delta_a \mid \alpha + \beta)}
(\sum\limits^{\infty}_{k=0} P_k (2s^{(0)}) P_k (2s^{(0)}) P_k (-2
\tilde{\frac{{\partial}}{\partial s^{(0)}}})-1))\times \\[3mm]
&\tau_\alpha (t+s)
\tau_{\beta} (t-s) +\\[3mm]
+ &\frac{1}{2} \sum\limits^m_{a=1} \epsilon(\delta_a , \alpha+\beta)
\sum\limits^{\infty}_{k=0} P_k (2s^{(a)}) P_k
(-2\tilde{\frac{{\partial}}{\partial s^{(a)}}}) -(-)^{(\delta_a \mid
  \alpha + \beta)} P_k (2s^{(-a)})\times\\[3mm]
& P_k (-2
\tilde{\frac{{\partial}}{\partial S^{(-a)}}})  \tau_{\alpha -
    \delta_a} (s+t) \tau_{\beta + \delta} (t-s) + \\[3mm]
+& \sum\limits^{m+n}_{a=m+1} \epsilon(\delta_a , \alpha + \beta)
\sum\limits^{\infty}_{k=0} P_k (2s^{(a)}) P_{k-1 + (\delta_a \mid
  \alpha - \beta)} (-\tilde{\frac{{\partial}}{\partial s^{(a)}}})\times\\[3mm]
&\tau_{\alpha - \delta_a} (s+t) \tau_{\beta + \delta_a} (t-s)+\\[3mm]
+& \sum\limits^{m+n}_{a=m+1} \epsilon (\delta_a , \alpha + \beta)
\sum\limits^{\infty}_{k=0} P_k (-2 s^{(a)}) P_{k-1 + (\delta_a \mid
  \beta - \alpha)} (\tilde{\frac{{\partial}}{\partial s^{(a)}}})\times\\[3mm]
 &\tau_{\alpha + \delta_a} (s+t) \tau_{\beta - \delta_a} (t-s) =0.
\end{array}
\]

where

\[
\tilde{\frac{{\partial}}{\partial s^{(a)}}} = \left\{
\begin{array}[h]{ll}
(\frac{\partial}{\partial S^{(a)}_1} , \frac{1}{2}
\frac{\partial}{\partial S^{(a)}_2} , \frac{1}{3}
\frac{\partial}{\partial S^{(a)}_3}, \ldots ) & a> m,\\[3mm]
(\frac{\partial}{\partial S^{(a)}_1} , \frac{1}{3}
\frac{\partial}{\partial S^{(a)}_3}, \frac{1}{5}
\frac{\partial}{\partial S^{(a)}_5}, \ldots ) & \mid a \mid \leq m.
\end{array}\right.
\]
Using Taylors formula we rewrite this once more:

\begin{equation}
  \label{3.6}
  \begin{array}[h]{ll}
 & \left( \frac{(-)^{(\delta_a \mid \alpha + \beta)} -1}{2} +
 \frac{\epsilon (n')}{2} (-)^{(\delta_a \mid \alpha + \beta)} \left(
 \sum\limits^{\infty}_{k=0} P_k (2s^{(0)} ) P_k (-2
\tilde{ \frac{{\partial}}{\partial u^{(0)}}}) -1 \right)
\right)\times\\[3mm]
 & D(s,u) \tau_{\alpha} (t+u) \tau_\beta (t-u)\mid_{u=0}\\[3mm]
+& \frac{1}{2} \sum\limits^{m}_{a=1} \epsilon (\delta_a , \alpha +
\beta) \sum\limits^{\infty}_{k=0} P_k (2s^{(a)}) P_k (-2
\tilde{\frac{{\partial}}{\partial u^{(a)}}}) - (-)^{(\delta_a \mid \alpha
  + \beta)} P_k (2s^{(-a)})\times\\[3mm]
& P_k (-2\tilde{ \frac{{\partial}}{\partial
  u^{(-a)}}})
  D(s,u) \tau_{\alpha - \delta_a} (t+u) \tau_{\beta + \delta_u} (t-u)
 \mid_{u=0} +\\[3mm]
 +& \sum\limits^{m+n}_{a=m+1} \epsilon(\delta_a , \alpha + \beta)
 \sum\limits^{\infty}_{k=0} P_k (2s^{(a)}) P_{k-1+(\delta_a \mid \alpha
   - \beta)} (-\tilde{\frac{{\partial}}{\partial u^{(a)}}}) D(s,u)\\[3mm]
 & \tau_{\alpha - \delta_a} (t+u) \tau_{b + \delta_a} (t-u)
 \mid_{u=0}+\\[3mm]
+ & \sum\limits_{a=m+1}^{m+n} \epsilon (\delta_a , \alpha + \beta)
\sum\limits^{\infty}_{k=0} P_k (-2s^{(a)}) P_{k-1 + (\delta_a \mid
  \beta - \alpha)} (\tilde{\frac{{\partial}}{\partial u^{(a)}}})
D(s,u)\times\\[3mm]
 & \tau_{\alpha + \delta_a} (t+u) \tau_{\beta - \delta_a} (t-u)
 \mid_{u=0} = 0 .
  \end{array}
\end{equation}

where

\[
D(s,u) = e^{\sum\limits^{m}_{a=1} \sum\limits^{\infty}_{r=1}
  s^{(a)}_{2r-1} \frac{\partial}{\partial u^{(a)}_{2r-1}}}e^{
\epsilon(n') \sum\limits^{\infty}_{r=0} S^{(0)}_{2r-1}
\frac{\partial}{\partial u^{(0)}_{2r-1}} } e^{\sum\limits^m_{a=1}
\sum\limits^{\infty}_{r=1} s^{(-a)}_{2r-1}
   \frac{\partial}{\partial u^{(-a)}_{2r-1}}}
 e^{\sum\limits^{m+n}_{a=m+1} s^{(a)}_{r} \frac{\partial}{\partial u^{(a)}_r}}
\]

In fact (\ref{3.6}) is a generating series of Hirota bilinear
equations (see the Introduction).

The equations for the DKP and modified DKP hierarchies are only given
if $n'$ is even, so $\epsilon (n')=0$. Then the two DKP hierarchies are
the set of equations for wich either $\alpha, \beta \in
\overline{\frak{L}}_0$ or $\alpha, \beta \in \overline{\frak{L}}_1$. The
modified
DKP is the set of equations for which $\alpha \in
\overline{\frak{L}}_{\overline{0}}$ and $\beta \in
\overline{\frak{L}}_{\overline{1}}$ (or the other way around which gives the
same set of equations). The equations for the KP or first modified KP
hierarchy appear only if $n' = m = \epsilon (n') = 0$ and one assumes
that $\tau_{\alpha} = 0$ for all $\alpha$ except $\alpha \in\frak{ M}_k$,
    respectively $\alpha \in\frak{ M}_k \cup\frak{ M}_{k+1}$. It is
straightforward
    to see that for those $\tau$'s the equations (\ref{3.6}) decouple
    into two sets of equations, which are in fact equivalent by
    Theorem (\ref{t1.14}).

We wil now give the bilinear equation for the wave
functions. Following T. Shiota~\cite{S} in which he describes
wave functions for the 2 component BKP hierarchy, we are lead in
this general case ($n, n'$ non-negative integers) to introduce two
wave functions. But before doing that it will be convenient to replace
in (\ref{3.4}) $t^{(a)}_{j}$ by $t^{(a)}_j+ \delta_{j 1}
x_0$ if $a\geq 0$ and by $t^{(a)}_j + \delta_{j1} x_1$ if $a<0$, then
of course $\tau_\gamma$ will be a function of all $t^{(a)}_j$'s and $x_0 ,
x_1$, we will use the notation $\tau_{\gamma} (x,t)$.

We will reformulate (\ref{3.2}) in such a way that we obtain
expressions involving matrix valued formal pseudodifferential
operators in $\partial_0 = \frac{\partial}{\partial x_0}$ and
$\partial_1 = \frac{\partial}{\partial x_1}$.

Introduce
\[
P^{\pm \sigma} (\alpha , x , t , \pm z) = (P^{\pm \sigma} (\alpha , x,
t, \pm z)_{jk})_{-n \leq j, k\leq n+m}
\]

for $\sigma = 0, 1$ as follows. Let $\lambda = +, -$ and assume
$j=1,2,\ldots , m+n$ if $\lambda = +$ and $j= m+1, m+2 , \ldots , m+n$
if $\lambda = -$, then define
\[
  \begin{array}[h]{ll}
&P^{\pm 0} (\alpha , x, t, \pm z)_{\lambda j + \delta_{\lambda,-} m,k}
=\\[3mm]
&\qquad\left\{
  \begin{array}[h]{cll}
\frac{\epsilon_{kj} e^{-\eta_k (t, \pm z)} \tau_{a \pm \lambda
    \delta_j \mp \delta_k} (x,t)}{\tau_{\alpha} (x,t)} & \mbox{if} &
k=1,\ldots ,m,\\[3mm]
\frac{\epsilon_{kj} z^{\lambda \delta_{jk} -1} e^{\mp \eta_k (t,z)}
  \tau_{a \pm \lambda \delta_j \mp \delta_k} (x,t)}{\tau_{\alpha}
  (x,t)} & \mbox{if}& k = m+1 , \ldots , m+n,\\[3mm]
-\frac{(1-\epsilon (n')) \tau_{a \pm \lambda \delta_j} (x,t) -
    \epsilon(n') e^{-\eta_0 (t, \pm z)} \tau_{\alpha \pm \lambda
    \delta_j} (x,t)}{\tau_{\alpha} (x,t)} & \mbox{if} & k=0,\\[3mm]
\frac{\epsilon_{j, m-k} z^{-\lambda \delta_{j,m-k} -1} e^{\pm
    \eta_{m-k} (t, -z)} \tau_{\alpha \pm \lambda \delta_j \pm
    \delta_{m-k}} (x,t)}{\tau_{\alpha} (x,t)} &\mbox{if} & k= -1, \ldots , -n,
  \end{array}\right.
  \end{array}
\]

\[
\begin{array}{ll}
&P^{\pm 0} (\alpha , x , t, \pm z)_{0,k} =\\[3mm]
&\qquad\left\{
\begin{array}[h]{cll}
\frac{e^{-\eta_k (t, \pm z)} \tau_{\alpha \mp \delta_k}
  (x,t)}{\tau_{\alpha} (x,t)} &\mbox{if}& k=1, \ldots , m,\\[3mm]
\frac{z^{-1} e^{\mp \eta_k (t,z)} \tau_{\alpha \mp \delta_k}
    (x,t)}{\tau_a (x,t)} &\mbox{if}& k=m+1, \ldots , m+n,\\[3mm]
1 - \epsilon(n') + \epsilon (n') \frac{e^{-\eta_0 (t, \pm
    z)}\tau_{\alpha} (x,t)}{\tau_{\alpha} (x,t)} & \mbox{if}&
k=0,\\[3mm]
-\frac{z^{-1} e^{\pm \eta_{m-k} (t,-z)} \tau_{a \pm \delta_{m-k}}
  (x,t)}{\tau_{a} (x,t)} &\mbox{if}& k=1,\ldots , -n,
\end{array}\right.
\end{array}
\]
\begin{equation}\label{3.7} \ \ \end{equation}

\[
P^{\pm 1} (\alpha , x , t, \pm z)_{jk} =
\left\{
  \begin{array}[h]{cll}
\frac{\epsilon_{jk} e^{-\eta_{-k} (t, \pm z)} \tau_{\alpha \pm\delta_j
   \mp \delta_k} (x,t)}{\tau_{a} (x,t)} &\mbox{if}& k=1, \ldots ,
m,\\[3mm]
\delta_{jk} &\mbox{if}& k=m+1 , \ldots, m+n,\\
            &\mbox{or}& k=-1, \ldots , -n,\\[3mm]
\frac{\tau_{\alpha \pm \delta_j} (x,t)}{\tau_{\alpha} (x,t)} &\mbox{if}&k=0,
  \end{array}\right.
\]

for $j=1,2,\ldots , m+n$ and

\[
P^{\pm 1} (\alpha , x, t, \pm z)_{0k} =
\left\{
  \begin{array}[h]{cll}
\delta_{k0} &\mbox{if}& k\neq 1,2,\ldots , m,\\[3mm]
\frac{e^{-\eta_{-k} (t, \pm z)} \tau_{\alpha \mp
    \delta_k} (x,t)}{\tau_{\alpha} (x,t)} &\mbox{if}& k=1, \ldots , m,
  \end{array}\right.
\]

\[
P^{\pm 1} (\alpha, x, t, \pm z)_{jk} =
\left\{
  \begin{array}[h]{cll}
\delta_{jk} &\mbox{if}& k\neq 0,\\[3mm]
\frac{\tau_{\alpha \mp} \delta_{m-j} (x,t)}{\tau_{\alpha} (x,t)}
&\mbox{if}& k=0,
  \end{array}\right.
\]

for $j=-1, -2, \ldots , -n$.

Next define $Q^{\pm \sigma} (t, \pm z)$, $R^{\pm \sigma} (\alpha ,
\pm z) \;\;(\sigma = 0,1)$ and
$S(z)$as follows:

\begin{equation}
  \label{3.4}
  \begin{array}[h]{lcl}
Q^{\pm 0} (t, \pm z) &=& \sum\limits^{m+n}_{j=1} e^{\pm \xi_j (t, z)}
E_{jj}+ (1 - \epsilon (n') + \epsilon(n') e^{\pm \xi_0 (t, z)}) E_{00}+\\[3mm]
& &+ \sum\limits^{-1}_{j=-n} e^{\mp \xi_{m-j} (t,-z)} E_{jj},\\[3mm]
Q^{\pm 1} (t, \pm z) &=& \sum\limits^m_{j=1} e^{\pm \xi_{-j} (t, z)}
E_{jj} + \sum\limits^{m+n}_{j = m+1} E_{jj} + \sum\limits^{-1}_{j=-n}
E_{jj},
\end{array}
\end{equation}
\begin{equation}
  \label{3.9}
  \begin{array}[h]{lcl}
R^{\pm 0} (\alpha, \pm z) &=& \sum\limits^m_{j=1} \epsilon (\delta_j ,
\alpha) E_{jj} + \sum\limits^{m+n}_{j=m+1} \epsilon(\delta_j , \alpha)
z^{\pm (\delta_j |\alpha)} E_{jj}\\[3mm]
 & & +(-)^{(\delta\mid \alpha)} E_{00} + \sum\limits^{-1}_{j=-n}
 \epsilon (\delta_{m-j} , \alpha)(-z)^{\mp (\delta_{m-j} \mid \alpha)}
E_{jj},\\[3mm]
  R^{\pm 1} (\alpha , \pm z) &=& \sum\limits^m_{j=1} \epsilon (\delta_j
  , \alpha) (-)^{(\delta_j \mid \alpha)} E_{jj} +
  \sum\limits^{m+n}_{j=m+1} E_{jj} + \sum\limits^0_{j=-n} E_{jj},
\end{array}
\end{equation}

\begin{equation}
  \label{3.10}
  S(z) = \frac{1}{2z} \sum\limits^m_{j=0} E_{jj} +
  \sum\limits^{m+n}_{j=m+1} E_{jj} - \sum\limits^{-1}_{j=-n} E_{jj}.
\end{equation}

And introduce the wave operators
\begin{equation}
  \label{3.11}
  W^{\pm \sigma} (\alpha , x, t, \pm z) = P^{\pm \sigma} (\alpha, x,
  t, \pm z) Q^{\pm \sigma} (t, \pm z) R^{\pm \sigma} (\alpha , \pm z)
\end{equation}
and the wave functions
\begin{equation}
  \label{3.12}
  \Psi^{\pm \sigma} (\alpha , x, t, z) = W^{\pm \sigma} (\alpha, x, t,
  \pm z) e^{\pm x_{\sigma} z} .
\end{equation}
Then (\ref{3.2}) is equivalent to the following bilinear identity of
the wave functions
\begin{equation}
  \label{3.13}
  \begin{array}[h]{c}
\mbox{Res}_{z=0} dz \Psi^{+ 0} (\alpha, x, t, z) S(z)^t \Psi^{-0}
(\beta , x', t', z) =\\[3mm]
\mbox{Res}_{z=0} dz \Psi^{+1} (\alpha , x, t, z) S(z)^t \Psi^{-1}
(\beta , x' , t' , z).
  \end{array}
\end{equation}

\section{Multicomponent BKP and DKP hierarchies; Sato and Lax
  equations} 

We will now rewrite the bilinear identity for the wave functions in
terms of formal pseudo-differential operators.\\

Let $N$ be some positive integer, $\partial = \frac{\partial}{\partial
  x}$, a formal $N\times N$ matrix pseudodifferential operator with
parameters $t=(t_j^{(a)})$ is an expression of the form
\begin{equation}
  \label{4.1}
  P(x,t,\partial )=\sum\limits_{j \in \mathbb{Z}} P_j (x,t) \partial^j ,
\end{equation}

where the $P_j$'s are $N\times N$ matrices over the algebra of formal
power series over $\mathbb{C}$ in the indeterminates $x$ and $t$. The
largest $M$ such that $P_M \neq 0$ is called the order of
$P(x,t,\partial)$ (we write ord$P(x,t,\partial)=M$, we sometimes allow
$M$ to be $\infty$). One defines the differential part of
$P(x,t,\partial)$ by $P(x,t,\partial)_+ = \sum_{j\geq 0} P_j(x,t)
\partial^j$ and let $P_- = P - P_+$. The vector space $\Psi$ over
$\mathbb{C}$ of all formal psuedodifferential operators decomposes in a
natural way ${\Psi} = {\Psi}_- \oplus
{\Psi}_+$ and turns into a algebra; the product is defined by

\[
\partial^k \circ P(x,t)\partial^\ell = \sum\limits_{j=0}^{\infty} \left(
\begin{array}[h]{c}
k \\ j
\end{array}\right) \frac{\partial^j P(x,t)}{\partial x^j}
\partial^{k+\ell -j} .
\]

One has a natural linear anti-involution $*$ on ${\Psi}$, it is
defined by
\[
(P(x,t) \partial^{\ell})^* = (-\partial)^{\ell} \circ\;^tP(x,t)
\]

It preserves the decomposition $\Psi = \Psi_- \oplus \Psi_+$.

The following Lemma will be very essential, for a proof see~\cite{KV}.

\begin{lemma}
  \label{l4.1}
Let $P(x,t,\partial)$ and $Q(x,t,\partial)$ be two formal
pseudodifferential operators. Then
\[
\mbox{Res}_{z=0} dz (P(x,t,\partial) e^{zx})( \:^{t} Q(x', t', \partial')
e^{-zx'}) = \sum\limits^{\infty}_{j=0} R_j (x,t,t') \frac{(x-x')^j}{j!}
\]

if and only if
\[
(P(x,t,\partial) \circ Q(x,t',\partial)^*)_- =
\sum\limits^{\infty}_{j=0} R_j (x,t,t') \partial^{-j-1}
\]
\end{lemma}

As a consequence of this lemma one has
\begin{corollary}
  \label{c4.2}
Let $P(x,t,\partial)$ and $Q(x,t,\partial)$ be two formal matrix
pseudodifferential operators. Then
\[
\mbox{Res}_{z=0} dz (P(x,t,\partial) e^{zx}) (\;^t Q(x', t' , \partial')
e^{-zx'}) = \sum\limits_j R_j (x,t) S_j (x' , t')
\]

if and only if
\[
(P(x,t,\partial) Q(x,t' ,\partial)^*)_- = \sum\limits_j R_j (x,t)
\partial^{-1} S_j (x,t') .
\]
\end{corollary}

Now let $\partial = \partial_0$ or $\partial_1$. Then
\[
\Psi^{\pm \sigma} (\alpha , x_0 , x_1 , t , z) = W^{\pm \sigma}
(\alpha , x_0 , x_1, t, \partial_{\sigma}) e^{\pm x_{\sigma} z}
\]

where

\[
W^{\pm \sigma} (\alpha , x_0 , x_1 , t, \partial_{\sigma}) = P^{\pm
  \sigma} (\alpha , x_0 , x_1 , t, \partial_{\sigma}) Q^{\pm \sigma}
(t, \partial_{\sigma}) R^{\pm \sigma} (t, \partial_{\sigma}).
\]

Using the above corollary we deduce that (\ref{3.13}) is
equivalent to
\begin{equation}
  \label{4.2a}
  \begin{array}[h]{c}
(W^{+0} (\alpha, x_0 , x_1, t, \partial_0) S(\partial_0) W^{-0}
(\beta, x_0 , x_1', t', \partial_0)^* )_{-}\\[3mm]
= \mbox{Res}_{z=0} dz W^{+1} (\alpha , x_0 , x_1 , t,z) S(z) e^{x_1z}
\partial_0^{-1}
\:^t W^{-1} (\beta , x_0 , x'_1, t' z) e^{-x'_1 z}
  \end{array}
\end{equation}
and
\begin{equation}
  \label{4.2b}
  \begin{array}[h]{c}
(W^{+1} (\alpha , x_0 , x_1 , t, \partial_1) S(\partial_1)
W^{-1}(\beta , x'_0 , x_1 , t' , \partial_1)^*)_{-}\\[3mm]
= \mbox{Res}_{z=0} dz W^{+0} (\alpha , x_0 , x_1 , t, z) S(z) e^{x_0
  z}
\partial^{-1}_1 \;^t W^{-0} (\beta, x'_0 , x_1 , t' , z)e^{-x_0z},
  \end{array}
\end{equation}

{}from which one deduces:

\[
\begin{array}[h]{l}
(W^{+0} (\alpha , x_0 , x_1 , t, \partial_o) S(\partial_0)
\partial^{-1}_{1} W^{-0} (\beta , x_0 , x_1, t' ,  \partial_0)^*)_- \\[3mm]
= (W^{+1} (\alpha , x_0 , x_1 , t,  \partial_1) S( \partial_1)
\partial_0^{-1} W^{-1} (\beta , x_0 , x_1 , t' ,  \partial_1 )^* )_-.
\end{array}
\]
Hence

\[
\begin{array}[h]{l}
(W^{+0} (\alpha , x_0 , x_1 , t,  \partial_0)S( \partial_0) W^{-0} (\beta
, x_0 , x_1 , t' ,  \partial_0 )^* )_- =\\[3mm]
\mbox{Res}_{ \partial_1} W^{+1} (\alpha , x_0 , x_1, t,  \partial_1)
S( \partial_1)  \partial_0^{-1} W^{-1} (\beta , x_0 , x_1 , t' ,
\partial' )^*
\end{array}
\]

and

\[
\begin{array}[h]{l}
(W^{+1} (\alpha , x_0 , x_1 , t ,  \partial_1) S( \partial_1) W^{-1}
(\mathbb{ }, x_0 , x_1 , t' ,  \partial_1)^* )_- =\\[3mm]
 \mbox{Res}_{ \partial_0} W^{+0} (\alpha , x_0 , x_1 , t,
\partial_0) S( \partial_0)  \partial_1^{-1} W^{-0} (\beta , x_0 , x_1
, t' ,  \partial_0)^*
\end{array}
\]
where
\[
\mbox{Res}_{\partial} \sum\limits_{j\in \mathbb{Z}}
P_j (x,t) \partial^k = P_{-1} (x,t).
\]

Notice that
\begin{equation}
  \label{4.3}
  \begin{array}[h]{lcl}
Q^{\pm \sigma} (t,  \partial_{\sigma})^* &=& Q^{\mp} (t,
\partial_{\sigma} )^{-1},\\[2mm]
R^{\pm \sigma} (\alpha ,  \partial_{\sigma} )^* &=& R^{\mp \sigma}
(\alpha,  \partial_{\sigma})^{-1} = R^{\mp} (-\alpha,  \partial_{\sigma}).
  \end{array}
\end{equation}

Unfortunately, we cannot prove that $P^{\pm \sigma} (\alpha , x_0 ,
x_1 , t,  \partial_{\sigma})$ is invertible for
arbitrary $m$. Hence  we do not know if it is possible to define the
operator $L_{\sigma} \equiv L_{\sigma} (\alpha , x_0 , x_1 , t,
\partial_{\sigma}) = P^{+\sigma} (\alpha , x_0 , x_1 , t,
\partial_{\sigma})\partial_{\sigma} P^{+\sigma} (\alpha , x_0 , x_1 ,
t, \partial_{\sigma})^{-1}$ such that
\[
L_{\sigma} \psi^{+\sigma}
(\alpha , x_0 , x_1 , t, z) = z \psi^{+\sigma} (\alpha , x_0 , x_1 , t
, z).
\]
For that reason {\it we assume from now on that} $m=0$. It is
then straightforward to check from the definition of $P^{\pm \sigma}
(\alpha , x_0 , x_1 , t, \partial_{\sigma})$ (see (\ref{3.7})) that
$P^{\pm \sigma} (\alpha , x_0 , x_1 , t, \partial_{\sigma})$ is indeed
invertible, and that
\begin{equation}
  \label{4.4}
  \begin{array}[h]{lcl}
P^{\pm 1} (\alpha , x_0 , x_1 , t , \partial_1 ) &=& 2I- P^{\pm
  0}_0 (\alpha , x_0 , x_1 , t),\\[2mm]
Q^{\pm 1} (t, \partial_1 )=R^{\pm 1} (\alpha , \partial_1 ) &=& I.
  \end{array}
\end{equation}

So from now on we assume that $x_0 = x_1$ and hence that there is only
one $\partial_{\sigma}$, viz $\partial_0$, and write from now on $x$
and $\partial$ instead of $x_0$ and $\partial_0$.

We also remove all superscripts $\sigma$ and write $T^{\pm} (\alpha ,
x, t)$ instead of $P^{\pm 1} (\alpha, x, t, \partial)$. Then the
bilinear identity of the wave function $\Psi^{\pm} (\alpha, x, t, z) =
W^{\pm} (\alpha, x, t, \partial) e^{\pm xz}$ takes the following form

\begin{equation}
  \label{4.5}
\begin{array}[h]{l}
  \mbox{Res}_{z=0} dz \Psi^+ (\alpha, x, t, z) S(z) \;^t \Psi (\beta ,
  x' , t' , z)\\[3mm]
= \mbox{Res}_{z=0} dz T^+ (\alpha, x, t) S(z) e^{xz\;t} T^- (\alpha, x'
, t') e^{-x'z}.
\end{array}
\end{equation}

Using Corollary {\ref{c4.2} one deduces
  \begin{equation}
    \label{4.6}
    \begin{array}[h]{l}
(W^+ (\alpha, x, t, \partial) S(\partial) W^- (\beta, x, t',
\partial)^*)_- =\\[3mm]
T^+ (\alpha, x,t) \frac{1}{2} E_{00} \partial^{-1} \;^t T^- (\beta, x, t'),
    \end{array}
  \end{equation}

{}from which we deduce, using (\ref{4.3} -\ref{4.4}), that
\begin{equation}
  \label{4.7}
 P^- (\alpha, x, t, \partial)^* = S(\partial)^{-1} P^+ (\alpha , x, t,
 \partial)^{-1} T^+ (\alpha , x, t) S(\partial) \;^t T^- (\alpha , x, t)
\end{equation}

and that
\begin{equation}
  \label{4.8}
  (P^+ (\alpha, x, t, \partial) R^+ (\alpha - \beta , \partial) P^+
  (\beta, x,t, \partial)^{-1} T({\beta, x, t}) S(\partial))_- =
  \frac{1}{2} T(\alpha , x, t) E_{00} \partial^{-1}.
\end{equation}

In order to get nice formulas we change the wave function. Define

\begin{equation}
  \label{4.9}
\begin{array}{lcl}
  \Phi^{\pm} (\alpha , x, t, z) &=&  V^{\pm} (\alpha , x, t, \partial)
  e^{\pm xz} \quad \mbox{where}\\[3mm]
V^{\pm} (\alpha , x, t, \partial) &=& K^{\pm} (\alpha , x, t,
\partial) R^{\pm} (\alpha , \partial) Q^{\pm} (t,\partial) \quad
\mbox{and}\\[3mm]
K^{\pm} (\alpha , x,t, \partial) &=& T^{\pm} (\alpha , x, t)^{-1}
P^{\pm} (\alpha , x, t, \partial).
\end{array}
\end{equation}

Notice that
\begin{equation}
  \label{4.10}
  \begin{array}[h]{lcl}
\Phi^{\pm} (\alpha, x, t, z) &=& P^{\pm}_0 (\alpha , x, t)
\;\;\Psi^{\pm} (\alpha, x, t, z),\\[3mm]
\Psi^{\pm} (\alpha, x, t, z) &=& (\frac{3}{2} I - \frac{1}{2}
K_0^{\pm} (\alpha, x, t)) \Phi^{\pm} (\alpha, x, t, z).
  \end{array}
\end{equation}

With the above definitions, formulas (\ref{4.5}-\ref{4.8}) turn
into
\begin{equation}
  \label{A}
  \mbox{Res}_{z=0} dz \Phi^+ (\alpha, x, t, z) S(\partial) \;^t \Phi^-
  (\beta , x' , t' , z) = \frac{1}{2} E_{00},
\end{equation}
\begin{equation}
  \label{B}
  (V^+ (\alpha, x, t, \partial) S(\partial) V^- (\beta , x, t,
  \partial))_- = \frac{1}{2} E_{00} \partial^{-1},
\end{equation}
\begin{equation}
  \label{C}
  K^- (\alpha, x, t, \partial)^* = S(\partial)^{-1} K^+ (\alpha, x, t,
  \partial)^{-1} S(\partial),
\end{equation}
\begin{equation}
  \label{D}
  (K^+ (\alpha , x, t, \partial) R^+ (\alpha - \beta, \partial) K^+
  (\beta, x, t, \partial)^{-1} S(\partial))_- = \frac{1}{2} E_{00}
\partial^{-1}.
\end{equation}

To obtain Sato's equation we differentiate (\ref{A}) by
$t^{(a)}_{\ell}$ and use lemma \ref{l4.1} and (\ref{C}):
\begin{equation}
  \label{4.9a}
  \frac{\partial K^+ (\alpha, x, t, \partial)}{\partial
    t^{(a)}_{\ell}} = - (K^+ (\alpha, x, t, \partial) N(\ell , a) K^+
  (\alpha, x, t, \partial)^{-1} S(\partial))_- S(\partial)^{-1} K^+
  (\alpha, x,t,\partial),
\end{equation}

where
\[
N({\ell, a)}) = \left\{
\begin{array}[h]{l}
\partial^{\ell} E_{00} \quad \mbox{if}\;\; a=0\;\;(\mbox{only if}\;
\epsilon (n') = 1),\\[3mm]
\partial^{\ell} E_{aa} - (-\partial)^{\ell} E_{-a, -a}.
\end{array}\right.
\]
To make the formulas nicer, we introduce a new decomposition of
$\Psi$, viz, let
\begin{equation}
  \label{4.10a}
  P_\geq =(P S(\partial))_+ S(\partial)^{-1} \;\; ,\;\; P_< = P-P_\geq
\end{equation}
then
\[
\Psi = \Psi_< \oplus \Psi_\geq
\]
and (\ref{D}) and (\ref{4.9a}) turn into
\begin{equation}
  \label{4.11}
  (K^+ (\alpha, x, t, \partial) R^+ (\alpha - \beta , \partial) K^+
  (\beta , x, t, \partial)^{-1})_< = E_{00},
\end{equation}
\begin{equation}
  \label{4.12}
  \frac{\partial K^+ (\alpha, x, t, \partial)}{\partial
    t^{(a)}_{\ell}} = -(K^+ (\alpha, x, t, \partial) N(\ell , a) K^+
  (\alpha, x,t,\partial)^{-1} )_{<} K^+ (\alpha, x, t, \partial).
\end{equation}

This new decomposition of $\Psi$ is very natural, it is based on the
following observations.

\begin{lemma}
  \label{l4.3}
Let $A=\sum^n_{j=-n} E_{j, -j}$, then
\[
K^+ (\alpha , x, t, \partial) = AK^- (\alpha, x, t, \partial) A
\]
\end{lemma}

The proof of this lemma is a straightforward verification of the
following two facts\break $T^+ (\alpha , x,t, \partial)
= AT^- (\alpha , x,t, \partial)A$ and $P^+ (\alpha,
x,t,\partial) = AP^- (\alpha, x, t, \partial) A$.

Let $D=S(\partial)A$, then
\begin{equation}
  \label{4.13}
  D=\frac{1}{2} E_{00} \partial^{-1} + \sum\limits^n_{j=1} (E_{j, -j}
  - E_{-j, j})
\end{equation}

and define the linear anti involution $i: \Psi \rightarrow \Psi$ by
\begin{equation}
  \label{4.14}
  i(P (x,t,\partial))= D\; P(x,t,\partial)^* D^{-1}
\end{equation}

Then $i$ preserves the decomposition $\Psi = \Psi_{<} \oplus
\Psi_{\geq}$ and
\begin{equation}
  \label{4.15}
  \begin{array}[h]{l}
i(K^+ (\alpha, x, t, \partial)) = K^+ (\alpha, x, t, \partial)^{-1},\\[3mm]
i(Q^+ (t,\partial)) = Q^+ (t, \partial)^{-1},\\[3mm]
i (R^+ (\alpha, \partial)) = R^+ (\alpha, \partial)^{-1},\\[3mm]
i(V^+ (\alpha, x, t, \partial)) = V^+ (\alpha, x, t, \partial)^{-1},\\[3mm]
i(K^+ (\alpha, x, t, \partial) N(\ell, a) K^+(\alpha, x, t,
\partial)^{-1}) = -K^+ (\alpha, x,t,\partial) N(\ell , a) K^+
  (\alpha, x, t,\partial)^{-1}.
  \end{array}
\end{equation}

The last line of (\ref{4.15}) is a consequence of the following

\begin{lemma}
  \label{l4.4}
Let $P(x,t\partial)$ be a pseudodifferential operator then
\[
i(V^+ (\alpha, x, t, \partial) P(x,t,\partial) V^+(\alpha,
x,t,\partial)^{-1} = \lambda V^+ (\alpha, x,t,\partial)
P(x,t,\partial) V^+ (\alpha, x,t,\partial)^{-1}
\]

if and only if $i(P(x,t,\partial)) = \lambda P(x,t,\partial)$.
\end{lemma}
Notice that the subspace
\[
\Psi^i = \{ P \in \Psi \mid i(P) = -P\}
\]

is a subalgebra of $\Psi$ with respect to the product $[A,B] = A\circ
B - B\circ A$, it will play an important role in the theorey of the
BKP hierarchy.

\begin{proposition}
  \label{p4.5}
Let $\alpha , \beta \in \hbox{supp}\; \tau$ such that $\alpha - \beta =
\delta_i$ or $-\delta_i$. Then $K(\alpha)\equiv
K^+(\alpha,x,t,\partial ) $ is completely determined by $K(\beta)\equiv
K^+(\beta,x,t,\partial )$.
\end{proposition}

{\bf Proof.}\\
For simplicity of notations we only consider the case that $\alpha -
\beta = \delta_i$. Then $R^+ (\alpha-\beta) = A\partial + B +
C\partial^{-1}$ where
\[
A = E_{ii}, \quad B=-E_{00} + \sum\limits^n_{j=1, j\neq i} \epsilon_{ji}
  E_{jj}, \quad C = -E_{-i, -i}.
\]

For $K(\alpha) = \sum\limits_{j=0}^{\infty} K(\alpha)^{(j)} =
K(\alpha)^{(0)} (I+\sum\limits^{\infty}_{j=1} W(\alpha)^{(j)})$,
its inverse is
\[
\begin{array}[h]{lcl}
K(\alpha)^{-1} &=& (I - W(\alpha)^{(1)} + \ldots )
K(\alpha)^{(0)^{-1}} \\[3mm]
&=& (I-K(\alpha)^{(0)^{-1}} K(\alpha)^{(1)} + \ldots ) K(\alpha)^{(0)^{-1}}
\end{array}
\]

It follows from (\ref{4.11}) that
\[
(K(\alpha) (A\partial + B + C\partial^{-1}) K(\beta)^{-1})_< = E_{00}.
\]

Now calculating
 $(K(\alpha) (A\partial + B +C\partial^{-1}) K(\beta)^{-1})_\geq$
we obtain
\[
\begin{array}[h]{l}
K(\alpha) (A\partial + B + C\partial^{-1}) K(\beta)^{-1} = \\[3mm]
K(\alpha)^{(0)} A[K(\beta)^{(0)^{-1}} \partial + \partial
(K(\beta)^{(0)^{-1}}) - W(\beta)^{(1)}K(\beta)^{(0)^{-1}} ] + K(\alpha)^{(1)}A
  K(\beta)^{(0)^{-1}} +\\[3mm]
K(\alpha)^{(0)} B K(\beta)^{(0)^{-1}} - (K(\alpha)^{(0)}
A[\partial(K(\beta)^{(0)^{-1}}) - W(\beta)^{(1)} K(\beta)^{(0)^{-1}}
]\\[3mm]
+ K(\alpha)^{(1)} A K(\beta)^{(0)^{-1}} + K(\alpha)^{(0)} B
K(\beta)^{(0)^{-1}} -1 ) E_{00}.
\end{array}
\]

Now multiply this from the left by $K(\alpha)^{(0)^{-1}}$ and from
the right by $K(\beta)$ and equating the coefficients of $\partial^m$
we find for $\partial^0$:

\begin{equation}
  \label{4.16}
  \begin{array}[h]{l}
\{ A[\partial(K(\beta)^{(0)^{-1}}) - W(\beta)^{(1)}
K(\beta)^{(0)^{-1}} ] + W^{(1)} (\alpha) A K(\beta)^{(0)^{-1}} +\\[3mm]
B K(\beta)^{(0)^{-1}} - K(\alpha)^{(0)^{-1}} \} E_{00} K(\beta)^{(0)}
= 0
  \end{array}
\end{equation}

and for $\partial^{-j}, \quad j=1,2,\ldots$:
\begin{equation}
  \label{4.17}
  \begin{array}[h]{l}
W(\alpha)^{(j+1)} A + W(\alpha)^{(j)} B + W(\alpha)^{(j-1)} C = \\[3mm]
A[ \partial (W(\beta)^{(j)}) + W(\beta)^{(j+1)} - W(\beta)^{(1)}
W(\beta)^{(j)} ] + W(\alpha)^{(1)} A W(\beta)^{(j)} +\\[3mm]
B W(\beta)^{(j)} - \{ A[\partial (K(\beta)^{(0)^{-1}}) -
W(\beta)^{(0)} K(\beta)^{(0)^{-1}} ] + W(\alpha)^{(1)} A
K(\beta)^{(0)^{-1}}\\[3mm]
+ B K(\beta)^{(0)^{-1}} - K(\alpha)^{(0)^{-1}} \} E_{00} K(\beta)^{(j)}.
  \end{array}
\end{equation}

{}From (\ref{4.16}) we deduce that
\begin{equation}
  \label{4.18}
  \begin{array}[h]{l}
\{ A[\partial (K(\beta)^{(0)^{-1}}) W(\beta)^{(1)}
K(\beta)^{(0)^{-1}}] + W(\alpha)^{(1)} A K(\beta)^{(0)^{-1}} + \\[3mm]
B K(\beta)^{(0)^{-1}} - K(\alpha)^{(0)^{-1}} \} E_{00} = 0.
  \end{array}
\end{equation}

Now substitute this in (\ref{4.17}) and one obtains
\begin{equation}
  \label{4.19}
\begin{array}{l}
  W(\alpha)^{(j+1)} A + W(\alpha)^{(j)} B + W(\alpha)^{(j-1)} C =\\[3mm]
A[\partial (W(\beta)^{(j)}) + W(\beta)^{(j+1)} - W(\beta)^{(1)}
W(\beta)^{(j)}] + W(\alpha)^{(1)} A W(\beta)^{(j)} + B W(\beta)^{(j)}.
\end{array}
\end{equation}

It is clear from (\ref{4.19}) that alle $W(\alpha)^{(j)}$, except
$(W(\alpha)^{(1)}_{\ell i}$, can be expressed in $W(\beta)$ and
  $(W(\alpha)^{(1)})_{\ell , i}\ (-n\leq \ell \leq n)$.

Now take the $(0,0)$-entry of (\ref{4.18}) one has
\[
(W(\alpha)^{(1)})_{0i} (K(\beta)^{(0)^{-1}})_{i0} = 2.
\]

Now notice that
\begin{equation}
  \label{4.20}
  \begin{array}[h]{l}
(K(\gamma)^{(0)^{-1}} - I ) = -(K(\gamma)^{(0)} - I) = -2
(P(\gamma)^{(0)} - I),\\[2mm]
(P(\gamma)^{(0)} - I)_{jk} = 0 \quad \mbox{for}\;\; k\neq 0 \quad
\mbox{and}\;\; j=k=0.
  \end{array}
\end{equation}

Using this we take the $(j,0)$-th entry of (\ref{4.18}) and deduce
{}from this that $(W({\alpha})^{(1)})_{0i}$ can be expressed in
$K(\beta)^{(0)}, W(\beta)^{(1)}$ and $K(\alpha)^{(0)}$. Hence using
(\ref{4.20}) it suffices to determine $(P(\alpha)^{(0)})_{j0} \quad
(j\neq 0)$. \\

Now for $j\neq 0$ one has:
\[
(P(\alpha)^{(0)})_{j0} = \left\{
\begin{array}[h]{ll}
- \frac{\tau_{\beta + \delta_i - \delta_j}}{\tau_{\beta + \delta_i}} &
j>0,\\[3mm]
- \frac{\tau_{\beta + \delta_i + \underline{\delta}_j}}{\tau_{\beta +
    \delta_i}} & j< 0.
\end{array}\right.
\]

Since $(P(\beta)^{(0)})_{i0} = - \tau_{\beta + \delta_i} /
\tau_{\beta} \neq 0$ one deduces from (\ref{3.7}) that
\[
(P(\alpha)^{(0)})_{j0} = \left\{
\begin{array}[h]{ll}
\epsilon_{ij} [ (P(\beta)^{(0)})_{i0} ]^{-1}
(P(\beta)^{(1-\delta_{ij})})_{ij} &j>0\\[3mm]
-\epsilon_{i,-j} [(P(\beta)^{0)})_{i0}]^{-1}
(P(\beta)^{(1+\delta_{i,-j})})_{ij} & j<0
\end{array}\right.
\]

Hence $K(\alpha)$ is completely determined by $K(\beta)$.
\hfill{$\Box$}

{}From now on we write $V(\alpha), K(\alpha) , R(\alpha)$, etc. instead
of $V^+ (\alpha, x, t, \partial), \  K^+(\alpha , x, t, \partial)$,
etc. Define the following operators:
\begin{equation}
  \label{4.21}
  \begin{array}[h]{lcl}
L(\alpha) &=& V(a) \partial^{-1} V(\alpha)^{-1}, \\[3mm]
M(\alpha) &=& V(\alpha) x V(\alpha)^{-1} ,\\[3mm]
N(\alpha, \beta) &=& V(\alpha) V(\beta)^{-1} ,\\[3mm]
C_{ij} (\alpha) &=& V(\alpha) E_{ij} V(\alpha)^{-1}.
  \end{array}
\end{equation}
where we write $x$ and $\partial$ for $xI_{2n+1}$ and $\partial I_{2n+1}$.

Notice, that several of the products appearing in (\ref{4.21}) consist
of products of infinite order pseudodifferential operators. Mulase
showed in~\cite{M1} that being careful enough, one can
indeed take products of infinite order operators. For more details see
{}~\cite{M1}, ~\cite{M2}. The above operators act on the wave function, as
follows
\begin{equation}
  \label{4.22}
  \begin{array}[h]{lcl}
L(\alpha) \Phi (\alpha, z) &=& \Phi (\alpha , z) z,\\[3mm]
M(\alpha) \Phi (\alpha , z) &=& \Phi (\alpha , z) x,\\[3mm]
N(\alpha , \beta) \Phi (\beta, z) &=& \Phi (\alpha,z),\\[3mm]
C_{ij} (\alpha) \Phi (\alpha , z) &=& \Phi (\alpha , z) E_{ij}.
  \end{array}
\end{equation}

It is easy to see that the operators $L(\alpha), N(\alpha , \beta)$
and $C_i (\alpha) := C_{ii} (\alpha)$ are well-defined, they are
finite order formal pseudodifferential operators:
\begin{equation}
  \label{4.23}
  \begin{array}[h]{lcl}
L(\alpha) &=& K(\alpha) \partial K(\alpha)^{-1},\\[3mm]
N(\alpha , \beta) &=& K(\alpha) R(\alpha - \beta) K(\beta)^{-1},\\[3mm]
C_i (\alpha) &=& K(\alpha) E_{ii} K(\alpha)^{-1}.
  \end{array}
\end{equation}

\begin{remark}
 It is important to notice that the operators $L(\alpha), M(\alpha)$
 and $C_{ij}(\alpha)$ have the same commutation relations as
 $\partial, x$ and $E_{ij}$ respectively.
\end{remark}

We deduce from (\ref{4.11}-\ref{4.12}) that
\begin{equation}
  \label{4.24}
  \begin{array}[h]{lcl}
N(\alpha, \beta) &=& N(\alpha, \beta)_{\geq} + E_{00},\\[3mm]
\frac{\displaystyle \partial V(\alpha)}{\displaystyle\partial
t^{(a)}_{\ell}}
&=& B(\alpha , \ell
, a) V(\alpha),
  \end{array}
\end{equation}
where
\begin{equation}
  \label{4.25}
  B(\alpha, \ell , a) = (K(\alpha) N(\ell , a) K(\alpha)^{-1})_{\geq}.
\end{equation}

This leads to
\begin{equation}
  \label{4.26}
  \begin{array}[h]{lcl}
\frac{\displaystyle\partial L (\alpha)}{\displaystyle\partial
t^{(a)}_{\ell}}
&=& [B (\alpha ,
\ell , a) , L(\alpha) ],\\[3mm]
\frac{\displaystyle\partial M(\alpha)}{\displaystyle\partial
t^{(a)}_{\ell}}
&=& [B (\alpha ,
\ell , a), M(\alpha) ],\\[3mm]
\frac{\displaystyle\partial N(\alpha , \beta)}{\displaystyle\partial
t^{(a)}_{\ell}}
&=& B(\alpha ,
\ell , a) N(\alpha , \beta) - N(\alpha , \beta) B(\beta , \ell , a),\\[3mm]
\frac{\displaystyle\partial C_{ij} (\alpha)}{\displaystyle\partial
t^{(a)}_{\ell}}
&=& [B
(\alpha , \ell , a) , C_{ij} (\alpha) ].
  \end{array}
\end{equation}
and to
\begin{equation}
  \label{4.27}
  \frac{\partial \Phi (\alpha, z)}{\partial t^{(a)}_{\ell}} = B(\alpha
    , \ell , a) \Phi (\alpha , z).
\end{equation}

The equations (\ref{4.26}) are called Lax equations, they are a direct
consequence of the Sato equation.

Let $\epsilon (n') = 0$ in this case a reduction to the DKP hierarchy
and KP hierarchy is possible. Notice that a BKP tau function $\tau =
\sum_{\alpha} \tau_{\alpha} e^{\alpha}$ is a DKP (resp. KP) tau
function if $\hbox{supp} \tau \subset \frak{N}_{\overline{0}}$ or $\hbox{supp}
\tau \subset
\frak{N}_{\overline{1}}$ (resp. $\hbox{supp} \tau \subset\frak{ M}_k$ for some
$k\in
\mathbb{Z}$). The corresponding wave functions $\Phi (\alpha ,
z)$ then have a simpler form, in fact $\Phi (\alpha , z) = \Psi
(\alpha, z)$ i.e., $\Psi (\alpha , z)_{i0} = \Psi(\alpha , z)_{0i} = 0$
for $i\neq 0$. If $\tau$ is a KP tau function and $\alpha \in \hbox{supp}
\tau$, then $\Phi (\alpha , z)$ has an even simpler form. One finds
that also $\Psi (\alpha ,z)_{ij} = 0$ for $i>0>j$ and $i<0<j$. And the
corresponding wave function is of the form
\[
\left(
  \begin{array}[h]{ccc}
                    & 0      & \\
\Psi^- (\alpha, -z) & \vdots & \quad 0  \\
                    & 0      &   \\
0\ldots 0           & 1      & 0\ldots 0 \\
                    & 0      &    \\
\quad 0             & \vdots &    \Psi^+ (\alpha, z)\\
                    & 0      &
  \end{array}
\right)
\]

Here $\Psi^+ (\alpha , z)$ is a wave function of the $n$-component KP
hierarchy
(see~\cite{KV}) and $\Psi^- (\alpha , z)$ is up to permutation of the
indices the adjoint wave function.

\section{The Orlov-Schulman-Adler-Shiota-van Moerbeke formula} 

In this section we will derive a generalization of the
Orlov-Schulman-Adler-Shiota-van Moerbeke (OSASM) formula for the BKP
hierarchy. A formula that connects the action of some $W$-algebra on
the tau functions to the so-called {\it additional symmetries} on the
wave function. This formula was stated for the (1-component) KP
hierarchy by Orlov and Schulman~\cite{OS} and proved by Adler, Shiota and van
Moerbeke (~\cite{ASV1},~\cite{ASV2} see also~\cite{V}). Operators from the
subspace $\Psi^i = \{
P \in \Psi \mid i(P) = -P\}$ of the space of formal pseudodifferential
operators, which form a subalgebra with respect to the Lie product
$[A,B]=A\circ B - B\circ A$, appear in this OSASM-formula.

Notice that the subspace generated by $L, L^{-1}, M$ and $C_{ij}$ is
isomorphic to $D(gl_{2n + 1})$, the algebra of differential operators
on $\mathbb{C}[t,t^{-1}]^{2n+1}$. The intersection $D(gl_{2n + 1}) \cap \Psi^i$
then
forms a subalgebra $D^{(2)} (gl_{2n + 1})$ of $D(gl_{2n +1})$. From now
    on let $\epsilon = \epsilon(n') = 0$ if $n'=0$ and $=1$ if
    $n'=1$. If $\epsilon = 0$ we need yet another subalgebra, viz
    $D^{(2)}(gl_{2n})$. We view this algebra as the subalgebra of
    $D^{(2)}(gl_{2n+1})$ where the $0$th row and column are zero. A
    central extension $W^{(2)}_{1+\infty} (gl_{2n + \epsilon}) =
    D^{(2)} (gl_{2n+\epsilon}) \oplus \mathbb{C} c$ of
    $D^{(2)}(gl_{2n+\epsilon})$ can also be formulated in terms of the
    fermions.

Let $p, q \in \{ 0,1,2, \ldots , n\}, \nu, \sigma \in \{ +, -\}$ and define

\begin{equation}
  \label{5.1}
  X^{\nu p , \sigma q} (y,w) = : \psi^{\nu p} (\nu y) \frac{\psi^{\sigma q}
  (-\sigma w)}{w^{\delta_{q , 0}}}: ,
\end{equation}

where $p,q \neq 0$ if $\epsilon = 0$ and if $\epsilon = 1$ and $p=0$
or $q=0$ then we assume that $\nu = +, \sigma = +$
respectively. Notice that
\begin{equation}
  \label{5.2}
  X^{\nu p, \sigma q} (y,w) = \psi^{\nu p} (\nu y) \frac{\psi^{\sigma
      q}(-\sigma w)}{w^{\delta_{q,0}}} -\frac{\delta_{pq} \nu}{y-w}
  (\delta_{p,0} + (1-\delta_{p,0}) \delta_{\nu , -\sigma}).
\end{equation}
We write
\begin{equation}
  \label{5.3}
  \begin{array}[h]{lcl}
X^{\nu p, \sigma q} (y,w) &=& \sum\limits^{\infty}_{\ell =0}
\frac{\displaystyle(y-w)^{\ell}}{\displaystyle \ell !} X^{\ell + 1, \nu p,
\sigma q} (w),\\[3mm]
X^{\ell + 1, \nu p, \sigma q} (w) &=& : \frac{\displaystyle\partial^{\ell}
  \psi^{\nu p} (\nu w)}{\displaystyle\partial w^{\ell}}
\frac{\displaystyle\psi^{\sigma q}
  (-\sigma w)}{\displaystyle w^{\delta_{q0}}} :\\[3mm]
&=& \sum\limits_{k\in \mathbb{Z}} X^{\ell + 1, \nu p , \sigma q}_k
w^{-k-\ell -1}.
  \end{array}
\end{equation}
We can now calculate the Lie bracket of the elements $X^{\ell + 1, \nu
  p, \sigma q}_k$ and $X^{n+1, \mu r, \lambda s}_m$, for that purpose
we use the folowing. The operator $X^{\ell + 1 , \nu p , \sigma q}_k$
acts on the Clifford algebra by derivations via the adjoint
representation $(a \in C\ell V)$:
\[
X^{\ell + 1, \nu p , \sigma q}_k (a) = [X^{\ell + 1, \nu p , \sigma q}_k
  , a].
\]

By identifying $(j\in \mathbb{Z} , \nu = +, -)$:
\begin{equation}
  \label{5.4}
  \begin{array}[h]{lcl}
\nu^j \psi^{-\nu p}_{j+\frac{1}{2}} &=& t^j e_{\nu p} ,\\[3mm]
(-)^j \psi^0_j &=& \sqrt{2} t^j e_0 \quad \mbox{if}\;\; \epsilon =1,
  \end{array}
\end{equation}

where $e_j, \ -n \leq j\leq n$, is a basis of $\mathbb{C}^{2n + \epsilon}$
(we assume that $j\neq 0$ if $\epsilon =0$), we can identify, via
(\ref{5.4}), the elements $X_n^{\ell + 1, \nu p, \sigma q}$ with
certain elements in $W^{(2)}_{1+ \infty} (gl_{2n+\epsilon})$. One
easily verifies that
\begin{equation}
  \label{5.5}
  \begin{array}[h]{lcl}
X_k^{\ell + 1, \nu p , \sigma q} &=& \sum\limits_{j\in \mathbb{Z}}
[j]_{\ell} \nu^j (-\sigma)^{-j -k-1+\delta_{q0}} : \psi^{\nu
  p}_{-j-\frac{1-\delta_{p0}}{{2}}} \psi^{\sigma
  q}_{j+k+\frac{1-\delta_{q0}}{2}} :\\[3mm]
&=& -(-\sigma)^{1-\delta_{q0}}
s(q,p)
t^{k+\ell}
  (\frac{\partial}{\partial t})^{\ell} E_{-\sigma q , \nu p} \\[3mm]
& & -(-\nu)^{1-\delta_{p0}}
s(p,q)
  t^{\delta_{p0}} (\frac{\partial}{\partial t})^{\ell} (-t)^{\ell + k
  - \delta_{q0}} E_{-\nu p , \sigma q} ,
  \end{array}
\end{equation}
where $[j]_{\ell} = j(j-1)\ldots (j-\ell + 1)$ and
$s(p,q)=(\sqrt{2})^{\delta_{p0}-\delta_{q0}}$.

Using Wicks theorem we can calculate the commutation relations between
$X^{\ell + 1, \nu p, \sigma q}_k$ and $X_m^{\ell + 1, \mu r, \lambda
  s}$. One finds that one has a central extension:
\begin{equation}
  \label{5.6}
  \begin{array}[h]{l}
[X^{\ell +1, \vee p, \sigma q}_k, X_m^{n+1 , \mu r, \lambda s}]=\\[3mm]
\quad X_k^{\ell + 1 , \nu p , \sigma q} X_m^{n+1 , \mu r, \lambda s} -
X_m^{n+1, \mu r , \lambda s} X_k^{n+1 , \nu p, \sigma q}
+\alpha (X_k^{\ell + 1, \nu p, \sigma q} , X_m^{n+1, \mu p , \lambda s}).
  \end{array}
\end{equation}

The first part of the left-hand-side of (\ref{5.6}) can be determined
using (\ref{5.5}). The two cocycle $\alpha$ is determined as
follows. The constant term of the OPE
\[
: \psi^{\nu p} (\nu y) \frac{\psi^{\sigma q}(-\sigma
  w)}{w^{\delta_{q,0}}} : \; \; : \psi^{\mu r} (\mu u)
\frac{\psi^{\lambda s} (-\lambda v)}{v^{\delta{s,0}}} :\;\; ,
\]

where we assume that $\nu , \sigma , \mu , \lambda = +$ whenever
$p,q,r,s =0$ respectively, is equal to
\[
\begin{array}[h]{l}
\frac{\nu \delta_{ps}}{y-v} (\delta_{p0} + (1-\delta_{p0})\delta_{\nu
  , -\lambda}) \frac{\mu \delta_{qr}}{w-u} (\delta_{q0} +
(1-\delta_{q0}) \delta_{\sigma , -\mu})\\[3mm]
-\frac{ \nu \delta_{pr}}{y+u} (-\delta_{p0} u + (1-\delta_{p0})
\delta_{\nu , -\mu}) \frac{\lambda \delta_{qs}}{w+v}
(\frac{\delta_{q0}}{v} + (1-\delta_{q0}) \delta_{\sigma , -\lambda}).
\end{array}
\]

Now differentiate this $\ell$ times to $y$ and  $n$ times to $u$
and put $w = y$ and $v = u$,   we thus find
\[
\begin{array}[h]{l}
\delta_{ps} \delta_{qr} \nu \mu (\delta_{p0} + (1-\delta_{p0})
\delta_{\nu , -\lambda}) (\delta_{q0}+(1-\delta_{q0}) \delta_{\sigma ,
  -\mu}) \frac{\displaystyle (-)^{\ell} \ell ! n!}
{\displaystyle(y-u)^{\ell + n + 2}}\\[3mm]
-\delta_{pr} \delta_{qs} \nu \lambda (\frac{\delta_{q0}}{u} + (1-
\delta_{q0}) \delta_{\sigma , -\lambda})\times\\[3mm]
 ((-\delta_{p0} u+ (1-
\delta_{p0}) \delta_{\nu , -\mu}) \frac{\displaystyle(-)^{\ell +n} (\ell +
  n)!}{\displaystyle(y+u)^{\ell + n + 2}}
-\frac{\displaystyle\delta_{p0} n(-)^{\ell +n-1} (\ell
  + n -1)!}{\displaystyle(y+u)^{\ell + n + 1}} ).
\end{array}
\]

{}From which we conclude that
\begin{equation}
  \label{5.7}
  \begin{array}[h]{l}
\alpha (X_k^{\ell + 1, \nu p , \sigma q} , X_m^{n+1 , \mu r, \lambda
  s}) =\\[3mm]
\delta_{ps} \delta_{qr} \nu \mu(\delta_{p0} + (1-\delta_{p0})
\delta_{\nu , -\lambda})(\delta_{q0} + (1-\delta_{q0}) \delta_{\sigma
  , -\mu})(-)^{\ell} \ell ! n! \left(
\begin{array}[h]{c}
k+\ell\\
\ell + n + 1
\end{array}\right) \delta_{k, -m}\\[3mm]
-\delta_{pr} \delta_{qs} \nu\lambda(\delta_{q0} + (1-\delta_{q0})
\delta_{\sigma , -\lambda}) \{-\delta_{p0} (\ell + n +1)! \left(
\begin{array}[h]{c}
k+\ell + 1\\
\ell + n + 1
\end{array}\right) + \\[3mm]
(1- \delta_{p0}) \delta_{\nu , \mu} (\ell + n)! \left(
\begin{array}[h]{c}
k+\ell \\
\ell + n + 1
\end{array}\right) \} (-)^{k+\ell +1} \delta_{k, -m -\delta_{p0} +
\delta_{q0}}.
  \end{array}
\end{equation}

In a similar way as for pseudodifferential operators we have a linear
anti-involution $*$ on $D(gl_{2n + 1})$, let $A\subset gl_{2n+1}$
then we define
\begin{equation}
  \label{5.8}
  (t^k (\frac{\partial}{\partial t})^{\ell} A)^* =
  (\frac{\partial}{\partial t})^{\ell} (-t)^k \;^t\! A.
\end{equation}

Let (compare with $D$, see (\ref{4.13}))
\begin{equation}
  \label{5.9}
  C=\frac{1}{2} E_{00} t^{-1} + \sum\limits^{n}_{j=1} E_{j, -j} -E_{-j
    , j},
\end{equation}

then we also have a linear anti-involution on $D(gl_{2n+1})$ that is
compatible with $i$ on $\Psi$, for notational convenience we shall
denote this involution also by $i$; let $X\in D(gl_{2n+1})$ then
\begin{equation}
  \label{5.10}
  i(X) = C^{-1} X^* C.
\end{equation}

We can extend $i$ to $W_{1+\infty} (gl_{2n+1})$, by defining $i(c) =
-c$. Now notice that the elements (\ref{5.5}) satisfy

\begin{equation}
  \label{5.11}
  i(X_k^{\ell + 1, \nu p, \sigma q})=-X_k^{\ell + 1, \nu p, \sigma q}.
\end{equation}

Let $\alpha \in {\cal L}$, in analogy with (\ref{5.3}) and (\ref{5.5})
we define
\begin{equation}
  \label{5.12}
\begin{array}{lcl}
  Y^{\nu p , \sigma q} (\alpha, y, w) &=& \sum\limits^{\infty}_{\ell =
    0} \sum\limits_{k \in \mathbb{Z}}
\frac{(y-w)^{\ell}}{\ell !} w^{-k-\ell -1}
 \{ (-\sigma)^{1-\delta_{q0}}
s(q,p)
 M(\alpha)^{\ell} L(\alpha)^{k+\ell} C_{\nu p , -\sigma q}(\alpha) +\\[3mm]
  &&  (-\nu)^{1-\delta_{p0}}
s(p,q)
      (-L(\alpha))^{k+\ell - \delta_{q0}} M(\alpha)^{\ell}
      L(\alpha)^{\delta_{p0}} C_{\sigma q , -\nu p} (\alpha)\} ,\\[3mm]
Z^{\nu p , \sigma q} (\alpha , y , w) &=& P^{(0) -1} (\alpha) Y^{\nu p ,
  \sigma q} (\alpha , y, w) P^{(0)-1} (\alpha) .
\end{array}
\end{equation}
Then
\begin{equation}
  \label{5.13}
  \begin{array}[h]{lcl}
&&Y^{\nu p , \sigma q} (\alpha , y, w) \Psi(\alpha, z)=\\
&=&\sum\limits^{\infty}_{\ell = 0} \sum\limits_{k\in \mathbb{Z}}
\frac{(y-w)^{\ell}}{\ell !} w^{-k -\ell -1}
\{ (-\sigma)^{1-\delta_{q0}}
s(q,p)
z^{k+\ell}
(\frac{\partial}{\partial z})^{\ell}
\Psi (\alpha ,z) E_{\nu p , -\sigma q} +\\[3mm]
&&(-\nu)^{1-\delta_{p0}}
s(p,q)
z^{\delta_{p0}} (\frac{\partial}{\partial z})^{\ell}
(-z)^{k+\ell - \delta_{q0}} \Psi (\alpha , z)
E_{\sigma q, -\nu p} \}\\[3mm]
&=& (-\sigma)^{1-\delta_{q0}}
s(q,p)
\delta(w-z) \Psi (\alpha , y) E_{\nu p , -\sigma q} +\\[3mm]
&&(-\nu)^{1-\delta_{p0}}
s(p,q)
z^{\delta_{p0}} e^{(y-w)\frac{\partial}{\partial  z}}
w^{-\delta_{q0}} \delta (w+z) \Psi(\alpha, -w) E_{\sigma q , -\nu p}\\[3mm]
&=& (-\sigma)^{1-\delta_{q0}}
s(q,p)
\delta(w-z) \Psi (\alpha , y) E_{\nu p, -\sigma q} +\\[3mm]
&&(-\nu)^{1 - \delta_{p0}}
s(p,q)
(-y)^{\delta_{p0}} w^{-\delta_{q0}} \delta (y+z)
\Psi (\alpha , -w) E_{\sigma q , -\nu p}.
  \end{array}
\end{equation}

We have now given all the ingredients to determine the OSASM-formula.
The proof of this formula is based on the following simple
observation. Let $v \in V$ then
\begin{equation}
  \label{5.14}
  (v\otimes 1) S = 1\otimes v - S (v\otimes 1).
\end{equation}

{}From (\ref{5.14}) we deduce that $(u, v \in V)$
\begin{equation}
  \label{5.15}
  S(uv \tau \otimes \tau) + u\tau \otimes v\tau - v\tau \otimes u\tau
  = \frac{1}{2} uv\tau \otimes \tau .
\end{equation}

Now replacing $u$ by $\psi^{\nu p} (\nu y)$ and $v$ by $\psi^{\sigma
  q} \frac{(-\sigma w)}{w^{\delta_{q0}}}$ the equation (\ref{5.15})
still holds. Next using (\ref{5.2}) we obtain

\begin{equation}
  \label{5.16}
  \begin{array}[h]{l}
SX^{\nu p, \sigma q} (y,w) \tau \otimes \tau + \psi^{\nu p} (\nu y)
\tau \otimes \frac{\displaystyle \psi^{\sigma q} (-\sigma
w)}{\displaystyle w^{\delta_{q0}}}-
\frac{\displaystyle \psi^{\sigma q} (-\sigma w)}{\displaystyle w^{\delta_{q0}}}
\tau \otimes
\psi^{\nu p} (\nu y) \tau =\\
\frac{1}{2} X^{\nu p, \sigma q} (y,w) \tau
\otimes \tau .
  \end{array}
\end{equation}

Again we assume that $\nu , \sigma = +$ whenever $p$, respectively
$q=0$. In the same way as in section 4 we use the vertex operators for
these fermionic fields and replace all $t^{(a)}_{1}$ by $t_1^{(a)} +
x$. We then take the coefficient of
\[
e^{\alpha + \lambda \delta_j} \otimes e^{\beta - \mu \delta_k}, \quad
\lambda , \mu =+,-,\ 0\leq j, k\leq n .
\]

{}From now on we assume that $\delta_0 = 0$ whenever $\epsilon =
0$. Since $X^{\nu p, \sigma q} (y,w)$ is not well-defined on $\tau_\gamma$
we define
\begin{equation}
  \label{5.17}
  \begin{array}[h]{c}
\mathbb{X}^{\nu p} (z) \tau_{\gamma} (x,t)=((1-\delta_{p,0})
\epsilon(\delta_p,\gamma) z^{(\nu \delta_p \mid \gamma)} +
\delta_{p,0} \frac{ (-)^{(\delta \mid \gamma)}}{ \sqrt{2}}
e^{\nu \xi_p (t,z)} e^{-\nu \eta_p (t,z)}
t_{\gamma} (x,t) e^{\nu zx}\\[3mm]
\mathbb{X}^{\nu p, \sigma q} (y,w) \tau_{\gamma} (x,t)=((1-\delta_{p,0})
\epsilon (\delta_p , \gamma + \delta_q)(\nu y)^{(\nu d_p \mid \gamma +
  \sigma \delta_q)} + \delta_{p,0}\frac{(-)^{(\delta \mid \gamma +
  \sigma \delta_q)}}{ \sqrt{2}})\times\\[3mm]
e^{\nu \xi_p (t,\nu y)} e^{-\nu \eta_p (t,\nu y)}
\frac{\displaystyle \mathbb{X}^{\sigma q}
  (-\sigma w)}{\displaystyle w^{\delta_{q0}}} \tau_{\gamma} (x,t) e^{yx}.
  \end{array}
\end{equation}

Using all this (\ref{5.16}) turns into
\begin{equation}
  \label{5.18}
  \begin{array}[h]{c}
\mbox{Res}_{z=0} dz \{ \frac{1-\epsilon}{2z}
(-)^{(\delta\mid \alpha + \lambda \delta_j)}
\mathbb{X}^{\nu p, \sigma q} (y,w)
\tau_{\alpha + \lambda \delta_j - \nu \delta_p - \sigma \delta_q}
(x,t) e^{xz} \otimes
(-)^{(\delta \mid \beta - \mu \delta_k)}
\tau_{\beta - \mu \delta_k } (x,t) e^{-xz}\\[3mm]
+\frac{\epsilon (-)^{(\delta \mid \alpha + \lambda \delta_j})}{\sqrt{2}}
  e^{\xi_0 (t,z)} e^{-\eta_0 (t,z)} \mathbb{X}^{\nu p, \sigma q} (y,w)
    \tau_{\alpha + \lambda \delta_j - \nu \delta_p - \sigma \delta_q}
    (x,t) e^{xz} \otimes
\frac{\mathbb{X}^0 (-z)}{z} \tau_{\beta - \mu \delta_k}\\[3mm]
+\sum\limits^n_{a=1} \epsilon (\delta_a , \alpha + \lambda \delta_j)
z^{(\delta_a \mid \alpha + \lambda \delta_j)} e^{\xi_a (t,z)}
e^{-\eta_a (t,z)} \mathbb{X}^{\nu p , \sigma q} (y,w)
\tau_{\alpha + \lambda \delta_j - \delta_a - \nu \delta_p - \sigma  \delta_q}
(x,t) e^{xz}\\[3mm]
\otimes \mathbb{X}^{-a} (z) \tau_{\beta + \delta_a - \mu \delta_k}
-\sum\limits^n_{a=1} \epsilon (\delta_a , \alpha + \lambda \delta_j)
(-z)^{-(\delta_a \mid \alpha + \lambda \delta_j)}\times \\[3mm] e^{-\xi_a (t,
-z)}
e^{\eta_a (t, -z)} \mathbb{X}^{\nu p , \sigma q} (y,w)
\tau_{\alpha + \lambda \delta_j + \delta_a - \nu \delta_p - \sigma\delta_q}
(x,t) e^{xz}
\otimes \mathbb{X}^a (-z) \tau_{\beta - \delta_a - \mu \delta_k} \}\\[3mm]
+ \mathbb{X}^{\nu p} (\nu y) \tau_{\alpha + \lambda \delta_j - \nu \delta_p}
(x,t)\otimes \frac{\mathbb{X}^{\sigma q} (-\sigma w)}{w^{\delta_{q0}}}
\tau_{\beta - \mu \delta_k - \sigma \delta_q} (x,t)\\[3mm]
- \frac{\mathbb{X}^{\sigma q} (-\sigma w)}{w^{\delta_{q0}}}
\tau_{\alpha + \lambda \delta_j - \sigma \delta_q} (x,t) \otimes
\mathbb{X}^{\nu p} (\nu y) \tau_{\beta - \mu \delta_k - \nu \delta_p}
(x,t)\\[3mm]
= \frac{1}{2} \mathbb{X}^{\nu p , \sigma q} (y,w)
\tau_{\alpha + \lambda \delta_j - \nu \delta_p - \sigma \delta_q}
(x,t) \otimes \tau_{\beta - \mu \delta_k} (x,t) .
  \end{array}
\end{equation}

Now divide by $\tau_{\alpha} \otimes \tau_{\beta}$ and rewrite terms
like
\[
\begin{array}[h]{c}
\frac{\displaystyle e^{-\eta_a (t,z)} \mathbb{X}^{\nu p, \sigma q} (y,w)
\tau_{a + \lambda
    \delta_j - \delta_a - \nu \delta_p - \sigma \delta_q}
  (x,t)}{\displaystyle \tau_{\alpha} (x,t)} = \\[3mm]
e^{-\eta_a (t,z)} \left(
\frac{\displaystyle \mathbb{X}^{\nu p, \sigma q} (y,w)
\tau_{\alpha + \lambda \delta_j -
    \delta_a - \nu \delta_p - \sigma \delta_q} (x,t)}{\displaystyle \tau_{a +
    \lambda \delta_j - \delta_a} (x,t)} \right)
\frac{\displaystyle e^{-\eta_a (t,z)} \tau_{\alpha + \lambda \delta_j -
\delta_a}
  (x,t)}{\displaystyle \tau_a (x,t)} .
\end{array}
\]

Remove the tensor symbols $\otimes$ and write $x' , t'$ etc. for the
second component of the tensor product. Then (\ref{5.18}) is
equivalent to
\begin{equation}
  \label{5.19}
  \begin{array}[h]{ll}
& \mbox{Res}_{z=0} dz \{ \frac{1-\epsilon}{2z}
  \frac{\mathbb{X}^{\nu p , \sigma q} (y,w) \tau_{\alpha + \lambda \delta_j -
      \nu \delta_p -\sigma \delta_q} (x,t)}{\tau_{\alpha+\lambda
      \delta_j} (x,t)}
  \Psi^+ (\alpha, x,t,x)_{\lambda j,0}
  \Psi^- (\beta , x' , t' , z)_{\mu k,0}\\[3mm]
+& \frac{\epsilon}{2z} e^{-\eta_0 (t,z)} \left(
   \frac{\mathbb{X}^{\nu p , \sigma q} (y,w)
   \tau_{\alpha + \lambda \delta_j - \nu \delta_p - \sigma \delta_q}
   (x,t)}{\tau_{\alpha + \lambda \delta_j - \nu \delta_p} (x,t)}
   \right) \Psi^+ (\alpha , x, t, z)_{\lambda j,0}
   \Psi^- (\beta, x' , t' , z)_{\mu k,0}\\[3mm]
+& \sum\limits^n_{a=1} e^{-\eta_a (t,z)}\left(
   \frac{\mathbb{X}^{\nu p , \sigma_q} (y,w) \tau_{\alpha + \lambda \delta_j -
       \delta_a - \nu \delta_p - \sigma \delta_q} (x,t)}
   {\tau_{a + \lambda \delta_j - \delta_a} (x,t)} \right)
   \Psi^+ (\alpha, x, t, z)_{\lambda j,a}
   \Psi^- (\beta , x' , t' , z)_{\mu k,a}\\[3mm]
-& \sum\limits^n_{a=1} e^{\eta_a (t,-z)} \left(
   \frac{\mathbb{X}^{\nu p , \sigma q} (y,w) \tau_{\alpha + \lambda \delta_j +
       \delta_a - \nu \delta_p - \sigma \delta_q} (x,t)}{\tau_{\alpha
       + \lambda \delta_j + \delta_a} (x,t)}\right)
   \Psi^+ (\alpha , x, t, z,)_{\lambda j,-a}
   \Psi^- (\beta , x' , t' , z)_{\mu k,-a}\\[3mm]
+& \frac{1}{(\sqrt{2})^{\delta_{p0}}}
   \frac{1}{(\sqrt{2}w)^{\delta_{q0}}} \frac{1}{z}
   \Psi^+ (\alpha, x, t, y)_{\lambda j, \nu p}
   \Psi^- (\beta , x' , t' , w)_{\mu k, -\sigma q}\\[3mm]
-& \frac{1}{(\sqrt{2})^{\delta_{p,0}}}
   \frac{1}{(\sqrt{2} w)^{\delta_{q0}}}
   \frac{1}{z} \Psi^+ (\alpha, x, t,-w)_{\lambda j, \sigma q}
   \Psi^- (\beta, x' , t' , -y)_{\mu k , -\nu p} \}\\[3mm]
=& \mbox{Res}_{z=0} dz \frac{1}{2z}
   \frac{\mathbb{X}^{\nu p , \sigma q} (y,w) \tau_{\alpha + \lambda \delta _j
       - \nu \delta_p - \sigma \delta_q} (x,t)}{\tau_{\alpha + \lambda
       \delta_j} (x,t)}
   \frac{\tau_{\alpha + \lambda \delta_j} (x,t)}{\tau_{\alpha} (x,t)}
   e^{xz}
   \frac{\tau_{\beta - \mu \delta_k} (x',t')}{\tau_{\beta} (x',t')} e^{-x'z}.
  \end{array}
\end{equation}

Now use (\ref{5.13}) and the observation that
\[
\mbox{Res}_{z=0} dz \frac{1}{z} f(\pm u) = \mbox{Res}_{z=0} dz \pm
  \delta (u \mp z) f(z),
\]

then (\ref{5.19}) turns into

\begin{equation}
  \label{5.20}
\begin{split}
    & \mbox{Res}_{z=0} dz  E_{\lambda j, \lambda j}
\{
(
   (1-\epsilon)
   \frac{\mathbb{X}^{\nu p , \sigma q} (y,w) \tau_{\alpha + \lambda \delta_j -
       \nu \delta_p - \sigma \delta_q} (x,t)}{\tau_{\alpha + \lambda
       \delta_j} (x,t)}\\[3mm]
     +& \epsilon e^{-\eta_0 (t,z)}
(   \frac{\mathbb{X}^{\nu p , \sigma q} (y,w) \tau_{\alpha + \lambda \delta_j -
       \nu \delta_p -\sigma \delta_q} (x,t)}{\tau_{\alpha + \lambda
       \delta_j} (x,t)}
)) \Psi^+ (\alpha , x, t, z)
     \; S(z) E_{00} \;^t \Psi^- (\beta , x' , t', z)\\[3mm]
     +& \sum\limits^n_{a=1} e^{-\eta_a (t,z)}
(
   \frac{\mathbb{X}^{\nu p , \sigma q} (y,w) \tau_{\alpha + \lambda \delta_j -
       \delta_a - \nu \delta_p - \sigma \delta_q} (x,t)}{\tau_{\alpha
       + \lambda \delta_j - \delta_a} (x,t)}
)
   \Psi^+ (\alpha , x, t, z)
     \;  S(z) E_{aa} \;^t \Psi^- (\beta , x' , t' , z)\\[3mm]
     +& \sum\limits^{n}_{a=1} e^{\eta_a (t,-z)}
(
   \frac{\mathbb{X}^{\nu p , \sigma q}(y,w) \tau_{\alpha + \lambda \delta_j +
       \delta_a - \nu \delta_p - \sigma \delta_q} (x,t)}{\tau_{\alpha
       + \lambda \delta_j + \delta_a}(x,t)}
)
     \Psi^+ (\alpha, x, t, z)
     \;  S(z) E_{-a, -a} \;^t \Psi^- (\beta , x' , t' , z)\\[3mm]
     +& Z^{\nu p , \sigma q} (y,w) \Psi^+ (\alpha , x, t, z) \;^t \Psi^-
     (\beta , x' ,t' , z)
\}\\[3mm]
     =& \mbox{Res}_{z=0} dz E_{\lambda j , \lambda j}
    \frac{\mathbb{X}^{\nu p , \sigma q} (y,w) \tau_{\alpha + \lambda \delta_j -
       \nu \delta_p - \sigma \delta_q}(x,t)}{\tau_{\alpha + \lambda
       \delta_j} (x,t)}
      T^+ (\alpha  , x, t, z) \frac{E_{00}}{2z} \;^t T^- (\beta , x' , t' , z).
\end{split}
\end{equation}

Next take $\beta = \alpha$ and define

\[
\begin{array}{l}
U^{\lambda j}_{\mu a} (z) = \sum\limits^{\infty}_{k=0}
U^{\lambda j}_{\mu a} (k) z^{-k} =\\
\left\{
  \begin{array}[h]{ll}
e^{-\mu \eta_a (t,z)} \left(
\frac{\mathbb{X}^{\nu p , \sigma q}(y,w) \tau_{a + \lambda \delta_j - \mu
    \delta_a - \nu \delta_p - \sigma \delta_q} (x,t)}{\tau_{a +
    \lambda \delta_j - \mu \delta_a} (x,t)} \right)
    & \mbox{if}\;\; a\neq 0,\\[3mm]
(1-\epsilon)  \frac{\mathbb{X}^{\nu p , \sigma q} (y,w) \tau_{\alpha +
    \lambda \delta_j - \nu \delta_p - \sigma \delta_p}
  (x,t)} {\tau_{\alpha + \lambda \delta_j} (x,t)}+ &  \\[3mm]
\epsilon e^{-\eta_0 (t,z)}\left(
  \frac{\mathbb{X}^{\nu p , \sigma q} (y,w) \tau_{\alpha + \lambda \delta_j -
      \nu \delta_p - \sigma \delta_q} (x,t)} {\tau_{\alpha + \lambda
      \delta_j} (x,t)}\right) & \mbox{if}\;\; a=0.
  \end{array}
\right.
\end{array}
\]

Again we assume that $\mu = +$ when $a=0$. Now using corollary
\ref{c4.2} we deduce from (\ref{5.20}) that
\[
\begin{array}[h]{ll}
 & E_{\lambda j, \lambda j} (
   Z^{\nu p , \sigma q} (\alpha , y, w)
   P^+ (\alpha) S(\partial) P^- (\alpha)^*
+ \sum\limits^n_{a=-n} \sum\limits^{\infty}_{k=0}
   U^{\lambda j}_{a} (k) P^+ (\alpha) E_{aa}\partial^{-k} S(\partial)
   P^- (\alpha)^* )_-\\[3mm]
=& E_{\lambda j \lambda j}
   \frac{\mathbb{X}^{\nu p , \sigma q} (y,w) \tau_{a + \lambda \delta_j - \nu
       \delta_p - \sigma \delta_q} (x,t)}
       {\tau_{a + \lambda \delta_j} (x,t)}
   T^+ (\alpha) \frac{E_{00} \partial^{-1}}{2} \;^t T^- \alpha.
\end{array}
\]

Substitute (\ref{4.7}) and one obtains.

\[
\begin{array}[h]{ll}
 & E_{\lambda j, \lambda j} ( Z^{\nu p , \sigma q} (\alpha , y, w)
   T^+ (\alpha) S(\partial) \;^t T^- (\alpha) +\\[3mm]
+& \sum\limits^n_{a=-n} \sum\limits^{\infty}_{k=0}
   U^{\lambda j}_a (k) P^+ (\alpha) E_{aa} \partial^{-k}
   P^+ (\alpha)^{-1} T^+ (\alpha) S(\partial) \;^t T^- (\alpha)
   )_-\\[3mm]
=& E_{\lambda j , \lambda j}
   \frac{\mathbb{X}^{\nu p , \sigma q} (y,w) \tau_{\alpha + \lambda \delta_j -
       \nu \delta_p - \sigma \delta_q}  (x,t)}{\tau_{\alpha + \lambda
       \delta_j} (x,t)} T^+ (\alpha)
  \frac{E_{00} \partial^{-1}}{2} \;^t T^- (\alpha) .
\end{array}
\]

We can mulitply from the right with $(^t\! T (\alpha)^-)^{-1}$, then
the above formula is equivalent to
\[
\begin{array}[h]{c}
E_{\lambda j, \lambda j} T^+ (\alpha)
(Y^{\nu p , \sigma q} (\alpha , y, w)_< +
\sum\limits^n_{a=-n} \sum\limits^{\infty}_{k=0}
U^{\lambda j}_a (k) (K^+ (\alpha) E_{aa} \partial^{-k}
(K^+ (\alpha))^{-1})_<\\[3mm]
-\frac{\mathbb{X}^{\nu p , \sigma q} (y,w) \tau_{a + \lambda \delta_j - \nu
    \delta_p - \sigma \delta_q} (x,t)}{\tau_{\alpha + \lambda
    \delta_j} (x,t)} E_{00} ) =0.
\end{array}
\]

Now notice that
\[
\begin{array}[h]{lcl}
E_{00} T (\alpha)^+                  &=& E_{00},\\[2mm]
E_{\lambda j, \lambda j} T^+ (\alpha) &=& E_{\lambda j, \lambda j} +
\frac{\tau_{\alpha + \lambda \delta_j} (x,t)}{\tau_{\alpha} (x,t)}
E_{\lambda j , 0},
\end{array}
\]

{}from which one concludes that
\begin{equation}
  \label{5.21}
  \begin{array}[h]{c}
E_{\lambda j , \lambda j} (Y^{\nu p , \sigma q} (\alpha , y , w)_< +
\sum\limits^n_{a=-n} \sum\limits^{\infty}_{k=0}
U^{\lambda j}_a (k) (K^+ (\alpha) E_{aa} \partial^{-k}
(K^+ (\alpha))^{-1} )_<\\[3mm]
-\frac{\mathbb{X}^{\nu p , \sigma q} (y,w) \tau_{\alpha + \lambda \delta_j -
\nu
  \delta_p - \sigma \delta_q} (x,t)}{\tau_{\alpha + \lambda \delta_j} (x,t)}
E_{00}) =0.
  \end{array}
\end{equation}

Finally, act with the left-hand-side of formula (\ref{5.21}) on
$\Phi^+ (\alpha, x, t, z)$ and one obtains the OSASM-formula.

\begin{theorem}
  \label{t5.1}
Let $0\leq j \leq n,\ \lambda = +, -$ and $\lambda = +$ when $j=0$,
moreover assume that $\delta_0 = 0$. Then one has the following
OSASM-formula for the BKP hierarchy:
\[
\begin{array}[h]{c}
E_{\lambda j, \lambda j} Y^{\nu p , \sigma q} (\alpha , y , w)_< \Phi^+
(\alpha , x, t, z) =\\[3mm]
- E_{\lambda j , \lambda j} ((1+\epsilon (e^{-\eta_0 (t,z)} -1))
(\frac{\mathbb{X}^{\nu p, \sigma q} (y,w) \tau_{\alpha + \lambda\delta_j - \nu
    \delta_p - \sigma \delta_q} (x,t)}{\tau_{\alpha + \lambda
    \delta_j} (x,t)}) \Phi^+ (\alpha , x, t, z) E_{00}\\[3mm]
-\frac{\mathbb{X}^{\nu p, \sigma q} (y,w) \tau_{\alpha + \lambda\delta_j - \nu
    \delta_p - \sigma \delta_q} (x,t)}{\tau_{\alpha + \lambda
    \delta_j} (x,t)} E_{00} \Phi^+(\alpha , x , t, z)) +\\[3mm]
+ \sum\limits^n_{a=1} (e^{-\eta_a (t,z)} (
\frac{\mathbb{X}^{\nu p, \sigma q} (y,w) \tau_{\alpha + \lambda\delta_j
-\delta_a - \nu
    \delta_p - \sigma \delta_q} (x,t)}{\tau_{\alpha + \lambda
    \delta_j -\delta_a} (x,t)}) \Phi^+ (\alpha , x, t, z)
E_{aa}\\[3mm]
- \frac{\mathbb{X}^{\nu p, \sigma q} (y,w) \tau_{\alpha + \lambda\delta_j
    -\delta_a - \nu
    \delta_p - \sigma \delta_q} (x,t)}{\tau_{\alpha + \lambda
    \delta_j - \delta_a} (x,t)}) E_{aa} \Phi^+ (\alpha , x, t,
z))\\[3mm]
+ \sum\limits^n_{a=1} (e^{\eta_a (t, -z)}
(\frac{\mathbb{X}^{\nu p, \sigma q} (y,w) \tau_{\alpha + \lambda\delta_j +
    \delta_a - \nu
    \delta_p - \sigma \delta_q} (x,t)}{\tau_{\alpha + \lambda
    \delta_j + \delta_a} (x,t)}) \Phi^+ (\alpha, x, t, z) E_{-a ,
  -a}\\[3mm]
-\frac{\mathbb{X}^{\nu p, \sigma q} (y,w) \tau_{\alpha + \lambda\delta_j +
\delta_a - \nu
    \delta_p - \sigma \delta_q} (x,t)}{\tau_{\alpha + \lambda
    \delta_j + \delta_a} (x,t)} E_{-a, -a} \Phi^+ (\alpha , x, t, z)).
\end{array}
\]
\end{theorem}
\ \\
\noindent {\Large \bf  Acknowledgements}

\ \\
Part of this research was done while JvdL was financially supported by
the ``Stichting Fundamenteel Onderzoek der Materie (F.O.M.). During
that time he also visited MIT. He would like to thank the Mathematics
Department of MIT for the kind hospitality.

\end{document}